\documentclass{nature}
\usepackage{graphicx}
\usepackage{caption}
\usepackage{bm} 
\usepackage{verbatim}
\usepackage{subfigure}
\usepackage{longtable}  
\usepackage{amssymb}
\usepackage{amsmath}
\usepackage{float}
\usepackage{txfonts}
\usepackage{hyperref}
\usepackage{upgreek}
\usepackage{lineno}
\usepackage[super,comma,sort&compress]{natbib}
\usepackage{color}

\newcommand\ionpat[2]{#1$\;${\scshape{#2}}}

\def\aj{\rmfamily{AJ~}}           
\def\apj{\rmfamily{ApJ~}}         
\def\apjl{\rmfamily{ApJ~}}        
\def\apjs{\rmfamily{ApJS~}}       

\def\aap{\rmfamily{A\&A~}}        

\def\araa{\rmfamily{ARA\&A~}}     
\def\mnras{\rmfamily{MNRAS~}}     
\def\pasp{\rmfamily{PASP~}}       
\def\pasj{\rmfamily{PASJ~}}       
\def\nat{\rmfamily{Nature~}}      
\def\ao{\rmfamily{Appl.~Opt.~}}   

\def\ssr{\rmfamily{Space~Sci.~Rev.~}}

\def\temp{11,750--14,750\,K}
\def\hostz{1.49}

\def\draft{false}

\def\hostz{1.49}

\def\sbalongarc{$F125W\approx25$\,mag\,arcsec$^{-2}$}

\def\velnearcausticsim{1000\,km\,s$^{-1}$}

\def\iclarcsecdensity{$6.3^{+10.3}_{-3.3}\times10^6\,{\rm M}_{\odot}\,{\rm kpc}^{-2}$}
\def\salpeterarcsecdensity{$1.9^{+0.6}_{-0.6}\times10^7\,{\rm M}_{\odot}\,{\rm kpc}^{-2}$}
\def\chabrierarcsecdensity{$1.1^{+0.3}_{-0.3}\times10^7\,{\rm M}_{\odot}\,{\rm kpc}^{-2}$}

\def\clusterarcsecdensity{$2\times10^9\,{\rm M}_{\odot}\,{\rm kpc}^{-2}$}

\def\bresolinlf{$\alpha=-2.53\pm0.08$}

\def\brighterpeak{$\sim4$}
\def\hostz{$1.49$}
\def\temp{11,000--14,000\,K}

\def\perdixiapyxoffset{$0.10''$}
\def\perdixdate{3 January 2017}
\def\iapyxdate{30 October 2016}

\def\brighterpeak{$\sim4$}

\def\lumdensity{$120\,{\rm L}_{\odot}$\,pc$^{-2}$}

\def\numobsaprsixteen{$N_{\rm obs}=37$}
\def\numobsaprseventeen{$N_{\rm obs}=50$}
\def\sepresolved{0.06\,pc}

\def\persistentmag{27.7\,mag}
\def\brightmlmag{26\,mag}

\def\kcorrectionVJ{$-1.10$\,mag}
\def\lsoneoffset{$7.9\pm0.5$\,kpc}

\def\abslumconstraint{$M_V = -9.0 \pm 0.75\,{\rm (sys)}$} 

\def\simbrightness{26.5\,{\rm mag}}

\def\captionmaysixteenpeak{Flux at LS1's position during Lev16A after subtracting flux present in 2011 imaging. Fluxes are measured from difference images created by subtracting exposures acquired in 2016 from template images taken in 2011. The zeropoint of listed fluxes is 25 AB, and no correction for Galactic extinction is applied.}

\def\captionunderlyingknot{Photometry of LS1 measured from HFF imaging (2013--2014), and archival near-UV {\it HST} imaging. The zeropoint of listed fluxes is 25 AB, and a correction for Galactic extinction is applied.}

\def\captionadjacentarc{Photometry of underlying lensed arc adjacent to LS1. The zeropoint of listed fluxes is 25 AB, and fluxes are corrected for Galactic extinction.}

\def\captionphottab{Photometry at locations of LS1/Lev 2016, Lev 2016, and Lev 2017 of {\it HST} imaging acquired 2004--2017. The zeropoint of listed fluxes is 25 AB, and no correction for Galactic extinction is applied. Values in brackets in LS1/Lev 2016 are estimates of star's WFC3 $F125W$ flux converted using the star's SED. For LS1/Lev16A, fluxes are the sum of flux measured from deep coaddition and that measured from a difference image created by subtracting each image from the deep coaddition. Fluxes at the positions of Lev16B and Lev 2017A are measured from difference imaging.}

\def\captiontab{Comparison between model light curves and observed 2004-2017 light curves for LS1 / Lev16A and Lev16B.
Measured $\langle \chi^2 \rangle$ statistics provide evidence about the binary fraction and IMF of the stellar population responsible for the intracluster light (ICL), distinguish among stellar evolution and supernova models \cite{woosleyhegerweaver02,fryerbelczynskiwiktorowicz12,speramapellibressan15}, and disfavor the possibility that 1--3\% of dark matter consists of 30\,M$_{\odot}$ PBHs. 
To interpret differences in $\langle\chi^2\rangle$ values, we fit simulated light curves, and compute the difference $\Delta$$\langle\chi^2\rangle$ values between the $\langle\chi^2\rangle$ values of the generative (``true'') model and of the best-fitting model.
For 68\% of simulated light curves, $\Delta$$\langle\chi^2\rangle \lesssim 13$, and for 95\% of simulated light curves, $\Delta$$\langle\chi^2\rangle \lesssim 25$.
These simulations assume that our estimates for the galaxy cluster's magnification and stellar-mass density are correct.
The $\Sigma$ column denotes whether light curves were computed with a low (L) or high (H) stellar-mass density. The ``Type'' column specified whether the stars are single (S), or have the mass-dependent binary fraction and mass ratios of stars determined at low redshift (B) \cite{duchenekraus13}. The ``IMF'' column specifies whether a Chabrier (``Cha") or Salpeter (``Sal'') IMF was used to assemble the ICL stellar population.  
}

\title{Extreme magnification of a star at redshift 1.5 by a galaxy-cluster lens}

\author{Patrick L. Kelly$^{1,2}$, Jose M. Diego$^{3}$, Steven Rodney$^{4}$, Nick Kaiser$^{5}$, Tom Broadhurst$^{6,7}$, Adi Zitrin$^{8}$, Tommaso Treu$^{9}$, Pablo G. P\'erez-Gonz\'alez$^{10}$, Takahiro Morishita$^{9,11,12}$, Mathilde Jauzac$^{13,14,15}$, Jonatan Selsing$^{16}$, Masamune Oguri$^{17,18,19}$, Laurent Pueyo$^{20}$, Timothy W. Ross$^{1}$, Alexei V. Filippenko$^{1,21}$, Nathan Smith$^{22}$, Jens Hjorth$^{16}$, S. Bradley Cenko$^{23,24}$, Xin Wang$^{9}$, D. Andrew Howell$^{25,26}$, Johan Richard$^{27}$, Brenda L. Frye$^{22}$, Saurabh W. Jha$^{28}$, Ryan J. Foley$^{29}$, Colin Norman$^{30}$, Marusa Bradac$^{31}$, Weikang Zheng$^{1}$, Gabriel Brammer$^{20}$, Alberto Molino Benito$^{32}$, Antonio Cava$^{33}$, Lise Christensen$^{16}$, Selma E. de Mink$^{34}$, Or Graur$^{35,36,37}$, Claudio Grillo$^{38,16}$, Ryota Kawamata$^{39}$, Jean-Paul Kneib$^{40}$, Thomas Matheson$^{41}$, Curtis McCully$^{25,26}$, Mario Nonino$^{42}$, Ismael Perez-Fournon$^{43,44}$, Adam G. Riess$^{30,20}$, Piero Rosati$^{45}$, Kasper Borello Schmidt$^{46}$, Keren Sharon$^{47}$, \& Benjamin J. Weiner$^{22}$}

\begin{document}
\spacing{2}

\maketitle

\begin{affiliations}
\item Department of Astronomy, University of California, Berkeley, CA 94720-3411, USA
\item School of Physics and Astronomy, University of Minnesota, 116 Church Street SE, Minneapolis, MN 55455, USA
\item IFCA, Instituto de F\'isica de Cantabria (UC-CSIC), Av. de Los Castros s/n, 39005 Santander, Spain
\item Department of Physics and Astronomy, University of South Carolina, 712 Main St., Columbia, SC 29208, USA
\item Institute for Astronomy, University of Hawaii, 2680 Woodlawn Drive, Honolulu, HI 96822-1839, USA
\item Department of Theoretical Physics, University of the Basque Country, Bilbao 48080, Spain
\item IKERBASQUE, Basque Foundation for Science, Alameda Urquijo, 36-5 48008 Bilbao, Spain
\item Physics Department, Ben-Gurion University of the Negev, P.O. Box 653, Beer-Sheva 8410501, Israel
\item Department of Physics and Astronomy, University of California, Los Angeles, CA 90095
\item Departamento de Astrof\'isica, Facultad de critical curve. F\'isicas, Universidad Complutense de Madrid, E-28040 Madrid, Spain
\item Astronomical Institute, Tohoku University, Aramaki, Aoba, Sendai 980-8578, Japan
\item Institute for International Advanced Research and Education, Tohoku University, Aramaki, Aoba, Sendai 980-8578, Japan
\item Centre for Extragalactic Astronomy, Department of Physics, Durham University, Durham DH1 3LE, U.K.
\item Institute for Computational Cosmology, Durham University, South Road, Durham DH1 3LE, U.K.
\item Astrophysics and Cosmology Research Unit, School of Mathematical Sciences, University of KwaZulu-Natal, Durban 4041, South Africa
\item Dark Cosmology Centre, Niels Bohr Institute, University of Copenhagen, Juliane Maries Vej 30, DK-2100 Copenhagen, Denmark 
\item Research Center for the Early Universe, University of Tokyo, Tokyo 113-0033, Japan
\item Department of Physics, University of Tokyo, 7-3-1 Hongo, Bunkyo-ku, Tokyo 113-0033, Japan
\item Kavli Institute for the Physics and Mathematics of the Universe (Kavli IPMU, WPI), University of Tokyo, 5-1-5 Kashiwanoha, Kashiwa, Chiba 277-8583, Japan
\item Space Telescope Science Institute, 3700 San Martin Dr., Baltimore, MD 21218, USA
\item Miller Senior Fellow, Miller Institute for Basic Research in Science, University of California, Berkeley, CA  94720, USA
\item Steward Observatory, University of Arizona, 933 N. Cherry Ave., Tucson, AZ 85721, USA
\item Astrophysics Science Division, NASA Goddard Space Flight Center, Mail Code 661, Greenbelt, MD 20771, USA
\item Joint Space-Science Institute, University of Maryland, College Park, MD 20742, USA
\item Las Cumbres Observatory, 6740 Cortona Dr., Suite 102, Goleta, CA 93117, USA
\item Department of Physics, University of California, Santa Barbara, CA 93106-9530, USA
\item Univ Lyon, Univ Lyon1, ENS de Lyon, CNRS, Centre de Recherche Astrophysique de Lyon UMR5574, F-69230, Saint-Genis-Laval, France
\item Department of Physics and Astronomy, Rutgers, The State University of New Jersey, Piscataway, NJ 08854, USA
\item Department of Astronomy and Astrophysics, UCO/Lick Observatory, University of California, 1156 High Street, Santa Cruz, CA 95064, USA
\item Department of Physics and Astronomy, The Johns Hopkins University, 3400 N. Charles St., Baltimore, MD 21218, USA
\item Department of Physics, University of California, Davis, 1 Shields Avenue, Davis, CA 95616, USA
\item Instituto de Astronomia, Geof\'isica e Ci\^encias Atmosf\'ericas, Universidade de S\~ao Paulo, 05508-090, S\~ao Paulo, Brazil
\item Department of Astronomy, University of Geneva, 51, Ch. des Maillettes, CH-1290 Versoix, Switzerland
\item Anton Pannekoek Institute for Astronomy, University of Amsterdam, NL-1090 GE Amsterdam, the Netherlands
\item Harvard-Smithsonian Center for Astrophysics, 60 Garden Street, Cambridge, MA 02138
\item Department of Astrophysics, American Museum of Natural History, Central Park West and 79th Street, New York, NY 10024, USA
\item NSF Astronomy and Astrophysics Postdoctoral Fellow
\item Dipartimento di Fisica, Universit\'a degli Studi di Milano, via Celoria 16, I-20133 Milano, Italy
\item Department of Astronomy, Graduate School of Science, The University of Tokyo, 7-3-1 Hongo, Bunkyo-ku, Tokyo 113-0033, Japan
\item Laboratoire d'Astrophysique, Ecole Polytechnique Federale de Lausanne (EPFL), Observatoire de Sauverny, CH-1290 Versoix, Switzerland
\item National Optical Astronomical Observatory, Tucson, AZ 85719, USA
\item INAF, Osservatorio Astronomico di Trieste, via Bazzoni 2, 34124 Trieste, Italy
\item Instituto de Astrofisica de Canarias (IAC), E-38205 La Laguna, Tenerife, Spain
\item Universidad de La Laguna, Dpto. Astrofisica, E-38206 La Laguna, Tenerife, Spain
\item Dipartimento di Fisica e Scienze della Terra, Universit\'a degli Studi di Ferrara, via Saragat 1, I-44122, Ferrara, Italy
\item Leibniz-Institut fur Astrophysik Potsdam (AIP), An der Sternwarte 16, 14482 Potsdam, Germany
\item University of Michigan, Department of Astronomy, 1085 South University Avenue, Ann Arbor, MI 48109-1107, USA
\end{affiliations}

\begin{abstract}
Galaxy-cluster gravitational lenses can magnify background galaxies by a total factor of up to $\sim50$. Here we report an image of an individual star at redshift $z=1.49$ (dubbed ``MACS\,J1149 Lensed Star 1 (LS1)'') magnified by $>2000$. A separate image, detected briefly 0.26$''$ from LS1, is likely a counterimage of the first star demagnified for multiple years by a $\gtrsim3$\,M$_{\odot}$ object in the cluster. For reasonable assumptions about the lensing system, microlensing fluctuations in the stars' light curves can yield evidence about the mass function of intracluster stars and compact objects, including binary fractions and specific stellar evolution and supernova models. Dark-matter subhalos or massive compact objects may help to account for the two images' long-term brightness ratio.
\end{abstract}

The pattern of magnification arising from a foreground strong gravitational lens changes with distance behind it. 
At each specific distance behind the lens, the locations that are most highly magnified are connected 
by a so-called caustic curve. 
Near the caustic curve in the source plane, magnification changes rapidly.
Over a distance of only tens of parsecs close to the MACS\,J1149 galaxy cluster's caustic at $z=1.5$, for example, magnification falls from a maximum of $\sim5000$ to only $\sim50$. 
Since the sizes of even compact galaxies are hundreds of parsecs, their total magnifications cannot exceed $\sim50$.

However, a well-aligned individual star adjacent to the caustic of a galaxy cluster could, in theory, become 
magnified by a factor of many thousands \cite{miraldaescude91}. 
When a galaxy cluster's caustic curve is mapped from the source plane defined at a specific redshift to the image plane on the sky, it is called the critical curve. 
Consequently, a highly magnified star should be found close to the foreground galaxy cluster's critical curve.

\section{A Lensed Blue Supergiant at Redshift $z=1.49$}

In {\it Hubble Space Telescope (HST)} Wide Field Camera 3 (WFC3) infrared (IR) imaging taken on 29 April 2016 to construct light curves of the multiple images of Supernova (SN)~Refsdal \cite{kellyrodneytreu15,rodneystrolgerkelly16,kellyrodneytreu16,oguri15,sharonjohnson15,diegobroadhurstchen16,jauzacrichardlimousin16,grillokarmansuyu16,kawamataoguriishigaki16,treubrammerdiego16}, we detected an unexpected change in flux of an individual point source (dubbed ``MACS\,J1149 Lensed Star 1 (LS1)'') in the MACS\,J1149 galaxy-cluster field \cite{ebelingbarrettdonovan07}.
As shown in Fig.~\ref{fig:m1149}, the unresolved blue source lies close to the cluster's critical curve at its host galaxy's redshift of $z=1.49$\cite{oguri15,sharonjohnson15,diegobroadhurstchen16,jauzacrichardlimousin16,grillokarmansuyu16,kawamataoguriishigaki16,treubrammerdiego16}. 
Fig.~\ref{fig:critcurveunc} shows that, while the location of the critical curve differs by $\sim0.25''$ among lens models, the blue source is no farther than $\sim0.13''$ from the critical curves of all publicly available, high-resolution models. 

The MACS\,J1149 galaxy cluster lens creates two partial, merging images of LS1's host galaxy separated by the cluster critical curve, as well as an additional full image. 
As shown in Extended Data Fig.~\ref{fig:fullimageloc}, LS1's predicted position inside the third, full image is near the tip of a spiral arm. 
According to our lens model, LS1 is \lsoneoffset\ from the nucleus of the host galaxy.
The multiply imaged SN~Refsdal exploded at a different position in the same galaxy \cite{smithebelinglimousin09,zitrinbroadhurst09,yuankewleyswinbank11,karmangrillobalestra16}.

At the peak of the microlensing event in May 2016 (Lensing Event ``Lev16A''), LS1 was a factor of $\sim4$ times brighter than it appeared in archival {\it HST} imaging during 2013--2015. 
Fig.~\ref{fig:sed} shows that the additional flux we measured at LS1's position has a spectral energy distribution (SED) statistically consistent with the source's SED during 2013--2015. As shown in Fig.~\ref{fig:sed}, model spectra of mid-to-late B-type stars at $z=1.49$ with photospheric temperatures of \temp\ \cite{castellikurucz04} provide a good match to the SED of LS1 ($\chi^2=12.9$ for 6 degrees of freedom; $\chi^2_\nu=2.15$), given that our model does not account for changes in magnification between the epochs when observations in separate filters were obtained. SED fitting finds probability peaks at $\sim8$ and $\sim35$\,Myr (see Extended Data Fig.~\ref{fig:sed_adjacent_region}) for the age of the arc underlying LS1's position.

A lensed luminous star provides a perhaps unexpected explanation (and yet the only reasonable one we could find) for the transient's variable light curve and unchanging SED.
Except for finite-source effects, gravitational lensing will magnify a star's emission at all wavelengths equally.
Therefore, as we observe for LS1, the SED of a lensed background star should remain the same, even as it appears brighter or fainter owing to changes in its magnification.
By contrast, the SEDs of stellar outbursts and supernovae change as they brighten by the factor of $\sim4$ observed in May 2016. 

As shown in Fig.~\ref{fig:sed}, LS1's SED exhibits a strong Balmer break, which indicates that the lensed object has a relatively  high surface gravity.  Stars, including blue supergiants, exhibit spectra with a strong Balmer break, but stellar outbursts and explosions have low surface gravity and lack a strong Balmer break.  The temperature of \temp\ inferred from fitting the Balmer break 
is also substantially larger than that of almost all H-rich transients during outburst such as luminous blue variables (LBVs).
While Lyman absorption of a background active galactic nucleus (AGN) at $z\approx9$ could potentially produce a continuum break 
at $\sim9500$\,\AA, the AGN's flux blueward of the break would be almost entirely absorbed, and additional images would be expected. 

Our ray-tracing simulations, which are described in detail in Ref.~\citenum{diegokaiserbroadhurst17},  show that the MACS\,J1149 galaxy cluster's gravitational potential effectively increases the Einstein radii of individual stars in the intracluster medium by a factor of $\sim100$ along the line of sight to LS1.  Consequently, even though intracluster stars account for $\lesssim1$\% of the cluster's mass along the line of sight to LS1, overlapping caustics arising from intracluster stars should densely cover the source plane of the host galaxy at $z=1.49$, as demonstrated by our simulation plotted in Extended Data Fig.~\ref{fig:close_to_caustic}. By contrast, Galactic microlensing magnification can be fully modeled using the caustic of a single star or stellar system.

The ray-tracing simulations show that a star at LS1's location should experience multiple microlensing events over a period of a decade with typical magnifications of $10^3$--$10^4$. 
In Fig.~\ref{fig:lightcurve}, we display the 2004--2017 light curve of LS1 constructed from all optical and IR {\it HST} observations of the field, and we show ray-tracing simulations that can describe LS1's light curve.

\section{A Separate Microlensing Event at a Different Position}
A foreground gravitational lens made of smoothly distributed matter should form a pair of images of a static background source at equal angular offsets from the critical curve.
However, only a single, persistent point source is apparent near the critical curve in {\it HST} imaging taken during the period 2004--2017. We initially considered the possibility that LS1 happens to be sufficiently close to the galaxy cluster's caustic that its pair of images have a small angular separation
unresolved in {\it HST} data.
As we continued to monitor the MACS\,J1149 cluster field, however, we detected an unexpected new source (``Lev16B'') on \iapyxdate\ 
offset by $0.26''$ from LS1. 
We measure magnitudes of $F125W=25.78\pm0.12$\,AB ($\lambda_{\rm pivot} = 1.25$\,$\upmu$m) and $F160W=26.16\pm0.22$\,AB ($\lambda_{\rm pivot} = 1.54$\,$\upmu$m). The $F125W-F160W$ color (which corresponds approximately to rest-frame $V-R$) of the new source is consistent with that of LS1, which has $F125W-F160W=-0.11\pm0.10$\,mag\,AB. 

\begin{figure}
\centering
\includegraphics[draft=\draft,angle=0,width=6.5in]{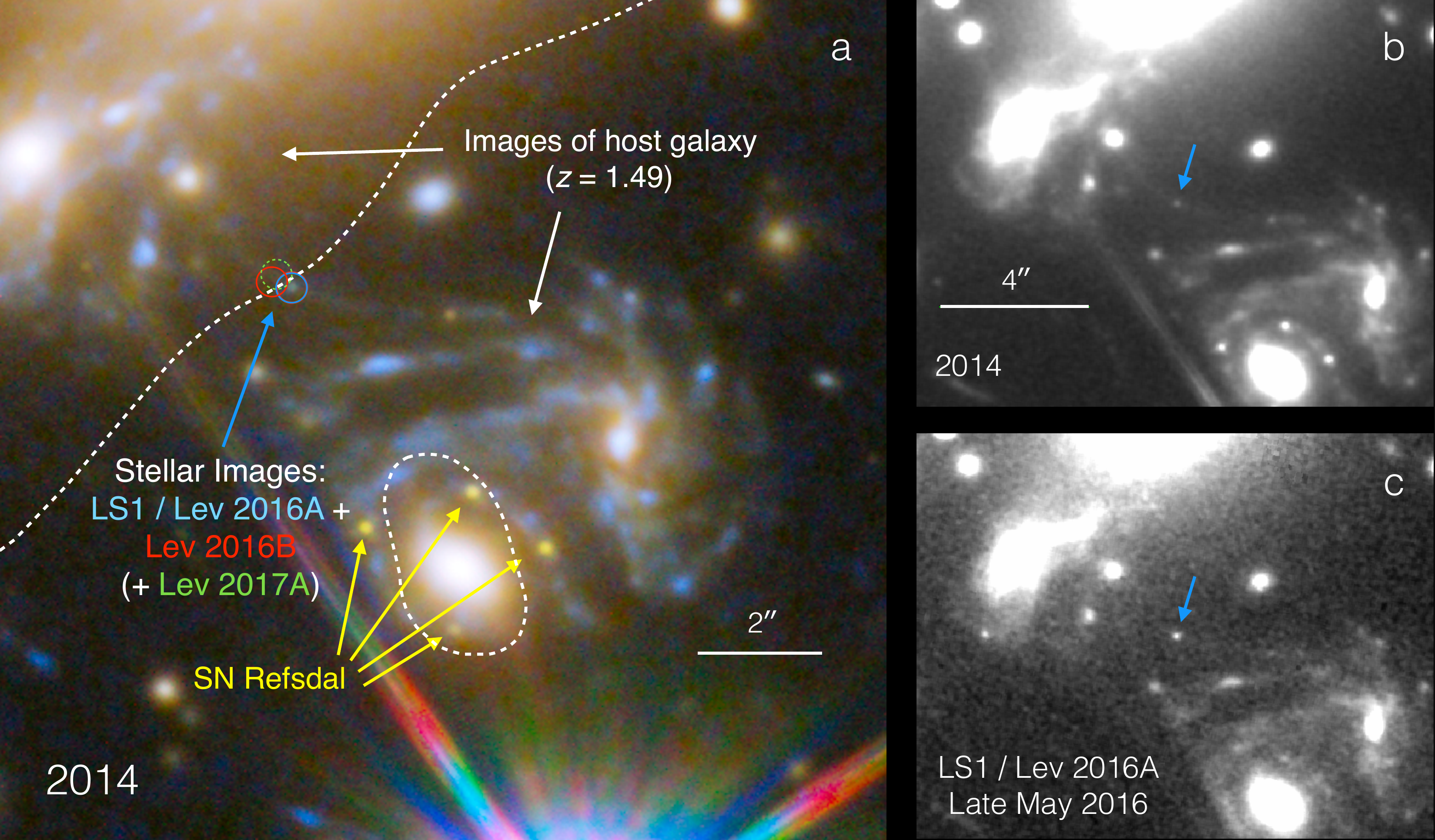}
\caption{Locations of lensing events coinciding with background spiral galaxy near the MACS\,J1149 galaxy cluster's critical curve.
Left panel shows the positions of magnified images of stars LS1 / Lev16A and Lev16B as well as candidate Lev 2017A close to ($\lesssim0.13''$) the critical curve, where magnification rises rapidly. Dashed line shows the location of the critical curve from the CATS cluster model \cite{jauzacrichardlimousin16}. The Einstein Cross formed from yellow point sources consists of images of SN~Refsdal \cite{kellyrodneytreu15}.  Right panel shows the field in 2014, and in May 2016 when LS1 exhibited a microlensing peak.} 
\label{fig:m1149}
\end{figure}

\begin{figure}
\centering
\includegraphics[draft=\draft,angle=0,width=3.25in]{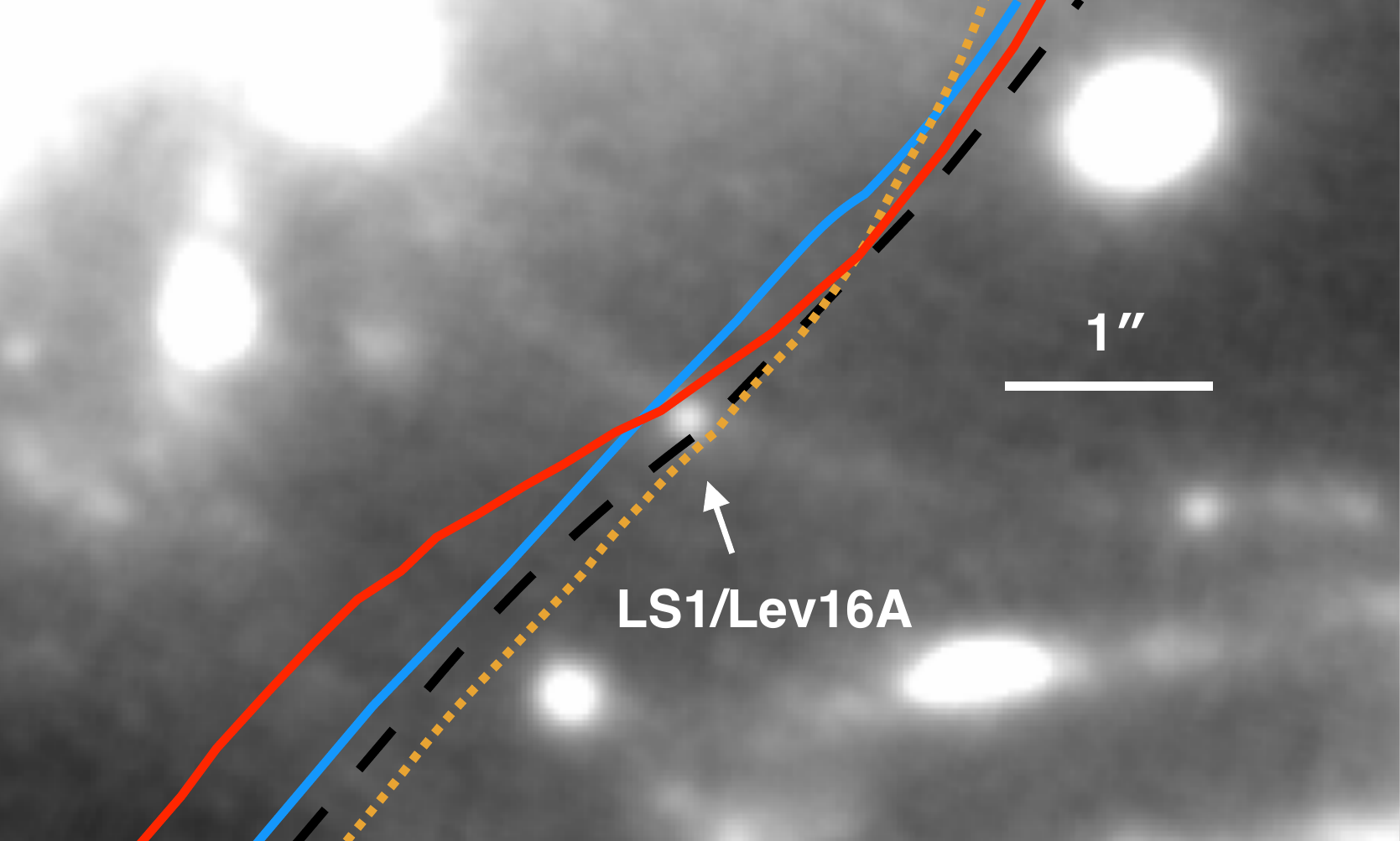}
\caption{Proximity of LS1/Lev16A to the MACS\,J1149 galaxy-cluster critical curve for multiple galaxy-cluster lens models. Critical curves for models with available high-resolution lens maps including Ref.\,\citenum{jauzacrichardlimousin16} (CATS; solid red), Ref.\,\citenum{zitrinfabrismerten15} (short-dash orange), Ref.\,\citenum{kawamataoguriishigaki16} (solid blue), and Ref.\,\citenum{keeton10} (long-dash black) are superposed on the {\it HST} WFC3-IR F125W image.
Although predictions for the location of the critical curve near LS1 disagree by $\sim0.25''$, LS1 lies within $\sim0.13''$ of all of these models' predictions.
\label{fig:critcurveunc}}
\end{figure}

\begin{figure}
\centering
\includegraphics[draft=\draft,angle=0,width=6.5in]{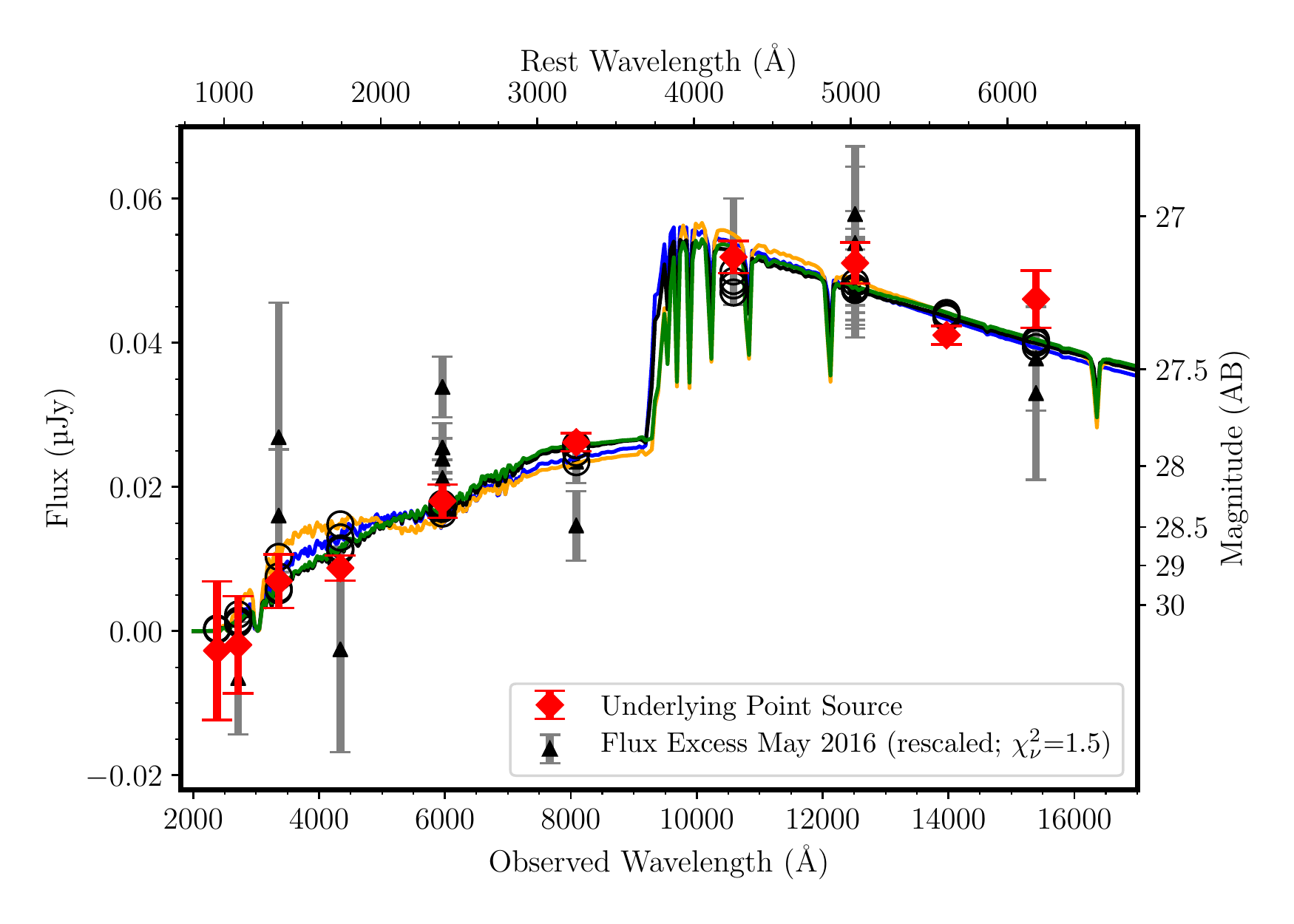} 
\caption{The SEDs of LS1 measured in 2013--2015 (red) and of the rescaled, excess flux density at LS1's position close to its May 2016 peak (Lev16A; black) are consistent.
Rescaling the SED of the flux excess to match to that of the 2013--2015 source yields $\chi^2_\nu=1.5$, indicating that they are statistically consistent with each other despite a flux density difference of a factor of \brighterpeak.
The SED shows a strong Balmer break consistent with the host-galaxy redshift of \hostz, and stellar atmosphere models \cite{castellikurucz04} of a mid-to-late B-type star provide a reasonable fit. 
The blue curve has $T=11,180$\,K, $\log g=2$, $A_V=0.02$, and $\chi^2 = 16.3$; the orange curve has $T=12,250$\,K, $\log g=4$, $A_V=0.08$, and $\chi^2 = 30.6$; the black curve has $T=12,375$\,K, $\log g=2$, $A_V=0.08$, and $\chi^2 = 12.9$; and the green curve has $T=13,591$\,K, $\log g=4$, $A_V=0.13$, and $\chi^2 = 16.5$.
Black circles show the expected flux density for each model. 
\label{fig:sed}}
\end{figure}

We consider that the new source could either be the counterimage of LS1, or a different lensed star. 
As can be seen in Fig.~\ref{fig:m1149}, the pair of images of LS1's host galaxy that meet at the critical curve appear flipped relative to each other. 
These images are said to have opposite parity, a property of lensed images set by the sign of the determinant of the lensing magnification matrix. 
Assuming they are counterimages, Lev16B and LS1/Lev16A would have negative and positive parity, respectively. 

We have found from our ray-tracing simulations that the parity of an image of a lensed background star strongly affects its microlensing variations\cite{diegokaiserbroadhurst17}.
Extended Data Fig.~\ref{fig:close_to_caustic} shows that, while an image of a background star on LS1/Lev16A's side of the critical curve always has magnification of $\gtrsim300$, its counterimage on Lev16B's side has extensive regions of much lower magnification ($\sim30$) in the source plane. 
If LS1 fell in such a low-magnification region on Lev16B's side for much of the period 2004--2017, that could explain why LS1/Lev16B was not detected except on 30 October 2016, as shown in Extended Data Fig.~\ref{fig:lightcurves_Lev_2016B_Lev_2017A}. 
A $\gtrsim3$\,M$_{\odot}$ object, such as a stellar binary system, or a neutron star or black hole, can cause an image of a star to have low magnification for sufficiently long periods on Lev16B's side of the critical curve. 

\section{Properties of Lensed Star 1}
If we assume that LS1/Lev16A and Lev16B are counterimages, 
then our model predicts each has an average magnification of 600. Different cluster models, however, show a factor of $\sim2$ disagreement about the magnification at LS1's position\cite{oguri15,sharonjohnson15,diegobroadhurstchen16,jauzacrichardlimousin16,grillokarmansuyu16,kawamataoguriishigaki16}. 
LS1 had {\it F125W}\ $\approx28.15$\,mag in 2004--2008, corresponding to an absolute magnitude of \abslumconstraint\ for a magnification of $\sim600$ per image. 

Post-main-sequence stars in the Small Magellanic Cloud (SMC) that have $U-B$ and $B-V$ colors approximately matching those of LS1 ($-0.40$ and $-0.05$\,mag, respectively) have luminosities reaching $M_V\gtrsim-8.8$\,mag \cite{dachs70}. 
The two statistically significant peaks in May 2016 could correspond to a projected separation for a binary star system of $\sim25$\,AU for a transverse velocity of 1000\,km\,s$^{-1}$ (see Methods). 

If LS1 instead consists of an unresolved pair of images, then the lensed star would need to have an offset of $\lesssim0.06$\,pc of the caustic curve to be unresolved in {\it HST} imaging (see Methods). Its total magnification would be $\sim10000$, corresponding to a star with $M_V\approx-6$\,mag. 

\section{Monte Carlo Simulation of Stellar Population Near Galaxy Cluster's Caustic}
We next perform simulations that allow us to estimate the probabilities (a) that LS1/Lev16A and Lev16B are counterimages of each other, and (b) of discovering a lensed star in {\it HST} galaxy-cluster observations.
We use measurements of the arc underlying LS1's position to estimate the number and luminosities of stars near the galaxy cluster's caustic.
For different potential stellar luminosity functions, we calculate the number of 
expected bright lensed stars and microlensing events.

The underlying arc extends for $\sim0.2''$ ($\sim340$\,pc in the source plane) along the galaxy cluster's critical curve. 
If LS1/Lev16A and Lev16B are counterimages, then the lensed star is offset from the caustic by 2.2\,pc in the source plane according to our lensing model.
In our simulation, we populate the source plane region within 0.4$''$ of the critical curve, or 21.9\,pc from the caustic, with stars.

We first need to infer the total luminosity in stars in the $21.9\,{\rm pc} \times 340$\,pc region adjacent to the galaxy cluster's caustic.
Gravitational lensing conserves the surface brightness, and we use the arc's \sbalongarc\  surface brightness to compute its absolute rest-frame $V$ surface brightness, which yields an estimate for the luminosity density of \lumdensity.

The next step is to place stars in the $21.9\,{\rm pc} \times 340$\,pc region adjacent to the caustic (within 0.4$''$ of the critical curve), whose area of 7100 pc$^2$ should enclose a total luminosity of $8.5 \times 10^5\,{\rm L}_{\odot}$.
We consider power-law luminosity functions where the number of stars with luminosity between $L$ and $L+dL$ is proportional to $L^{-\alpha}\,dL$. 
For luminosity functions with $-1.5 \leq \alpha \leq 3$, we normalize the
luminosity function so that the integrated luminosity equals $8.5 \times 10^5\,{\rm L}_{\odot}$ and compute the expected number of stars in each 0.1\,L$_{\odot}$ interval. 
We draw from a Poisson distribution to determine the number of stars in each luminosity bin, and assign each star a random position within 21.9\,pc of the caustic.

We next compute the average magnification $\bar{\mu}$ of each star. 
For a lens consisting of only smooth matter, the predicted magnification at an offset $R$ in parsecs from the caustic is $\bar{\mu} = 880\,/ \sqrt{R}$.
Our ray-tracing simulations find that the average magnification deviates from this prediction closer than $\sim1.3$\,pc from the caustic curve (0.1$''$ from the critical curve) due to microlensing.
To estimate $\bar{\mu}$ for stars closer than $0.10''$ to the critical curve, we interpolate in the image plane between $\bar{\mu} = 5000$ at the critical curve and $\bar{\mu} = 680$
at an offset of $0.10''$.

Our next step is to estimate the number of bright microlensing peaks ($F125W\leq$\brightmlmag) we expect to find in existing {\it HST} observations of the MACS\,J1149 galaxy-cluster field. 
LS1 is expected to have a transverse velocity of order 1000\,km\,s$^{-1}$ relative to the cluster lens (see Methods), which corresponds approximately to LS1/Lev16A's two-week peak duration\cite{miraldaescude91}. 
If {\it HST} observations taken within a period of 10\,days are counted as a single observation, then there were \numobsaprseventeen\ observations of the MACS\,J1149 field
in all optical and IR wide-band filters through 13 April 2017, and \numobsaprsixteen\ observations through 15 April 2016 just before the detection of LS1.

After taking into account stellar microlensing, the fraction of the source plane where the magnification exceeds $\mu$ is (Kaiser et al., in preparation)
\begin{equation}
f_S(\mu,\bar{\mu})\approx2.5\times10^{-4}\bigg( \frac{\kappa}{3\times10^{-3}}\bigg) \bigg(\frac{\bar{\mu}}{500}\bigg)^3\bigg(\frac{\mu}{10^4}\bigg)^{-2},
\label{eqn:ml}
\end{equation}
where $\mu$ is the total amplification, $\kappa$ is the surface density of stars making up the intracluster light (ICL) in units of the critical density, and $\bar{\mu}$ is the expected magnification if the cluster consisted entirely of smoothly distributed matter. 
Eq.~\ref{eqn:ml} does not apply at offsets smaller than $\sim1.3$\,pc from the cluster caustic where the optical depth for microlensing exceeds unity.  Our ray-tracing simulations indicate, however, that the formula should provide a reasonable first-order approximation at smaller distances from the caustic when we use our estimate of the average magnification $\bar{\mu}$ near the critical curve \cite{diegokaiserbroadhurst17}.
The number of expected microlensing events with magnification exceeding $\mu$ for each star will be $N_{\rm obs} \times f_S(\mu,\bar{\mu})$.

The simulations provide support for the hypothesis that Lev16A and Lev16B are counterimages of LS1.
Extended Data Fig.~\ref{fig:fractiondetectedeventsstar} shows that, 
if a star has an average apparent $F125W$ brightness of at least \persistentmag\ similar to LS1, then 
it will be responsible for $\gtrsim99$\% of $F125W\leq$\brightmlmag\ events.
Likewise, Extended Data Fig.~\ref{fig:lumprob} shows that observing a bright lensed star sufficiently close to the caustic that its images are unresolved ($\lesssim0.06$\,pc from the caustic) is a factor of ten less probable than observing a resolved pair of bright images of a lensed star.

In nearby galaxies, the bright end of the luminosity function has a power law index of $\alpha \approx 2.5$. 
Young star-forming regions such as 30 Doradus, however, can have shallower functions where $\alpha \approx 2$.
Extended Data Fig.~\ref{fig:lumprob} suggests that the probability of observing
a persistent bright lensed star ($F125W\leq$\persistentmag) in the underlying arc may be $\sim10$\% in existing {\it HST} observations, given a shallow stellar luminosity function
where $\alpha\approx2$.
For the steeper mean luminosity function ($\alpha\approx2.5$) measured for nearby galaxies \cite{bresolinkennicuttferrarese98}, we find a probability of 0.01--0.1\%.
The probability of observing at least one bright ($F125W\leq$\brightmlmag) microlensing event is 
$\sim3$\% for  $\alpha\approx2$, and $\sim0.1$\% for $\alpha\approx2.5$.
We have repeated our simulation using the distribution
of stars in 30 Doradus in the Large Magellanic Cloud (LMC), which yields similar probabilities as for the case where $\alpha\approx2$.

To estimate to first order the probability of finding a lensed star in all existing {\it HST} galaxy-cluster observations, we make the simplifying assumption that all strong lensing arcs have properties similar to that underlying LS1.
Of the total time used to image cluster fields with {\it HST}, only at most $\sim10$\% has been used to observe MACS\,J1149. Each of several dozen galaxy cluster fields monitored by {\it HST} contains $\sim4$ giant arcs \cite{xupostmanmeneghetti16}.
Consequently, to take into account all {\it HST} galaxy-cluster observations,
we need to multiply our above Monte Carlo probabilities by an approximate factor of $10 \times 4 = 40$. 
This suggests that the probability of finding a lensed star may be reasonable, but only if the 
average stellar luminosity function at high redshift is shallower than $\alpha \approx 2.5$.

We note that we detected a new potential source (Lev 2017A) which has a $\sim4\sigma$ significance in the WFC3-IR imaging acquired on \perdixdate, although the significance is only $\sim2.5\sigma$ considering all {\it HST} imaging and the independent apertures adjacent to the critical curve. 

\section{Multiple Limits on the Physical Size of Lensed Star 1}
Each bright microlensing peak must correspond to light from an individual star in the source plane, given the small area of high magnification adjacent to the microcaustics of intracluster stars. Additional considerations provide evidence that the persistent source, LS1, is too compact to be a typical stellar cluster, and is instead a single stellar system (e.g., an individual star or a binary). If  LS1 consists of two unresolved counterimages at the location of the critical curve, then 
LS1 must be more compact than $\sim0.06$\,pc given the upper limit on its angular size ($\lesssim 0.040''$; see Methods).

If Lev16A and Lev16B are instead mutual counterimages, the limit on LS1's angular size constrains it to have a physical dimension perpendicular to the caustic of $\lesssim1$--$2$\,pc, which is significantly smaller than the typical size of a stellar cluster.

The absence of a persistent image at Lev16B's position places a stronger potential limit on LS1's size.
To explain the lack of a peristent counterimage at Lev16B's location, all stars in a hypothetical stellar association at LS1's position would need to fall in a region of low magnification on the Lev16B side of the critical curve. Ray-tracing simulations indicate that LS1 would need to be smaller than $\sim0.1$\,pc. A hypothetical dark matter (DM) subhalo, however, could also potentially demagnify a counterimage at Lev16B's position.

\section{Inferences from LS1/Lev16A and Lev16B Assuming They Are Mutual Counterimages}
We next compare the {\it HST} light curve for LS1/Lev16A and Lev16B with simulated light curves created for different assumptions about intracluster stellar population and the abundance of 30\,M$_{\odot}$ primordial black holes (PBHs). In our ray-tracing simulations, LS1/Lev16A and Lev16B are counterimages with average magnifications from the cluster of 600.

We assume (a) all intracluster light (ICL) stars are single, or (b) apply mass-dependent binary fractions and mass ratios  \cite{duchenekraus13}. We use a stellar-mass density of \iclarcsecdensity\ for a Chabrier initial mass function (IMF), or higher densities estimated in an improved analysis of \chabrierarcsecdensity\ and \salpeterarcsecdensity\ for Chabrier and Salpeter IMFs, respectively. 
The most massive star that is still living found in the intracluster medium (ICM) at $z=0.54$ is assumed to have $M=1.5$\,M$_{\odot}$.
In Extended Data Figs.~\ref{fig:binarysims}, \ref{fig:imfs}, and \ref{fig:addingpbhs}, we plot the simulated light curves for an $R=100\,{\rm R}_{\odot}$ lensed star 
where we adopt the ``Woosley02'' \cite{woosleyhegerweaver02}, ``Fryer12'' \cite{fryerbelczynskiwiktorowicz12}, or ``Spera15'' \cite{speramapellibressan15} models of stellar evolution and core-collapse physics. 

For steps of 50\,km\,s$^{-1}$ in the range 100--2000\,km\,s$^{-1}$, we stretch the simulated light curves and identify the regions that best match the data.
Table~\ref{tab:chisq} lists the average $\chi^2$ value for the 150 best matches $\langle\chi^2\rangle$.
To interpret differences in $\langle\chi^2\rangle$ values, we fit simulated light curves, and compute the difference $\Delta$$\langle\chi^2\rangle$ values between the $\langle\chi^2\rangle$ values of the generative (``true'') model and of the best-fitting model.
For 68\% of simulated light curves, $\Delta$$\langle\chi^2\rangle \lesssim 13$, and for 95\% of simulated light curves, $\Delta$$\langle\chi^2\rangle \lesssim 25$.

For stars with $-7.5 > M_V > -9.5$\,mag, a range consistent with the most luminous stars in the SMC and LMC given the uncertainty in magnification, 
models constructed using a prescription for the binary fraction \cite{duchenekraus13}
are favored over those where all stars are single (see Extended Data Fig.~\ref{fig:singlebinary}). The $\langle\chi^2\rangle$ statistics also favor the Fryer12 stellar model, and a Salpeter IMF over a Chabrier IMF (see Methods).  The fitting also provides evidence against models where 1\% and 3\% of DM consists of 30\,M$_{\odot}$ PBHs \cite{birdcholismunoz16}.
Within the confidence intervals, the differences remain robust when extending the upper $M_V$ limit to $-10.5$ mag.

Table~\ref{tab:chisq} also shows $\langle\chi^2\rangle$ values if we restrict the absolute magnitude to $-7.5<M_V<-8.5$ (for $\mu = 600$), although such a low luminosity would be difficult to reconcile with LS1's light curve. The Fryer12 model and a Salpeter IMF are still favored, but there is no preference for the binary prescription.

Although our confidence intervals assume that our estimates for the stellar-mass density and magnification are correct,
it may be reasonable to assume that differences in $\langle\chi^2\rangle$ values will be robust to modest errors in these parameters.
Our cluster model also does not include DM subhalos, which could affect the average fluxes of the images. 
Although our fits do not favor models where 30\,M$_{\odot}$ PBHs account for 1\% or 3\% of DM,
PBHs consisting of $\gtrsim 3$\% of DM could produce a slowly varying average magnification, and potentially explain the absence of flux at Lev16B's position.

\noindent\textbf{Supplementary Information} is linked to the online version of the paper at www.nature.com/nature.

\noindent\textbf{Acknowledgements} 
We express our appreciation to the Directors of the Space Telescope Science Institute, the Gemini Observatory, the GTC, and the European Southern Observatory for granting us discretionary time. We thank B. Katz, D. Kushnir, B. Periello, I. Momcheva, T. Royale, L. Strolger, D. Coe, J. Lotz, M. L. Graham, R. Humphreys, R. Kurucz, A. Dolphin, M. Kriek, S. Rajendran, T. Davis, I. Hubeny, C. Leitherer, F. Nieva, D. Kasen, J. Mauerhan, D. Kelson, J. M. Silverman, A. Oscoz Abaz, and Z. Levay for help with the observations or other assistance. 
The Keck Observatory was made possible with the
support of the W.M. Keck Foundation. 
NASA/STScI grants 14041, 14199, 14208, 14528, 14872, and 14922 provided financial support.
P.L.K., A.V.F., and W.Z. are grateful for assistance from the Christopher R. Redlich Fund, the TABASGO Foundation, and the Miller Institute for Basic Research in Science (U.C. Berkeley).
The work of A.V.F. was completed in part at the Aspen Center for Physics, which is supported by NSF grant PHY-1607611. 
J.M.D. acknowledges support of projects AYA2015-64508-P (MINECO/FEDER, UE), AYA2012-39475-C02-01, and the consolider project CSD2010-00064 funded by the Ministerio de Economia y Competitividad. 
P.G.P.-G. acknowledges support from Spanish Government MINECO AYA2015-70815-ERC and AYA2015-63650-P Grants. M.O. is supported by JSPS KAKENHI Grant Numbers 26800093 and 15H05892. 
M.J. acknowledges support by the Science and Technology Facilities Council [grant ST/L00075X/1].
R.J.F is supported by NSF grant AST-1518052 and Sloan and Packard Foundation Fellowships.
M.N. acknowledges support from PRIN-INAF-2014 1.05.01.94.02.
O.G. was supported by NSF Fellowship under award AST-1602595.
J.H. acknowledges support from a VILLUM FONDEN Investigator Grant [number 16599].
{\it HST} imaging was obtained at \url{https://archive.stsci.edu}.

\noindent\textbf{Author Contributions} 
P.K. planned and analyzed observations, wrote the manuscript, and developed simulations. P.K., S.R., P.G.P-G., T.M., M.J., J.S., A.V.F., J.H., M.G., D.H., B.L.F., M.B., W.Z., G.B, A.M.B., A.C., L.C., C.G., J-P.K., T.M., C.M., M.N., I.P.-F., A.G.R., P.R., K.B.S, and B.J.W. obtained follow-up {\it HST} and ground-based imaging. J.M.D. developed microlensing simulations. S.R., T.B., A.Z., T.T., P.G.P-G., M.J., M.O., A.V.F., N.S., J.H., B.L.F., and S.E.d.M. helped to prepare the manuscript. N.K. interpreted the microlensing events and derived analytic rate formula. T.B., A.Z., T.T., M.J., M.O., X.W., S.W.J., R.J.F., S.E.d.M., O.G., and B.J.W. aided the interpretation. P.G.P-G. modeled the arc's SED. T.M. modeled the ICL. M.O., J.R., R.K., and K.S. modeled the galaxy cluster. L.P., and C.N. considered the possibility that Icarus could exhibit diffraction effects. T.W.R. analyzed the microlensing simulations. N.S. aided interpretation of the star's SED. X.W. estimated the gas-phase metallicity at Icarus' location. M.N. extracted photometry of Icarus using a complementary pipeline.

\noindent\textbf{Author Information} Data used for this publication may be retrieved from the NASA Mikulski Archive for Space Telescopes (http://archive.stsci.edu). Reprints and permissions information is available at www.nature.com/reprints/. The authors declare no competing financial interests. Correspondence should be addressed to P.K. (patrick.l.kelly@gmail.com).

\clearpage

\spacing{1}

\begin{figure}
\centering
\includegraphics[angle=0,width=6.5in]{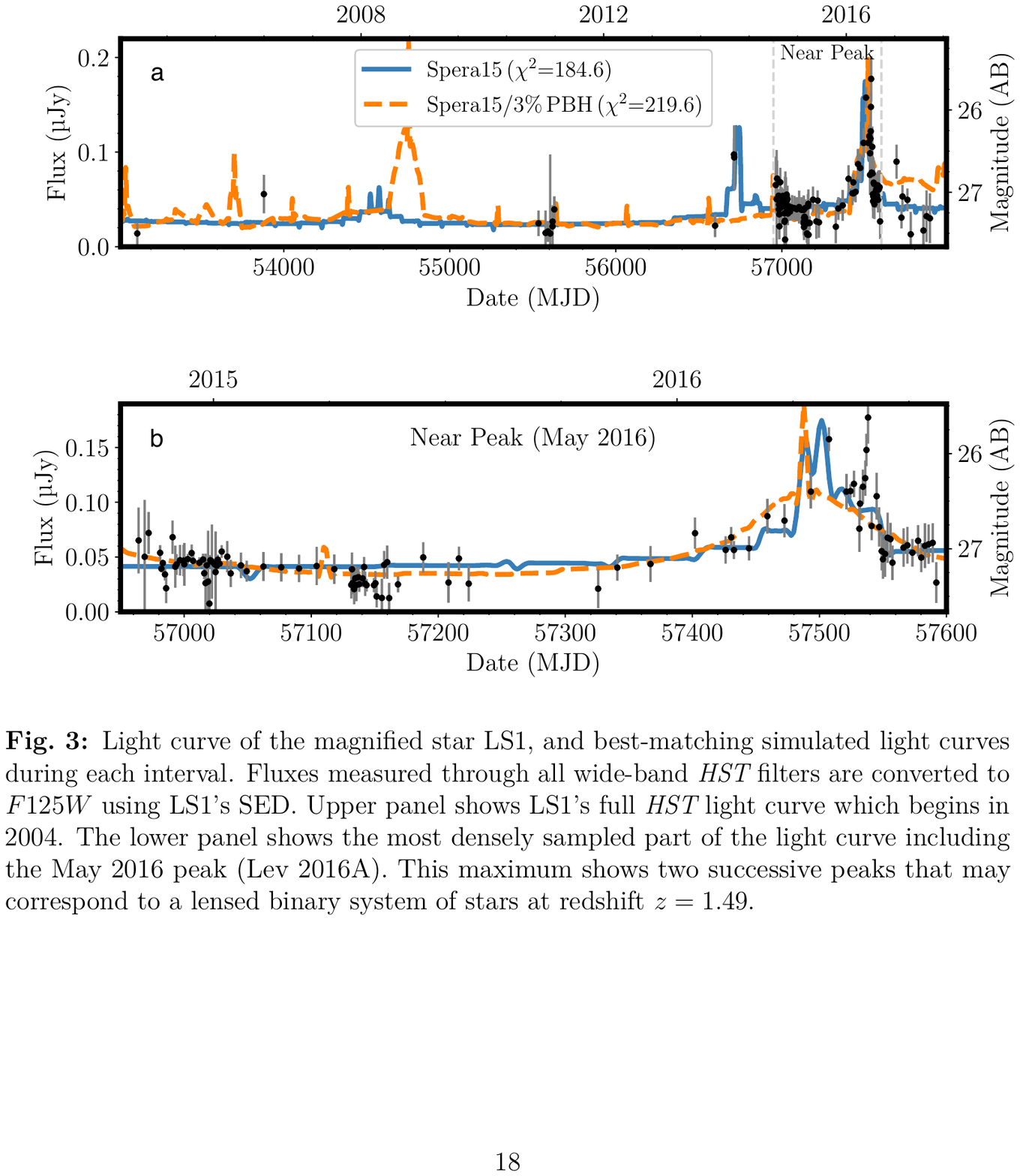}
\caption{Light curve of the magnified star LS1, and best-matching simulated light curves during each interval. Fluxes measured through all wide-band {\it HST} filters are converted to $F125W$ using LS1's SED. 
The upper panel shows LS1's full {\it HST} light curve which begins in 2004. 
The lower panel shows the most densely sampled part of the light curve including the May 2016 peak (Lev16A). This maximum shows two successive peaks that may correspond to a lensed binary system of stars at $z=1.49$.
\label{fig:lightcurve}}
\end{figure}

\begin{figure}
\centering
\includegraphics[draft=\draft,angle=0,width=6.25in]{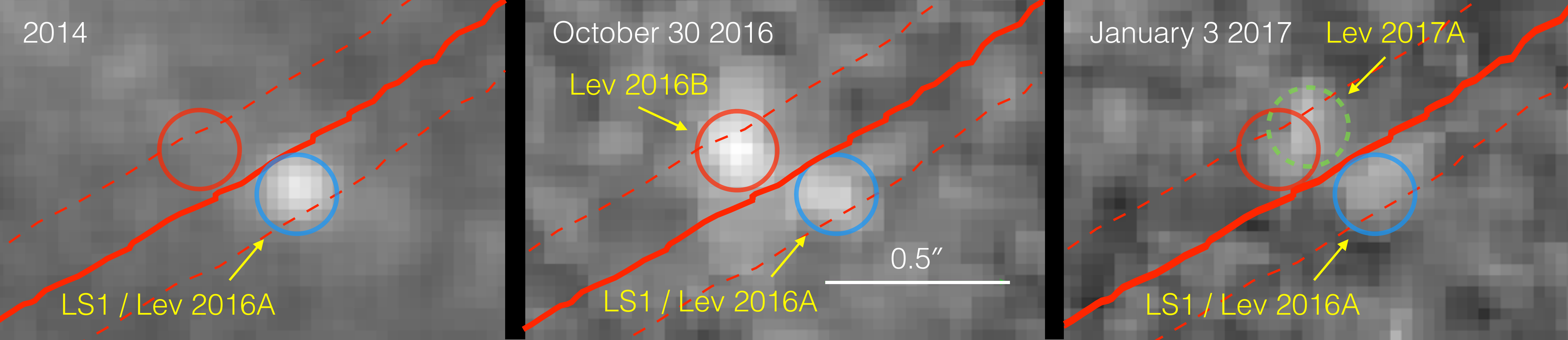}
\caption{Highly magnified stellar images located near the MACS\,J1149 galaxy cluster's critical curve. 
The left panel shows LS1 in 2014; we detected LS1 when it temporarily brightened by a factor of $\sim4$ in late-April 2016. 
The center panel shows the appearance of a new image dubbed Lev16B on 30 October 2016. 
The solid red curve marks the location of the cluster's critical curve from the CATS cluster model \citep{jauzacrichardlimousin16}, and the dashed lines show the approximate $1\sigma$ uncertainty from comparison of multiple cluster lens models \citep{oguri15,sharonjohnson15,diegobroadhurstchen16,jauzacrichardlimousin16,grillokarmansuyu16,kawamataoguriishigaki16}.
The position is  consistent with the possibility that it is a counterimage of LS1. 
The right panel shows the candidate named Lev 2017A with $\sim4\sigma$ significance detected on \perdixdate.
If a microlensing peak, Lev 2017A must correspond to a different star.
}
\label{fig:transient}
\end{figure}

\begin{table}
\centering
\footnotesize
\begin{tabular}{c|rllrrr|rllrrr}
\hline
&\multicolumn{6}{c|}{$-7.50 > M_V > -9.50$}&\multicolumn{6}{c}{$-7.50 > M_V > -8.50$}\\
&\multicolumn{6}{c|}{150 Best Matches (406\,yr)}&\multicolumn{6}{c}{150 Best Matches (406\,yr)}\\
&\multicolumn{1}{c}{$\langle$$\chi^2$$\rangle$}&\multicolumn{1}{c}{$\Sigma$}&\multicolumn{1}{c}{Model}&\multicolumn{1}{c}{PBH}&\multicolumn{1}{c}{T}&\multicolumn{1}{c|}{IMF}&\multicolumn{1}{c}{$\langle$$\chi^2$$\rangle$}&\multicolumn{1}{c}{$\Sigma$}&\multicolumn{1}{c}{Model}&\multicolumn{1}{c}{PBH}&\multicolumn{1}{c}{T}&\multicolumn{1}{c}{IMF}\\
\hline
&\multicolumn{12}{c}{Low Stellar-Mass Density}\\
{\bf Best}&356.0&L&Fryer12&&B&Cha&416.3&L&Fryer12&&B&Cha\\
&366.5&L&Woosley02&&B&Cha&462.1&L&Spera15&&S&Cha\\
&372.4&L&Spera15&&B&Cha&464.0&L&Woosley02&&S&Cha\\
&383.6&L&Woosley02&&S&Cha&488.9&L&Spera15&3\%&S&Cha\\
&392.4&L&Spera15&&S&Cha&516.5&L&Woosley02&&B&Cha\\
&403.0&L&Fryer12&&S&Cha&534.0&L&Spera15&&B&Cha\\
&406.8&L&Spera15&1\%&S&Cha&560.6&L&Fryer12&&S&Cha\\
{\bf Worst}&462.4&L&Spera15&3\%&S&Cha&567.2&L&Spera15&1\%&S&Cha\\
\hline
&\multicolumn{12}{c}{High Stellar-Mass Density}\\
{\bf Best}&347.4&H&Spera15&&B&Sal&412.1&H&Spera15&&B&Sal\\
{\bf Worst}&367.8&H&Spera15&&B&Cha&508.3&H&Spera15&&B&Cha\\
\hline
\end{tabular}
\caption{\captiontab}
\label{tab:chisq}
\end{table}

\spacing{2}

\clearpage

\section*{Methods:}
\subparagraph{{\it HST} Imaging.}
The {\it HST} observations include those from GO programs (and Principal Investigators) 12065 (M. Postman), 13459 (T.T.), 13504 (J. Lotz), 13790 and 14208 (S.R.), and 14041, 14199, 14528, 14872, and 14922 (P.K.).

\subparagraph{Constructing Light Curve.}
All optical and IR {\it HST} imaging of the MACS\,J1149 field with moderate depth has yielded a detection of LS1.
For each instrument and wide-band filter combination, we constructed a light curve for LS1.
We first measured LS1's flux in a deep template coaddition of {\it Hubble} Frontier Fields 
and early SN~Refsdal follow-up imaging.
The {\it Hubble} Frontier Fields program \cite{lotzkoekemoercoe17} acquired deep imaging of the MACS\,J1149 galaxy cluster 
between November 2013 and May 2015 in the ACS WFC $F435W$ ($\lambda_{\rm c} = 0.43$\,$\upmu$m), $F606W$ ($\lambda_{\rm c} = 0.59$\,$\upmu$m), and $F814W$ ($\lambda_{\rm c} = 0.81$\,$\upmu$m), and the WFC3 IR $F105W$ ($\lambda_{\rm pivot} = 1.05$\,$\upmu$m), $F125W$ ($\lambda_{\rm pivot} = 1.25$\,$\upmu$m), $F140W$ ($\lambda_{\rm pivot} = 1.39$\,$\upmu$m), and $F160W$ ($\lambda_{\rm pivot} = 1.54$\,$\upmu$m) wide-band filters.
The second step was to measure the differences in LS1's flux between the deep template coadded image and at each imaging epoch. We accomplished this latter step by subtracting the deep template coaddition from coadditions of imaging at each epoch, and measuring the change in LS1's flux from these resulting difference images.

To measure LS1's flux in each deep template coaddition, we first fit and then subtracted the ICL surrounding the BCG. We next measured the flux at LS1's position inside an aperture radius of $r=0.10$$''$ using the {\tt PythonPhot} package \cite{jonesscolnicrodney15}. To measure the uncertainty in the background from the underlying arc, we placed a series of four nonoverlapping apertures having $r=0.10$$''$ along it. We use the standard deviation of these aperture fluxes as an estimate of the uncertainty in the background. Aperture corrections were calculated from coadditions of the standard stars P330E and G191B2B \cite{rodneypatelscolnic15}. 

For each {\it HST} visit, we created a coadded image of all exposures acquired in each wide-band filter. 
We next subtracted the deep template coaddition from the visit coaddition to create a difference image.
Using the {\tt PythonPhot} package \cite{jonesscolnicrodney15}, we measured the flux inside of an $r=0.10$$''$ circular aperture
in the difference image.
We finally computed the total flux at each epoch by adding the flux measured
from the deep template coaddition and that measured from each difference image.

There is no source apparent at Lev16B's position in deep template coadditions.  Therefore, we do not add any flux measured from the
deep template coaddition to the light curve we construct for Lev16B, which is plotted in Extended Data Fig.~\ref{fig:lightcurves_Lev_2016B_Lev_2017A}.

\subparagraph{Estimating LS1's Color.}
LS1's brightness changed between the epochs when the deep template imaging was acquired by the {\it Hubble} Frontier Fields and SN~Refsdal follow-up programs.
However, the MACS\,J1149 cluster field was monitored using $F125W$ (and $F160W$) with a cadence of $\sim2$\,weeks after the discovery of SN~Refsdal in November 2014 \cite{kellyrodneytreu15}. 
We measured LS1's $F125W$ light curve from these data, and performed a fit to the light curve using using a third-order polynomial. 

We used the polynomial to estimate LS1's $F125W$ flux at the average epoch when the 
deep template imaging in each detector and filter was acquired. 
The ratio between the $F125W$ flux and that in the other filter (e.g., $F140W$)
provides a measurement of LS1's color ($F140W-F125W$, in this example). 
We restricted the {\it Hubble} Frontier Fields imaging to that taken between November 2014 and May 2015, since 
monitoring of SN~Refsdal is available to construct LS1's IR light curve beginning in November 2014.

\subparagraph{Creating a Combined Light Curve.}
We used our estimates of LS1's color to convert the light curves measured in
all available optical and IR filters to $F125W$ light curves, and combined them.
We also binned all $F125W$ observations 
to construct the combined light curve plotted in Fig.~\ref{fig:lightcurve} and used
to fit models.

\subparagraph{Constraints on the Age of Stellar Population in Underlying Arc}
To compute models using Flexible Stellar Population Synthesis (FSPS) \cite{conroy09,conroygunn10}, we adopt a simple stellar population with an instantaneous burst of star formation, and
include nebular and continuum emission. 
We use Python bindings (http://dan.iel.fm/python-fsps/current/) to FSPS to calculate simple stellar populations
with a Kroupa IMF \cite{kro01} and a Cardelli extinction law\cite{cardelli89} with $R_V=3.1$. 
These models use the Padova isochrones \cite{marigogirardi07,marigogirardibressan08}.

A recent analysis finds a solar oxygen abundance of 12 + log(O/H) $=8.69\pm0.05$\,dex
and a solar metallicity\cite{asplund09} of $Z=0.0134$. 
We therefore calculate models using $\log(Z/{\rm Z}_{\odot})=-0.35$, which corresponds to $Z=0.006$ and is best matched with the FSPS parameter ${\rm zmet}=15$.

To estimate the age and dust extinction of the adjacent stellar population along the arc, we 
use {\tt emcee} \cite{foremanmackeyhogglang13}, which is an implementation of a
Markov Chain Monte Carlo (MCMC) ensemble sampler. We adopt a uniform prior on the stellar 
population age from 0 to 3\,Gyr, and a uniform prior on the extinction $A_V$ from 0 to 2\,mag.

\subparagraph{Ability to Detect Pair of Images of a Star Adjacent to Cluster Caustic.} A possibility is that we do not observe a pair of images of LS1, because it 
is very close to the cluster caustic and the available {\it HST} imaging is
not able to resolve its two images. 
Here we calculate how close the star must be to the caustic. 
If LS1 lay very close to the critical curve, then Lev16B would correpond to the 
microlensing event of a different star, at an offset of $0.26''$ from the cluster critical curve.

For a star sufficiently close to the caustic, the pair of images will not be resolved with {\it HST}.
Our simulations suggest that demagnifying one of the two images will be unlikely
when the star is close to the cluster caustic, although they do not include expected dark-matter subhalos. The angular resolution of {\it HST} is greatest in the $F606W$ band ($\lambda_{\rm c} = 0.59$\,$\upmu$m) and almost as sharp in the $F814W$ band ($\lambda_{\rm c} = 0.81$\,$\upmu$m), and observations in these wide-band filters provide the best opportunity to test  
whether LS1 consists of two adjacent images. 
We use coadditions of imaging taken by the {\it Hubble} Frontier Fields program.
To determine the limit we can place on the separation of two possible images, we inject pairs of point sources having the same combined magnitude as the images made using the ACS WFC $F606W$ and ACS WFC $F814W$ exposures at each epoch.
 
The full width at half-maximum intensity (FWHM) of our ACS point-spread-function (PSF) models constructed from observations of the standard stars P330E and G191B2B  only agree within 10\% with the measured FWHM of the stars in our coadded images. 
Consequently, the measured FWHM of the injected pairs of PSFs should not correspond directly to what we 
would we measure in the ACS data.
Therefore, we compute the fractional increase in the measured FWHM with the increasing separation between the 
pair of injected PSFs. Next, we multiply the FWHM estimated from the stars in each image by this factor to compute limits from the ACS imaging.

We inject 100 fake stars with the same magnitude using models of the ACS WFC $F606W$ and $F814W$ PSF. 
After injecting the point sources in a grid, we use the IRAF task {\tt imexam} to estimate the
Gaussian FWHM (GFWHM) using the ``comma'' command from the simulated
data. The {\tt imexam} model we use has a Gaussian profile, and we specify three radius adjustments 
while the fit is optimized. Pixels are fit within 3\,pix ($0.03''$ pix$^{-1}$) of the center, and we use a background buffer of 5\,pix. 

These simulations show that any separation between the
two images greater than 0.035--$0.040''$ can be detected $>$$95$\% of the
time. The upper limit implies that the star would have to be closer
than \sepresolved\ to the cluster caustic. 
As we show in Extended Data Fig.~\ref{fig:lumprob}, the relative probability of a persistently bright ($F125W<$ \persistentmag)  
star being located within \sepresolved\ is $\lesssim10$\%.

DOLPHOT \cite{dolphin00} fit parameters for the images of LS1, Lev16B, and Lev 2017A fall in the range expected for point sources in the DOLPHOT reference manual \url{http://americano.dolphinsim.com/dolphot/dolphot.pdf}, although these criteria are not highly sensitive to a pair of images.

\subparagraph{Transverse Velocity of Star.}
We use the following expression [Equation~12 of Ref. \citenum{miraldaescude91}] for the apparent transverse velocity of a lensed source:
\begin{equation}
v_{\bot} =  \left| \frac{v_s - v_o}{(1 + z_s)} - \frac{D_s (v_l - v_0)}{D_l (1 + z_l)} \right|,
\end{equation}
where $D_l$ and $D_s$ are the angular-diameter distances of the lens and source, and $v_0$, $v_l$, and $v_s$ are (respectively) the transverse velocities of the observer, the lens, and the source with respect to the caustic. The expression only applies for a universe without spatial curvature.

Cosmological simulations have found that merging galaxy-cluster halos and subhalos have pairwise velocities of $\sim500$--1500\,km\,s$^{-1}$ with tails to lower and higher velocity \cite{watsonilievdaloisio13}.
Given the expected velocity of the lens, the peculiar velocities of Earth ($\sim400$\,km\,s$^{-1}$) and of the host galaxy relative to the Hubble flow, and 
the motion of the star ($<200$--300\,km\,s$^{-1}$) relative to its host galaxy, a typical transverse velocity should be 1000\,km\,s$^{-1}$. 
In our light-curve fitting analysis, we consider transverse velocities of 100--2000\,km\,s$^{-1}$.

\subparagraph{Intrinsic Luminosity of Lensed Star.} 
While extremely luminous stars are rare in the nearby universe, they require smaller magnification and can be at a greater distance from the caustic.
For a lens with a smooth distribution of matter, the magnification $\mu$ falls within the distance $d$ from the caustic as $\mu \propto 1/\sqrt{d}$. Therefore, the area $A$ in the source plane in which the magnification is greater than $\mu$ scales as $A(>\mu) \propto 1/\mu^2$. The observed flux $F$ of a lensed object is $F \propto L\mu$, where the object's luminosity is $L$.  Therefore, a star with luminosity $L$ appears brighter than $f$ inside an area $A(>f;L) \propto L^2$. 

\subparagraph{Galaxy-Cluster Lens Model.}
Prior to the identification of LS1 in late-April 2016, the cluster potential had been modeled using the codes  LTM \cite{zitrinbroadhurstumetsu09,zitrinfabrismerten15}, {\tt {\tt WSLAP+}} \cite{diegobroadhurstchen16}, {\tt GLAFIC} \cite{oguri10,oguri15,kawamataoguriishigaki16}, {\tt LENSTOOL} \cite{jullokneiblimousin07,sharonjohnson15,jauzacrichardlimousin16}, and {\tt GLEE} \cite{suyuhalkola10,suyuhenselmckean12,grillokarmansuyu16}.
These used several different sets of multiply imaged galaxies \cite{treubrammerdiego16}, which included new data from the Grism-Lensed Survey from Space (GLASS; PI Treu) \cite{schmidttreubrammer14,treuschmidtbrammer15}, MUSE (PI Grillo) \cite{grillokarmansuyu16}, and grism follow-up observations of SN~Refsdal (PI Kelly) \cite{kellybrammerselsing16}.

In Fig.~\ref{fig:transient}, we plot as an example the position of the critical curve from the CATS model \cite{jauzacrichardlimousin16} created using LENSTOOL \cite{jullokneiblimousin07}, showing that it passes close to the position of LS1. The critical curves of all of these models, however, pass within similarly small offsets from LS1's coordinates. 

For the simulation of the light curves of a star passing close to the cluster caustic,
we use the {\tt WSLAP+} model of the cluster mass distribution \cite{diegobroadhurstchen16}
and draw stars randomly from a Chabrier IMF \cite{cha03} until the stellar mass density
equals the value we estimate of \chabrierarcsecdensity\ for a Chabrier IMF and \salpeterarcsecdensity\ for a Salpeter IMF.

The {\tt WSLAP+} cluster lens model, which includes only smoothly distributed matter and cluster galaxies, yields several important relations describing the magnification near the critical curve, and the relationship between lensed $\theta$ and unlensed $\beta$ angles. 
The magnification $\bar{\mu}$ for a smooth cluster model [e.g., Eq. 35 of Ref. \citenum{gaudipetters02}] can be described as $\bar{\mu} = 155 / \theta$, where $\theta$ is the observed angular offset from the critical curve in arcseconds, and $\bar{\mu} = 19 / \sqrt{\beta}$ relates the unlensed angular position $\beta$ and $\bar{\mu}$. The angles $\beta$ and $\theta$ follow the relation $\beta = \theta^2 / 66.5$, and both $\beta$ and $\theta$ are in units of arcseconds.

We simulate the light curves of caustic-crossing events using a resolution of 1\,$\mu$arcsec per pixel over an area of $\sim83$\,pc\,$\times$\,6.5\,pc in the lens plane. This area is aligned in the direction where a background source moving toward the cluster caustic would appear to be moving. If the background star is moving with an apparent velocity of \velnearcausticsim\ in the source plane, its associated counterimage would take $\sim400$\,yr to cross the 83\,pc of the simulated region, which corresponds to $(1.2 \times 10^{-7})$$''$\,yr$^{-1}$.
The lensed star is given a transverse velocity of \velnearcausticsim\ in the source plane;
the resulting light curve can be streched to simulate different transverse velocities.
Owing to the high magnification, the counterimages' apparent motion in the image plane is large.

$N$-body simulations show that clusters of galaxies contain (and are surrounded by) a large number of subhalos. Smaller subhalos near the cluster center may not survive the tidal forces of the cluster and are easily disrupted. The larger surviving halos and smaller subhalos along the line of sight can produce small distortions in the deflection field that could in principle distort the critical curve (and caustic). Lens models of MACS\,J1149 do predict such distortions around the member galaxies. However, since the typical scales of the distortion in the deflection field are proportional to the square root of the mass of the lens, the distortions from the surviving halos are orders of magnitude larger than the scale of the distortion associated to the microlenses (from the ICL). Consequently, on the scales relevant for this work ($\sim0.2''$), the combined deflection field of the cluster plus the DM substructure can still be considered as a smooth distribution and the critical curve could still be well approximated by a straight line. 

For a cluster model populated with stars in the ICM, Extended Data Fig.~\ref{fig:train} shows the ``trains'' or multiple counterimages of a single background star near the cluster's caustic. Replacing the cluster's smoothly varying matter distribution with an increasing fraction of $\sim30$\,M$_{\odot}$ PBHs yields an increasingly long train, although its expected extent ($\sim3$\,milliarcsec) when PBHs account for 10\% of DM would be smaller than would be possible to detect in the {\it HST} imaging. 

\subparagraph{Initial Mass Function for Stellar and Substellar Objects.}
Strong lensing and kinematic \cite{treuaugerkoopmans10,augertreugavazzi10,spiniellokoopmanstrager11,cappellarimcdermidalatalo12}, as well as spectroscopic \cite{vandokkumconroy10,conroyvandokkum12} analyses of early-type galaxies have found 
evidence that the IMF of stars in early-type galaxies may be ``bottom-heavy'' -- 
a larger fraction of stars have subsolar masses than is observed in the Milky Way.
Spectroscopic evidence for a Salpeter-like bottom-heavy IMF in the inner regions of early-type galaxies comes from 
the strength of spectral features sensitive to the surface gravity of stars with $M \lesssim 0.3\,{\rm M}_{\odot}$  \cite{vandokkumconroy10,conroyvandokkum12}.
However, these two sets of diagnostics do not always show agreement in the same galaxies, and the discrepancy is not yet understood \cite{newmansmithconroy16}. 
In Extended Data Fig.~\ref{fig:imfs}, we show that a Salpeter IMF yields a substantially higher frequency of microlensing peaks than a Chabrier IMF.

In stellar kinematics and strong lensing, the DM is assumed to follow a simple parameteric (e.g., 
power-law) function near the galaxy center, while stellar matter is assumed to trace the optical emission.  
The total matter profile inferred from observations is decomposed into stellar and DM components, and
the $M_{\ast}/L$ ratio of the stellar component is used to place constraints on the IMF.

Substellar objects having masses below the H-burning limit ($M\approx0.08\,{\rm M}_{\odot}$)
are not generally included as a component of the stellar mass in kinematic and lensing analyses.
Substellar masses, however, should also trace the stellar mass distribution. 
The inferred $M_{\ast}/L$ ratios near the centers of elliptical galaxies are approximately twice as large as those expected for Milky-Way-like Chabrier IMFs, e.g., Refs.\citenum{treuaugerkoopmans10,augertreugavazzi10,spiniellokoopmanstrager11}.
If the stellar IMF in early-type galaxies has a Salpeter slope, the ratio of $\sim2$ would imply that 
a Salpeter IMF cannot extend to object masses significantly smaller than the H-burning limit \cite{barnabespiniellokoopmans13}.

Indeed, the integral of the Salpeter IMF from zero mass through the H-burning 
limit diverges, so the IMF of substellar objects must be less steep than Salpeter 
below the 0.08\,M$_{\odot}$. The integral of a Chabrier IMF in the range $0<M<0.10\,{\rm M}_{\odot}$ 
is $\lesssim10$\% of the integral in $0.10<M<100$\,M$_{\odot}$. 
High signal-to-noise-ratio spectra of NGC 1407 are best fit by a super-Salpeter IMF ($\Gamma = 1.7$; $dN/d \log{m} \propto m^{-\Gamma}$)
to the H-burning limit \cite{conroyvandokkumvillaume17}.

In the Milky Way, surveys of substellar objects find that their mass function is likely 
flat or declining with decreasing mass. 
IR imaging of the young Milky-Way cluster IC\,348 yields a population of brown dwarf stars
consistent with log-normal mass distribution \cite{alvesdeoliveiramorauxbouvier13}.
$\Gamma = 0.0$ and $\Gamma = -0.3$ provide a reasonable fit to the populations of objects with 
masses smaller than 0.1\,M$_{\odot}$ in IC\,348 and Rho\,Oph, respectively.
Analysis of the Pleiades open clusters to 0.03\,M$_{\odot}$ found a population
consistent with a log-normal distribution with $m_c = 0.25\,{\rm M}_{\odot}$
and $\sigma_{\rm log\,{\rm m}} = 0.52$ \cite{morauxbouvierstauffer03}.

For this analysis, we only include objects with initial masses greater than
0.01\,M$_{\odot}$.  For the light curves generated with a Chabrier IMF,
we assume that the IMF continues to this lower-mass cutoff. 
For the Salpeter light curves, the Salpeter form truncates at 0.05\,M$_{\odot}$; 
for lower initial masses, we assume that the number density of objects is
constant in logarithmic intervals.

\subparagraph{Mass Function of Surviving Stars and Compact Remnants in the ICM.}
For a given a star-formation history, GALAXEV computes the mass in surviving stars and in remnants using the Renzini93 prescription for the mapping between zero-age main-sequence masses and remnant masses (the initial--final mass function) \cite{renziniciotti93}. Dead stars with initial masses $M_i<8.5$\,M$_{\odot}$ become white dwarfs with mass 0.077\,M$_{\odot}+0.48\,M_i$; those with 8.5\,M$_{\odot} \leq M_i<40$\,M$_{\odot}$ become 1.4\,M$_{\odot}$ neutron stars; and those with $M_i \geq 40$\,M$_{\odot}$ become BHs with 0.5\,$M_i$.

We assume that the most massive surviving star found in the ICM at $z=0.54$ has a mass of 1.5\,M$_{\odot}$, approximately the expected value for a $\sim4.5$\,Gyr stellar population.
For stars with masses $\gtrsim1.5$\,M$_{\odot}$, we use three separate theoretical initial--final mass functions to compute the distribution of remnant masses.

The evolution of massive stars and the mass of their remnants is expected to 
depend on the stars' mass-loss rate, which is thought to vary significantly with their metallicity. 
Integral field-unit (IFU) spectroscopy of low-redshift galaxy clusters has
been able to place approximate constraints on the metallicity and age of the stars 
found in the ICM.  
IFU spectroscopy within $\sim75$\,kpc of the BCGs of the nearby Abell~85, Abell~2457, and II~Zw~108 galaxy clusters found that the ICL light can be best fit by a combination of 
substantial contributions from an old population ($\sim13$\,Gyr) with high metallicity ($Z\approx2\,{\rm Z}_{\odot}$) 
and from a younger population ($\sim5$\,Gyr) with 
low metallicity ($Z\approx0.5\,{\rm Z}_{\odot}$) \cite{edwardsalperttrierweiler16}.

\subparagraph{Light-Curve Fitting.}
To fit the LS1 / Lev16A and Lev16B light curves,
we identify the peaks in the simulated light curves that are 2$\sigma$ above each light curve's
mean magnification.
We next stretch the model light curves in time for transverse velocities 
in the range 100--2000\,km\,s$^{-1}$ in steps of 50\,km\,s$^{-1}$.
For each light curve and velocity, we find a best-fitting solution for a series of 
intervals in absolute magnitudes between $M_V = -7$ and $M_V = -10.5$\,mag in 
increments of 0.5\,mag.  For each interval and peak, we find the best-fitting value of $M_V$ 
within the upper and lower bounds in luminosity.

For the set of fits at each transverse velocity and each $M_V$ interval, 
we rank all peaks according to the $\chi^{2}$ values separately for  
LS1 / Lev16A and Lev16B.  We then pair these 
ranked lists of best-fitting peaks (i.e., the best-fitting peak for LS1 / Lev16A is matched with that for Lev16B, etc.), and add the $\chi^2$ values for each pair together.
Next, we identify the best $\chi^2$ values for all 
values of transverse velocity and ranges in absolute luminosity, and assemble
a list of these best $\chi^2$ values.
Our goodness-of-fit statistic $\langle \chi^2 \rangle$ is the average of the 150 best $\chi^2$ values.

\subparagraph{Interpreting the $\chi^2$ Statistic Using Simulated Light Curves.}
To interpret the $\langle \chi^2 \rangle$ values, we generate fake light curves for each of the models listed in Table~\ref{tab:chisq}.
The simulation for each model yields a magnification over a 406\,yr period for a transverse velocity of 1000\,km\,s$^{-1}$. 
For lensed stars with absolute magnitudes $M_V$ of $-8$, $-9$, and $-10$, we 
create simulated apparent light curves, and we append the light curve after reversing the temporal axis 
to create effectively an 812\,yr light curve. 

For each simulated light curve, we identify all peaks where the apparent $F125W$ AB magnitude is brighter than \simbrightness.
For each peak, we randomly select a transverse velocity drawn from a uniform distribution in the interval 100--2000\,km\,s$^{-1}$.
We use the cadence and flux uncertainties of the measured light curve of LS1 to generate a fake light curve.
We next shift the peak of the measured light curve of LS1 / Lev16A (or Lev16B) to match the peak of the simulated model light curve.
We then create a fake observation by sampling the simulated light curve at the same epochs as the actual measurements, and adding Gaussian noise matching the measurement uncertainties.

For each such simulated light curve, we compute the $\langle \chi^2 \rangle$ statistic using the full set of models.
The region of the simulated light curve used to created the fake data set is excluded from fitting.
As shown in Extended Data Fig.~\ref{fig:simstats}, the combined $\langle \chi^2 \rangle$ statistic we measure for LS1 / Lev16A and Lev16B falls inside of the expected range of values.  We note that a significant fraction of simulated light curves have average values of $\langle \chi^2 \rangle$ $>$1000, implying that they are not well-fit by other regions of the simulated light curves.

For all simulated light curve where the average $\langle \chi^2 \rangle$ value is within 100 of the value we measure for LS1 / Lev16A and Lev16B, we calculate the difference $\Delta$$\langle\chi^2\rangle$ values between the $\langle\chi^2\rangle$ values of the generative (``true'') model and of the best-fitting model.
For 68\% of simulated light curves, $\Delta$$\langle\chi^2\rangle \lesssim 13$, and, for 95\% of simulated light curves, $\Delta$$\langle\chi^2\rangle \lesssim 25$.

\subparagraph{Massive Stellar Evolution Models}
The fates of massive stars remain poorly understood owing to the complexity of massive stellar evolution and the physics of supernova (SN) explosions. Indirect evidence suggests that a fraction of massive stars may collapse directly to a black hole instead of exploding successfully \cite{smartteldridgecrockett09,gerkekochanekstanek15}, due to (for example) failure of the neutrino mechanism \cite{pejchathompson15}.
We compute light curves and magnification maps using three sets of predictions for the initial--final mass function \cite{woosleyhegerweaver02,fryerbelczynskiwiktorowicz12,speramapellibressan15}. 
As a first model, we adopt the initial--final mass function predicted by the solar-metallicity, single stellar evolution models \cite{woosleyhegerweaver02} (Woosley02) [Fig.~9 of Ref.\citenum{fryerbelczynskiwiktorowicz12}]. 
In the Woosley02 models, the prescription for driven mass-loss rate at solar metallicity causes 
stars with initial masses $\gtrsim33$\,M$_{\odot}$ to end their lives with 
significantly reduced He core masses, leading such stars to become BH remnants with 
masses no larger than 5--10\,M$_{\odot}$.
The Woosley02 mass-loss prescription uses theoretical models of radiation-driven winds
for OB-type stars with $T > 15,000$\,K, and empirical estimates for Wolf-Rayet stars \cite{hamannschoenbernerheber82} that have been adjusted downward by a factor of three to account for the effects of clumping in 
the stellar wind \cite{hamannkoesterke98}.
The mass-loss rate for single O-type stars during their main-sequence evolution may have been overestimated by a factor of 2--3, owing to unmodeled clumping in their winds \cite{smith14}.

A second (Fryer12) initial--final mass function was computed for single stars at subsolar metallicity ($Z=0.3$\,Z$_{\odot}$; $Z=0.006$) [the ``DELAYED'' curve in Fig.~11 of Ref. \citenum{fryerbelczynskiwiktorowicz12}].  These predictions use the StarTrack population synthesis code \cite{belczynskikalogerarasio08,belczynskibulikfryer10}.  
According to this model, BHs with masses up to 30\,M$_{\odot}$ form from the collapse of massive stars. 

Finally, we use a third initial--final mass function (Spera15) [Fig.~6 of Ref. \citenum{speramapellibressan15}] to calculate the masses of remnants for the stellar population making up the ICM. The Spera15 relation we adopt was computed using the PARSEC evolution tracks for stars with metallicity $Z=0.006$, and explosion models where the SN is ``delayed,'' occurring $\gtrsim0.5$\,s after the initial bounce.  According to the Spera15 initial--final mass relation we adopt, stars having initial masses $\gtrsim33$\,M$_{\odot}$ become BHs with masses within the range 20--50\,M$_{\odot}$.
Approximately 70\% of massive stars exchange mass with a companion, while $\sim1/3$ of stars will merge \cite{sanademink12}. It is also possible that the success of explosions is not related in a simple way to the stars' initial mass or density structure, given the potentially complex dependence of the critical neutrino luminosity for a successful explosion on these progenitor properties \cite{pejchathompson15}.

\subparagraph{Observations of Similar Cluster Fields with {\it HST}.}
Massive galaxy clusters have been the target of extensive {\it HST} imaging and grism-spectroscopy campaigns in the last several years, and these need to be taken into account when considering the probability of finding a highly magnified star microlensed by stars making up the ICL.  Detecting transients requires at least two separate observing epochs, which is possible only for programs designed with more than a single visit, or when archival imaging is available. In addition to smaller search efforts \cite{sharongalyammaoz10}, large programs have been the Cluster Lensing and Supernova survey with {\it Hubble} (CLASH) \cite{postmancoebenitez12}, the GLASS \cite{schmidttreubrammer14,treuschmidtbrammer15}, the {\it Hubble} Frontier Fields \cite{lotzkoekemoercoe17}, and the Reionization Lensing Cluster Survey (RELICS). Transients in {\it Hubble} Frontier Fields and GLASS imaging have been subsequently observed by the FrontierSN program.

With a total of 524 orbits, CLASH acquired imaging of 25 galaxy-cluster fields.  A search for transients in the CLASH imaging made use of template archival imaging when available. The survey acquired imaging over a period of three months of each cluster field with repeated visits in each filter. 
Each epoch had an integration time of 1000--1500\,s.
The systematic search for transients in CLASH imaging had near-infrared limiting magnitudes at each epoch of $F125W \approx 26.6$ and $F160W \approx 26.7$\,mag\,AB [see Tab. 1 of Ref.~\citenum{graurrodneymaoz14}].

The {\it Hubble} Frontier Fields program used 140 orbits to observe each of 6 galaxy-cluster fields (total of 840 orbits). For each cluster, ACS optical and WFC3 IR imaging split into in two campaigns, each of which lasted for approximately a month. These were separated from each other by a period of $\sim6$\,months to allow the telescope roll angles to differ by $\sim180^{\circ}$, to image a parallel field adjacent to the cluster with the same instruments. 
Each of six to twelve epochs in each WFC3 IR wideband filter for each cluster field had an integration time of $\sim5500$\,s.
The systematic search for transients in near-infrared {\it Hubble} Frontier Field imaging by the FrontierSN team (PI: Rodney) had near-infrared limiting magnitudes at each epoch of $F125W \approx 27.5$ and $F160W \approx 27.2$\,mag\,AB.

The GLASS survey \cite{schmidttreubrammer14,treuschmidtbrammer15} acquired WFC3 IR grism spectroscopy of ten galaxy-cluster fields over 140 orbits.  The survey acquired direct pre-imaging in the WFC3 IR $F105W$ and $F140W$ over four epochs 

The probability of observing a luminous star adjacent to a caustic will depend on the number of 
lensed galaxies that overlap a galaxy cluster's caustic.
For massive clusters, caustic curves coincide with the ICL, so magnified stars should 
exhibit microlensing fluctuations.
Earlier work has estimated that a $10^5$\,L$_{\odot}$ star ($M_V\approx-7.5$\,mag) in a giant arc (with $z=0.7$) and crossing the caustic of cluster Abell 370 ($z=0.375$), a {\it Hubble} Frontier Fields target, would remain brighter than $V=28$\,mag for 700\,yr, given a {\it V}-band limiting magnitude of 28 and a 300\,km\,s$^{-1}$ transverse velocity \cite{miraldaescude91}.

A study of {\it HST} imaging of CLASH galaxy-cluster fields\cite{xupostmanmeneghetti16} has found that each cluster field contains $4\pm1$ giant arcs with length $\geq6''$ and length-to-width ratio $\geq7$. Given that the host galaxy of LS1 would not be classified as a giant arc according to these criteria, it is likely that additional galaxies in each field may lie on the cluster caustic.

An approximate census finds that $\lesssim10$\% of {\it HST} galaxy-cluster observing time has
been used to image MACS\,J1149. Considering the full set of {\it HST} cluster observations, and the results of our Monte Carlo simulations for the single arc underlying LS1 (1--3\% for a shallow $\alpha\approx2$ luminosity function; 0.01--0.1\% for $\alpha\approx2.5$), the probability of observing at least one bright magnified star adjacent to a critical curve should be appreciable, in particular if the average luminosity function is more shallow at high redshifts than in the nearby universe.

\clearpage
\section*{Supplementary Information:}

\subparagraph{Simulation Using Stars Near Center of 30\,Doradus.}
We used the SIMBAD (http://simbad.u-strasbg.fr/simbad/) catalog to retrieve information for objects in the \ionpat{H}{ii} region 30\,Doradus in the LMC.  Stars were first selected from the catalog having $V$-band magnitudes and closer than $20'$ to its center ($\sim280$\,pc).
We include objects classified as stars, and exclude F,G,K, or M type stars unless they are classified as supergiants.
Using a distance of $49$\,kpc to the LMC, we calculated the stars' absolute magnitudes $M_V$, and their
distances from the center of the cluster.  For 10,000 trials, we randomly placed the cluster's CC within 20\,pc of the 30\,Doradus' center, and rotated the 
stars around the center of 30\,Doradus by an angle drawn from a uniform distribution.  We find a probability of $\sim1$\% of
finding a star with a persistent average brightness of at least \persistentmag,
and we find that such a star will also be responsible for $\gtrsim99$\% of $<$\brightmlmag\ microlensing events.
These probabilities are similar to those we estimated from our simulation where the
positions of luminous stars near the CC were drawn randomly from a uniform distribution

\subparagraph{Slope of Stellar Luminosity Function in 30\,Doradus.} 
We have also used the SIMBAD catalog of stellar sources to estimate the stellar luminosity function of bright stars in 30\,Doradus.
Placing stars in 0.5\,mag bins by their absolute $V$-band magnitudes, we
measure a power-law index of $\alpha\approx-2$. 
No correction is applied for crowding, or the binary fraction.

\subparagraph{Slope of the Stellar Luminosity Function in Nearby Galaxies.}
Stars found in OB associations in seven nearby galaxies observed with {\it HST} show a luminosity function of $\alpha =  2.53 \pm 0.08$ \cite{bresolinkennicuttferrarese98}.
The stellar luminosity function for stars more luminous than $M_V < -8.5$\,mag is not well constrained,
the number counts of the $M_V < -8.5$\,mag stars in this study are consistent with the
slope measured for stars with $-8.5 < M_V < -5$\,mag.
The slope of $\alpha =  2.53 \pm 0.08$ agrees approximately with a separate earlier analysis \cite{freedman85}, which studied the slope of the upper end of the stellar luminosity function of the bluest stars in nearby galaxies using ground-based imaging and found $\alpha = 2.68 \pm 0.08$, although the latter analysis extended only to $M_V\approx-9.5$\,mag [see Fig.~7 of Ref. \citenum{freedman85}]. 
A second census of the stellar population in galaxy M101 shows that it may host a small number of luminous stars with absolute magnitude $-10 \lesssim M_V \lesssim -11$ [see Fig.~7 of Ref. \citenum{grammerhumphreys13}]. 

The luminosity function of OB associations \cite{bastianercolanogieles07} can be well described by a power-law function having an index $\alpha\approx2$. The luminosity function of star-forming regions may become flatter in galaxies with higher star-formation rates and star-formation rate densities \cite{cookdalelee16}.

\subparagraph{Ground-Based Follow-up Campaigns.}
We observed the field with direct imaging with the Low Resolution Imaging Spectrometer (LRIS) \cite{okecohen95} on the Keck-I 10~m telescope on 6 May 2016 (PI Filippenko). Director's Discretionary programs with the GTC (PI P\'{e}rez Gonz\'{a}lez; GTC2016-052), the Very Large Telescope (PI Selsing; 297.A-5026), Gemini North (PI Kelly; GN-2016A-DD-8), and the Discovery Channel Telescope (PI Cenko) obtained follow-up imaging in optical bandpasses.

\subparagraph{Detection from the Ground with the Gran Telescopio Canarias.}
We obtained {\it i'}-band observations of the MACS\,J1149 field with the 10.4\,m Gran Telescopio Canarias (GTC) on 6 June 2016 and 7 June 2016, after the May 2016 peak. 
To estimate the flux at LS1's position, we extracted the flux inside several apertures with diameters 
within 1--2 times the the PSF FWHM, and applied aperture corrections to obtain integrated fluxes. 
The flux estimates for the different apertures agree within 0.10--0.15\,mag.
The {\it i'}-band AB magnitudes are $27.73\pm0.52$ on 57544.9381 MJD  
in conditions with $1.0''$ seeing and a total integration of 3000\,s, 
and $28.35\pm0.43$ on 57546.9445 MJD in $0.8''$ seeing and a total integration of 5430\,s.

\subparagraph{Metallicity of the Local Host-Galaxy Environment of LS1.}
The gas-phase oxygen abundance of LS1's host galaxy, including that within the immediate environment of SN~Refsdal, has been studied using multiple datasets. 
LS1 and SN~Refsdal have similar offsets from the host nucleus (within $\sim0.5$\,kpc), so the local metallicity near LS1's and SN~Refsdal's host-galaxy locations should have similar values.
Both the CATS model \cite{jauzacrichardlimousin16} ($\sim6.7$\,kpc and $\sim7.3$\,kpc, respectively) and the GLAFIC model \cite{kawamataoguriishigaki16} ($\sim7.9$\,kpc and $\sim8.2$\,kpc) find similar nuclear offsets for LS1 and SN~Refsdal. 

Analysis of Keck-II OSIRIS integral-field unit (IFU) spectra reported a $3\sigma$ upper limit of \mbox{12 + log(O/H)}\ $< 8.67$\,dex in the Pettini \& Pagel N2 calibration \cite{pettini04} for an \ionpat{H}{ii} region $\sim200$\,pc away from SN~Refsdal's site. 
From the same observations, the authors find a combined upper limit of \mbox{12 + log(O/H)}\ $< 8.11$\,dex from observations of nine \ionpat{H}{ii} regions at nuclear offsets between $\sim5$ and $\sim7$\,kpc \cite{yuankobayashikewley15}, which is similar to the offsets of LS1 and SN~Refsdal.

Recent work \cite{wangjonestreu17} has analyzed WFC3 grism spectra taken by GLASS \cite{schmidttreubrammer14,treuschmidtbrammer15} and follow-up observations of SN~Refsdal.
They fit the abundance measurement using a linear model,
\begin{equation}
{\rm 12 + log(O/H)} =  (-0.0666 \pm 0.0232) \times r + 8.82 \pm 0.039\ {\rm dex},
\end{equation}
where $r$ is the offset from the nucleus in kpc.
This yields an abundance at LS1's offset (assuming \lsoneoffset) of {\rm 12 + log(O/H)} $= 8.29\pm0.19$ dex. 
This analysis does not take into account the [\ionpat{N}{ii}] line when estimating the oxygen abundance, as it can be a biased tracer at $z > 1$ \cite{steidelrudiestrom14}.

Finally, while [\ionpat{N}{ii}] was not detected in the OSIRIS IFU spectra of the site of SN~Refsdal \cite{yuankobayashikewley15}, a 1\,hr Keck-II MOSFIRE integration yielded a [\ionpat{N}{ii}] detection. 
The [\ionpat{N}{ii}] line strength yields a PP04 N2 oxygen abundance of \mbox{12 + log(O/H) = $8.3\pm0.1$\,dex} \cite{kellybrammerselsing16}, which is in agreement with the above estimate made using the Maiolino calibration \cite{maiolinonagaograzian08} from WFC3 grism spectra.

Given the above grism as well as MOSFIRE [\ionpat{N}{ii}] metallicity estimates, we use an oxygen abundance of \mbox{12 + log(O/H) $\approx8.3$\,dex} as the metallicity of the massive stellar population near LS1's coordinates.
The Castelli \& Kurucz 2004 stellar atmosphere models\cite{castellikurucz04} are parameterised based on the  Grevesse \& Sauval 1998 \cite{grevessesauval98} solar oxygen abundance of 12 + log(O/H) = $8.83\pm0.06$\,dex. Therefore, we adopt 
$\log (Z/{\rm Z}_{\odot})=-0.5$ when drawing comparisons with the Castelli \& Kurucz ATLAS9 models.

\subparagraph{K-correction and Distance Modulus.}
We calculate {\it K}-corrections following Equation 2 of Ref. \citenum{kimgoobarperlmutter96},
\begin{equation}
K=2.5\times\log_{10}(1+z) + m_{F125W,{\rm syn}}^{\rm AB} - m_{V, {\rm syn}}^{\rm Vega},
\end{equation}  
where $z=1.49$, $m_{F125W,{\rm syn}}^{\rm AB}$ is the WFC3 $F125W$ synthetic magnitude of a redshifted model spectrum, and $m_{V,{\rm syn}}^{\rm Vega}$ is the synthetic Johnson {\it V}-band magnitude of the rest-frame model spectrum.  Here the {\it K}-correction $K_{\rm xy}$ is defined as 
\begin{equation}
m_{\rm y} = M_{\rm x} + dm + K_{\rm xy},
\end{equation}
where $m_{\rm y}$ is the observer-frame apparent magnitude in the {\rm y} band, $M_{\rm x}$ is the rest-frame absolute magnitude in the {\rm x} band, and $dm$ is the distance modulus.
Using the best-fitting spectral models, we calculate $K_{V,F125W}=$\kcorrectionVJ, and adopt $dm=45.21$\,mag at $z=1.49$ (with no correction for magnification).

\subparagraph{Stellar-Mass Density Along the Line of Sight to LS1.}
We computed two separate estimates of the stellar-mass density to LS1. 
The first estimate was the value we used when we created most of the simulated light curves, 
but it excluded light from the nearby brightest cluster galaxy (BCG). 
We computed a second, improved estimate that accounted for all intracluster light (ICL) along the line of sight.
The updated analysis yielded a density approximately 
twice as high as the initial value. 

{\bf Initial Estimate:} Galaxies with $m_{\rm F160W}<26$\,AB\,mag are selected and fit with single S\'{e}rsic profiles by using GALFIT \cite{penghoimpey02} in a postage stamp (300\,pix $\times$ 300\,pix). 
At the same time, the local sky background, assumed to be constant across the stamp, is fitted with the galaxy light profile.
After fitting all the galaxies in the field, we reconstruct the ICL map by using the estimated local sky background values. Overlapping pixels are stacked, weighted by the $\chi^2_\nu$ value from the fit.
The uncertainty is estimated from the original root-mean-square (RMS) map (published by the {\it Hubble} Frontier Fields team) and the systematic differences caused by changing the stamp size. We repeat this procedure for the ACS WFC $F435W$, $F606W$, $F814W$, and WFC3 IR $F105W$, $F125W$, $F140W$, and $F160W$ filter bands \cite{morishitaabramsontreu16}. A correction is applied for Galactic extinction \cite{sch98}. 

Stellar mass is estimated in each pixel using the Fitting and Assessment of Synthetic Templates (FAST) software tool \cite{kriekvandokkumlabbe09} with the BC03 \cite{bc03} stellar population model. FAST uses the Galaxy Spectral Evolution Library (GALAXEV; http://www.bruzual.org/) code to assemble composite stellar populations. We use the BC03 isochrones, a Chabrier IMF, and an exponentially declining star-formation history.
Stars have initial masses that are between 0.1 and 100\,M$_{\odot}$, and models are computed for metallicities of 0.004, 0.008, 0.02, and 0.05. 
The hot gas in the galaxy-cluster ICM is thought to destroy dust, and extinction from dust is assumed to be zero. We note, however, that allowing the dust extinction to be a free parameter would change the estimated stellar mass by $<1\%$.

{\bf Revised Estimate:} We calculated a second estimate for the stellar-mass density that includes the contribution of stellar light associated with the BCG.
We first constructed a total of eight apertures around the BCG shown in Extended Data Fig.\,\ref{fig:icl_apertures}.
The apertures' offsets from the BCG center and $F140W$ surface brightnesses are similar to those of LS1, and were selected to exclude point sources and cluster-member galaxies (except the BCG).

We estimate ACS WFC $F435W$, $F606W$, $F814W$, and WFC3 IR $F105W$, $F125W$, $F140W$, and $F160W$ fluxes within each aperture, 
and apply a correction for Galactic extinction \cite{sch98}. 
We next determine the ratio between the stellar mass ($M_{\ast}$) and the WFC3-IR $F140W$ flux ($L$) within each aperture. 
We estimate $M_{\ast}$ with FAST \cite{kriekvandokkumlabbe09} and the BC03 \cite{bc03} stellar population synthesis models.
We adopt a delayed exponentially declining star-formation history and include both subsolar and solar metallicity ($\sim0.02$ and 0.008) populations.
Separate model fits are made for Chabrier and Salpeter IMFs, and the stars in our BC03 population synthesis models have initial masses that are between 0.1 and 100\,M$_{\odot}$.
 
Within each aperture, the statistical uncertainties of the total WFC3-IR flux in each bandpass are $\lesssim0.5$\%.
Among fits within the apertures, the average $e$-folding time is $\sim600$\,Myr, and, at redshift $z=0.54$, the stellar population 
ages are, on average, $\sim4$\,Gyr. The uncertainty in $M_{\star}/L$ for each aperture is $\sim30$\%, which approximately
equals the standard deviation among the best-fitting estimates for all apertures. 

To estimate the stellar-mass density along the line of sight to LS1, we multiply the mean $M_{\ast}/L$ computed across all eight apertures by the average $F140W$ surface brightness in the two apertures adjacent to LS1. These apertures adjacent to LS1 may contain contamination in observer-frame optical bandpasses from the underlying, young lensed galaxy. 
Within the WFC3-IR $F140W$ bandpass, however, light from the cluster dominates.

For Chabrier and Salpeter IMFs, the stellar mass densities computed using the BC03 model are \chabrierarcsecdensity\ and \salpeterarcsecdensity, respectively. 
These revised estimates as well as our initial estimate for the density include remnants, whose masses are computed using the Renzini ``initial--final'' mass function \cite{renziniciotti93}. 
The total local projected mass density inferred from cluster models \cite{diegobroadhurstchen16,zitrinfabrismerten15,jauzacrichardlimousin16} is $\sim$ \clusterarcsecdensity. 

\spacing{1}

\clearpage
\renewcommand{\thefigure}{\arabic{figure}}
\setcounter{figure}{0}
\renewcommand{\figurename}{Extended Data Fig.}

\clearpage
\renewcommand{\thetable}{\arabic{table}}
\setcounter{table}{0}
\renewcommand{\tablename}{Extended Data Table}

\pagebreak

\begin{figure}[!htbp]
\centering
\includegraphics[draft=\draft,angle=0,width=2.8in]{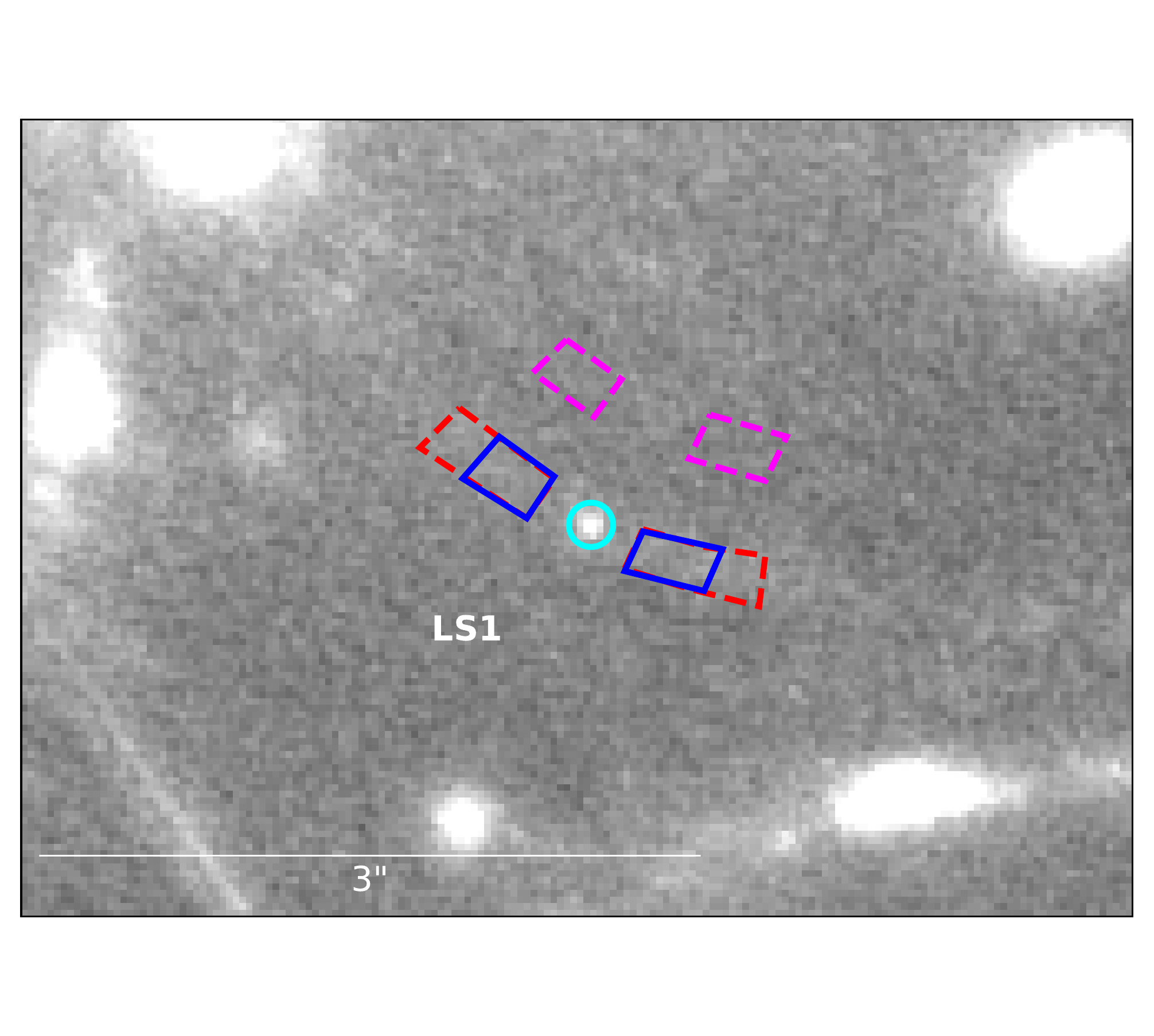} 
\includegraphics[draft=\draft,angle=0,width=3.0in]{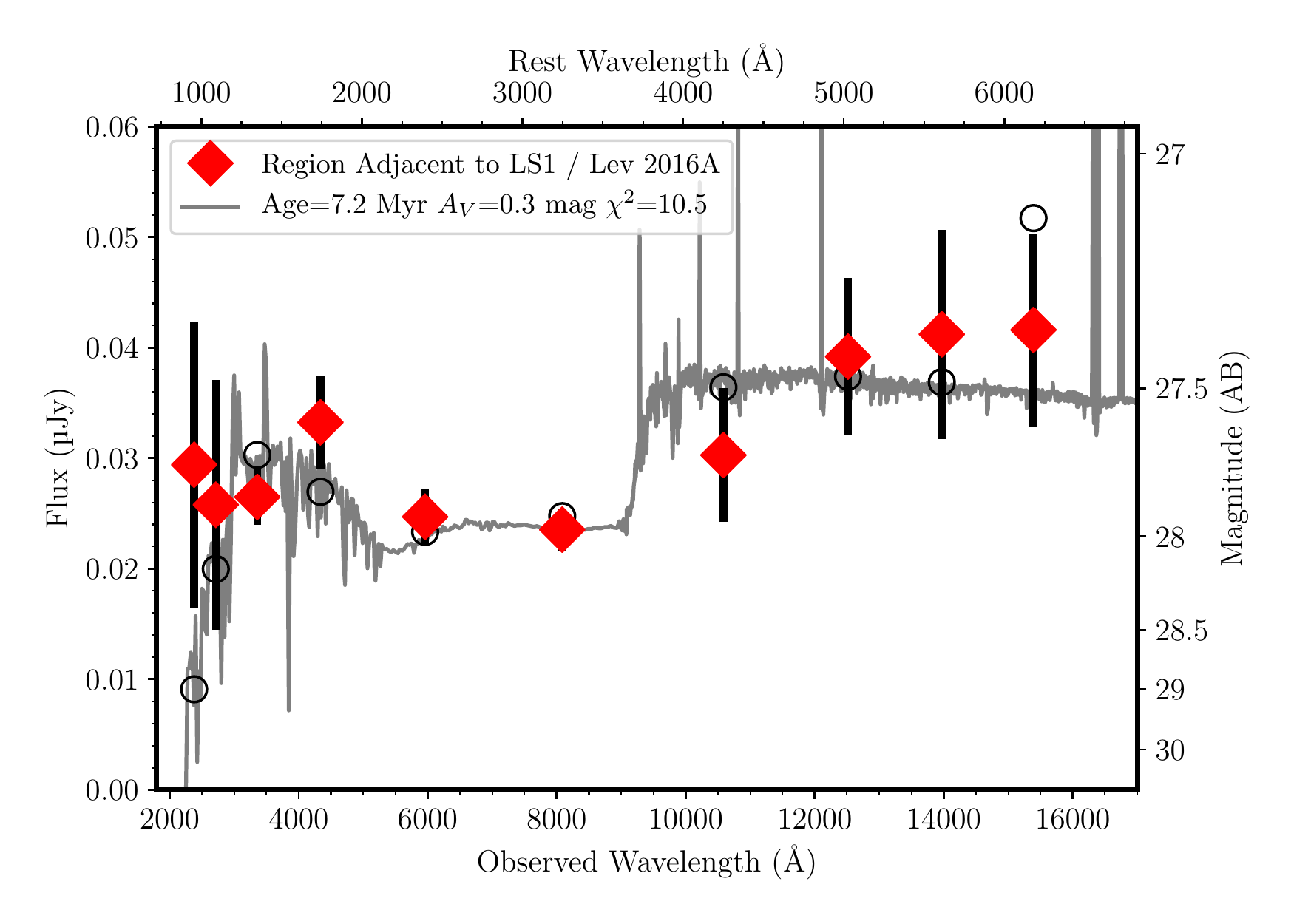}
\includegraphics[draft=\draft,angle=0,width=2.4in]{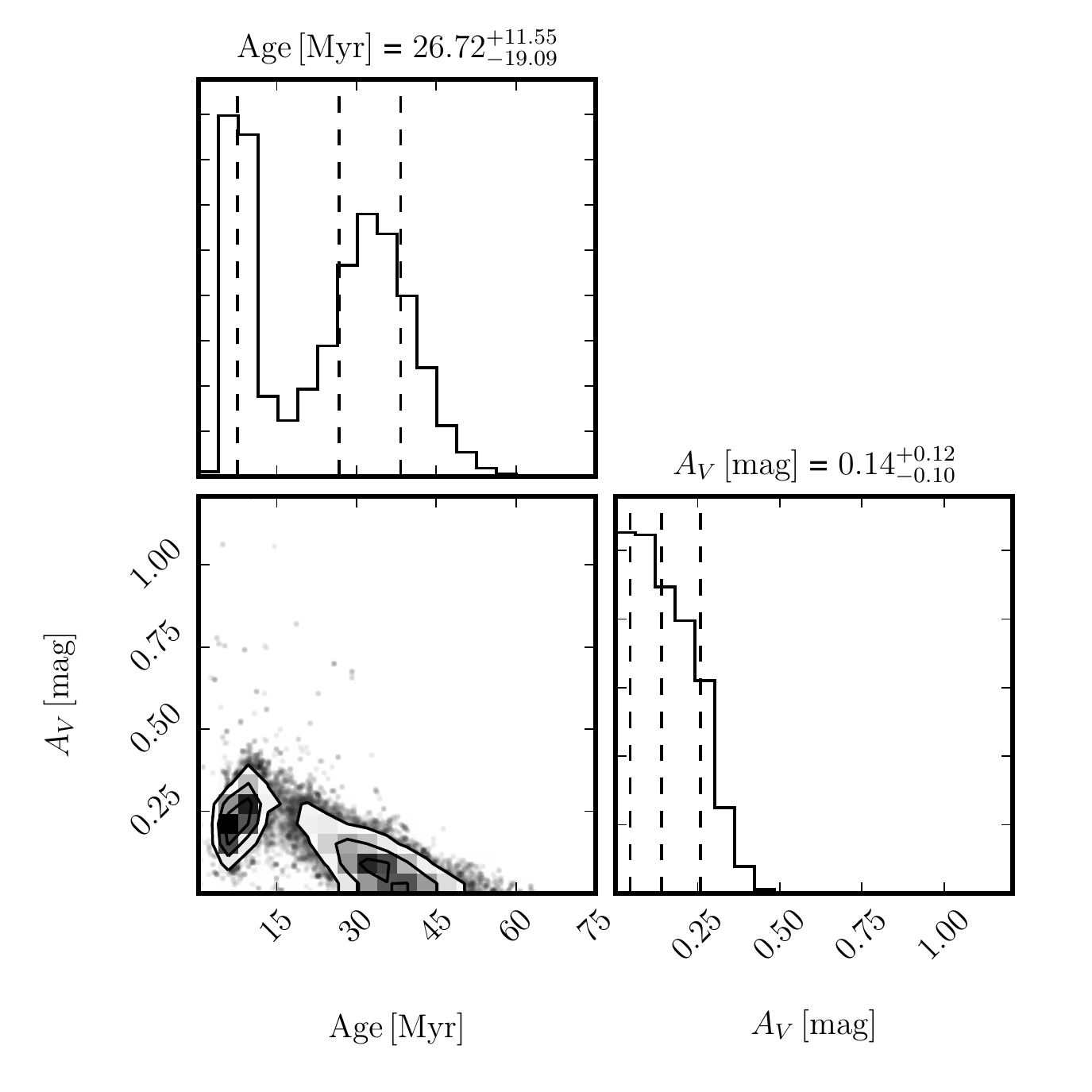}
\caption{Constraints on the age and dust extinction of the stellar population along the lensed arc adjacent to LS1.
LS1's flux was measured in the circular cyan aperture and the background measured inside of the red dashed aperture shown in upper-left panel. 
Upper-right panel plots the SED of the underlying arc measured inside the aperture outlined by a blue boundary,
after subtracting the background measured in the aperture outlined in magenta. 
Bottom panel shows the posterior probability distributions of the age and extinction $A_V$ of the stellar population. 
Spectra of the host galaxy favor a gas-phase metallicity of $Z\approx-0.3$.  
At such a metallicity, we find a bimodel posterior probability distribution with peaks at $\sim8$ and $\sim35$\,Myr. 
An age of $\sim8$\,Myr would be consistent with the age of a blue supergiant star. 
The stellar population synthesis model is constructed using the Padova isochrones \cite{marigogirardi07,marigogirardibressan08}, and we apply a Cardelli extinction law with $R_V = 3.1$ \cite{cardelli89}. 
\label{fig:sed_adjacent_region}}
\end{figure}

\pagebreak

\begin{figure}[!htbp]
\centering
\includegraphics[angle=0,width=6.5in]{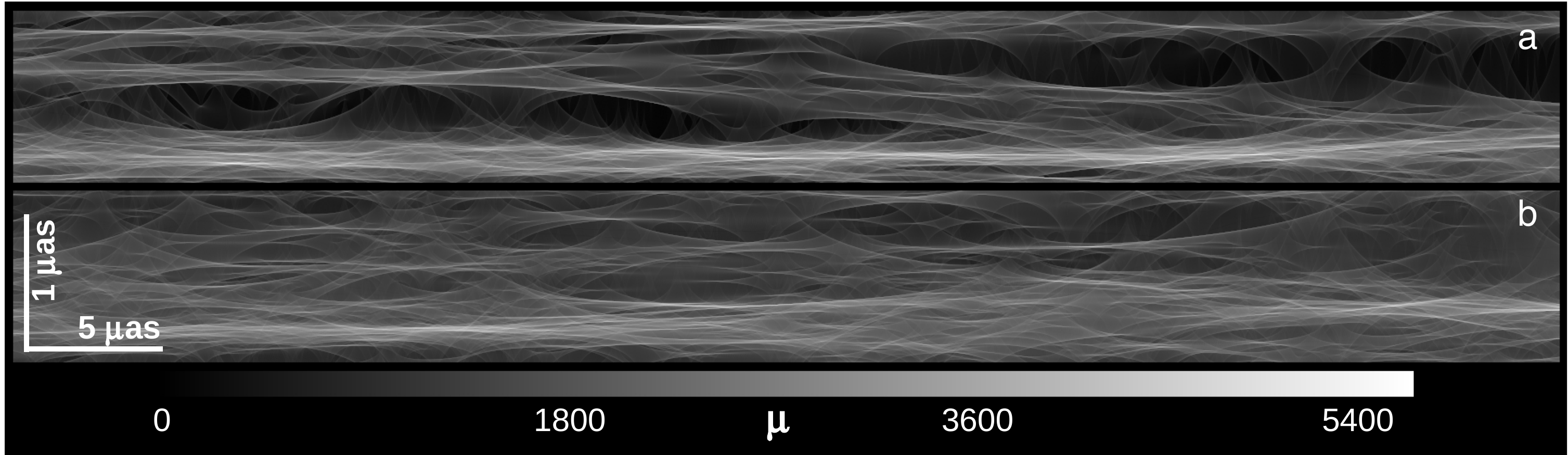}
\caption{Distinct magnification patterns for respective counterimages Lev16B (upper panel) and LS1/Lev16A (lower panel) of LS1 within the source-plane host galaxy at redshift $z=1.49$ from a ray-tracing simulation. 
Extensive regions of low magnification ($\lesssim$100) for negative-parity image Lev16A could explain why it is undetected in {\it HST} imaging acquired in all except a single epoch acquired from 2004 through 2017. The map for positive-parity image LS1/Lev16A lacks such regions of extensive low magnification, and it always detected in deep imaging. 
Plotted angular scale is in the source plane, and one \,$\upmu$arcsec in each panel corresponds to a physical $8.6 \times 10^{-3}$\,pc at redshift $z=1.49$.
If LS1 has an apparent transverse velocity of 1000\,km\,s$^{-1}$, it would travel 1\,$\upmu$arcsecond in 8.6 observer-frame years.
These ray-tracing simulations are realistic if Lev16B and LS1/Lev16A are mutual counterimages offset by 0.13$''$ on opposite sides of the galaxy cluster's critical curve in the image plane, and each of the counterimages has an average magnification of 600. 
The galaxy-cluster caustic, which is offset by 2.1\,pc from these maps, is oriented parallel to the horizontal axes of each panel. 
The different patterns of magnification correspond to the parity of the image; Lev16B has negative parity, while LS1/Lev16A has positive parity. 
Here we have created a random realization of foreground intracluster stars and remnants having a mass-density  (\salpeterarcsecdensity) matching that we infer for a Salpeter IMF.
\label{fig:close_to_caustic}}
\end{figure}

\pagebreak

\begin{figure}[!htbp]
\centering
\includegraphics[draft=\draft,angle=0,width=3.5in]{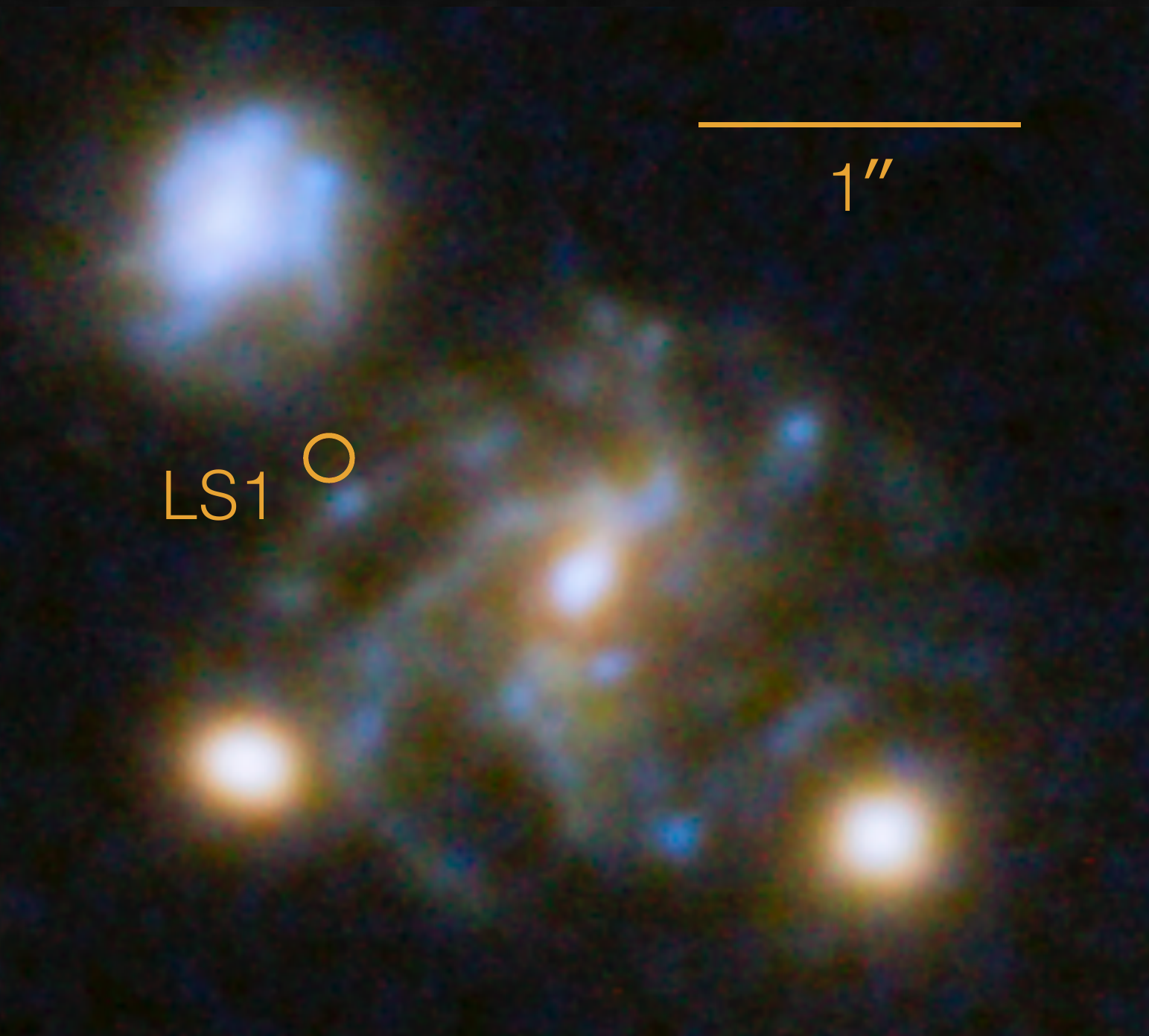}
\caption{Predicted position of LS1 in separate, full image of its host galaxy created by MACS\,J1149 galaxy-cluster lens. The cluster lens create three images of the host galaxy at redshift $z=1.49$.  We detected LS1  adjacent to the critical curve separating two partial, merging images which have opposite parity.  The third, full image shown here is at a greater distance from the cluster center, and it shows that LS1 lies close to the tip of a spiral arm. 
\label{fig:fullimageloc}}
\end{figure}

\pagebreak

\begin{figure}[!htbp]
\centering
\includegraphics[angle=0,width=5.0in]{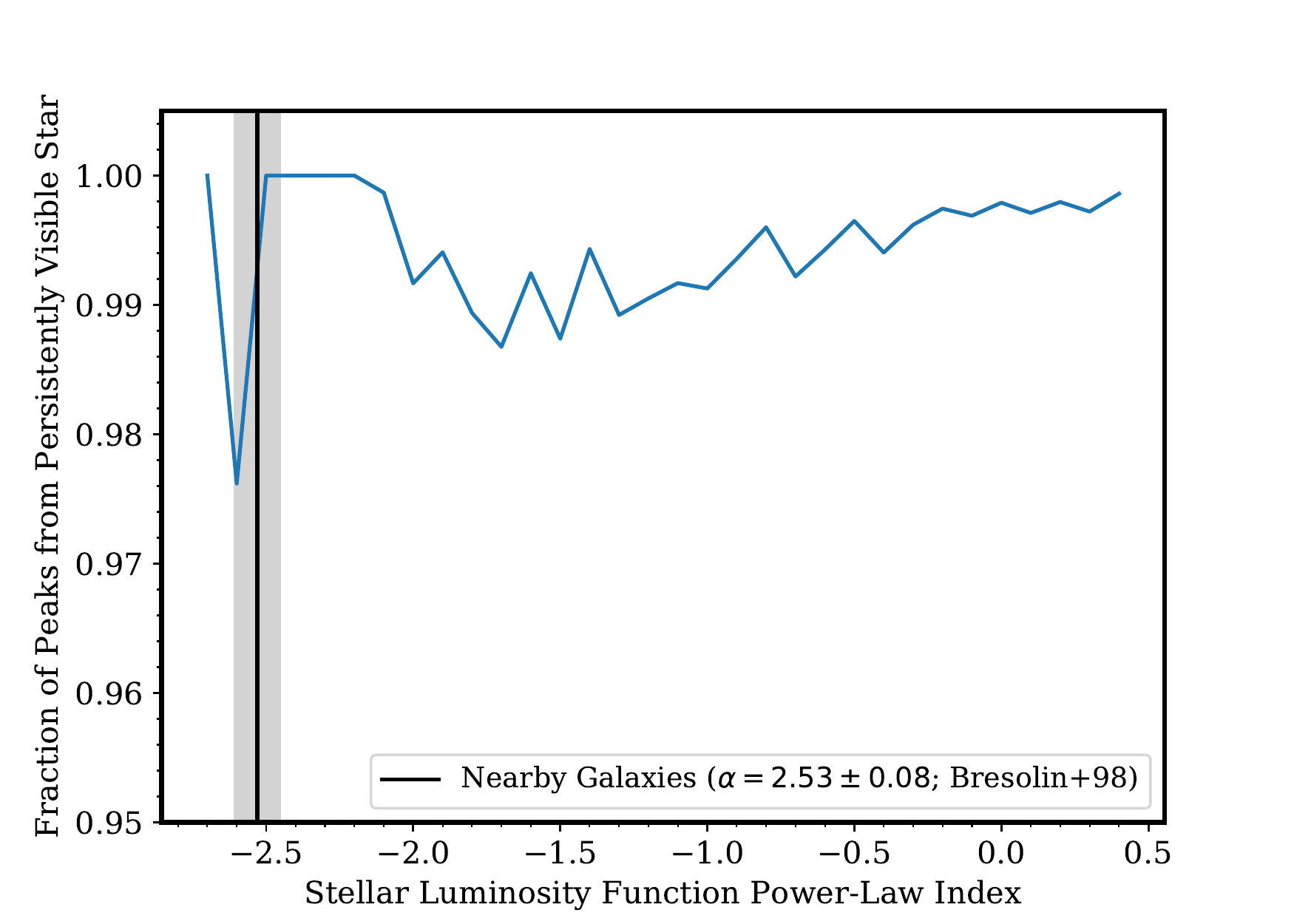}
\caption{When a lensed star has a bright average apparent magnitude ($F125W<$ \persistentmag), it will also be responsible for almost all bright microlensing peaks ($F125W<$ \brightmlmag) observed near the critical curve for simple assumptions.
Here we plot the fraction of bright peaks caused by the bright star against the index of the stellar luminosity function.  Since $\gtrsim99$\% of events likely arise from the luminous star, it is likely that Lev16B corresponds to the same star as LS1. 
However, this simulation randomly assigns positions to massive stars. 
To determine whether the observed clustering of massive stars could yield a greater probability that LS 1 / Lev16A and Lev16B are different stars, we carry out a simulation instead using the observed absolute magnitudes and positions of stars in the 30\,Doradus cluster in the LMC from the SIMBAD catalog, and find a similarly low probability.
\label{fig:fractiondetectedeventsstar}}
\end{figure}

\pagebreak

\begin{figure}[!htbp]
\centering
\includegraphics[draft=\draft,angle=0,width=6.5in]{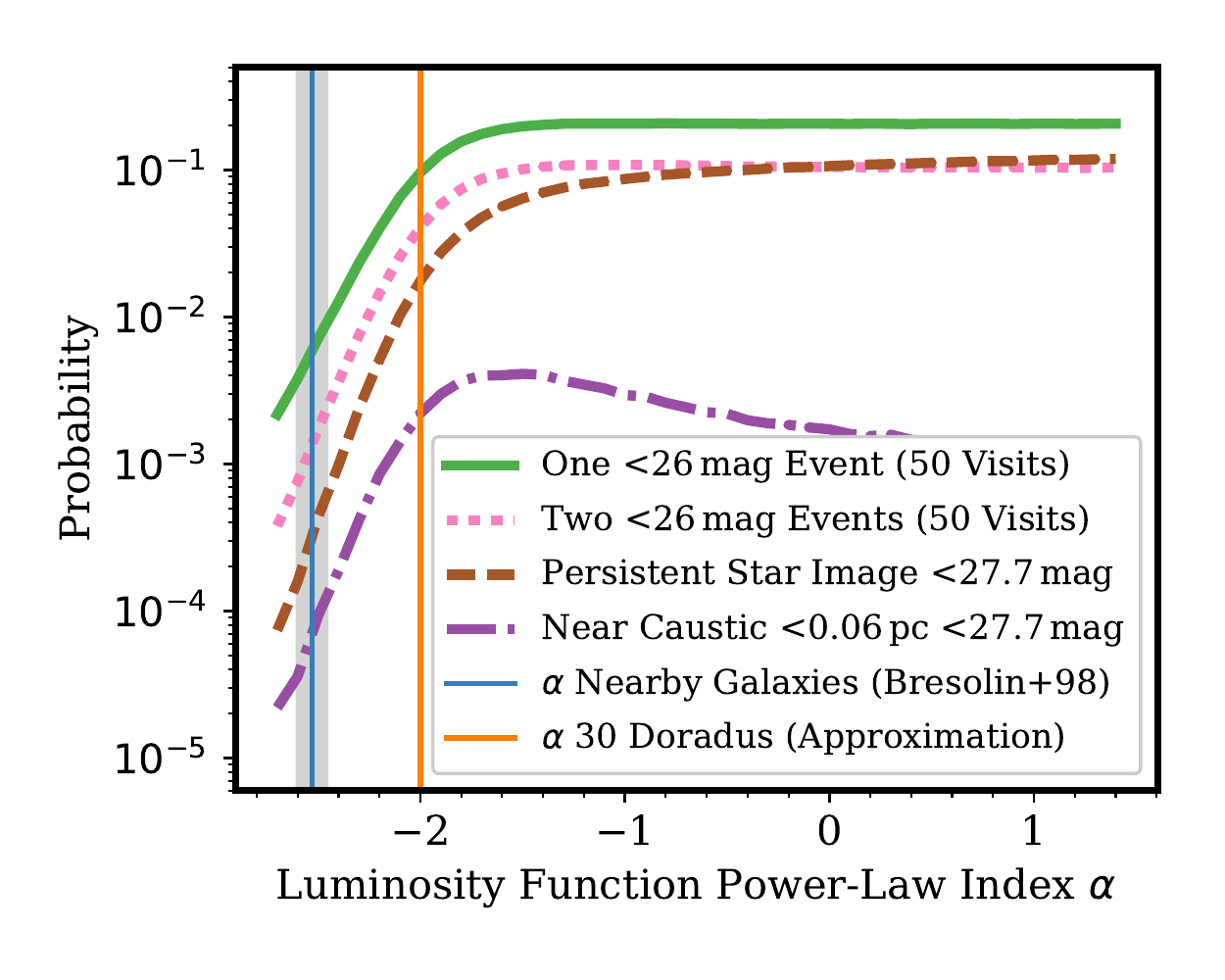}
\caption{Dependence of the probability of observing highly magnified stellar images on the stellar luminosity function of the underlying arc.
Panel shows probabilities of (a) bright microlensing events ($F125W< $\brightmlmag; solid green and dotted pink), (b) a persistently bright magnified star ($F125W< $\persistentmag) similar to that observed at LS1's position in 2004--2017 (dashed brown), and (c) a persistently bright magnified star ($F125W< $\persistentmag) within 0.06\,pc (dot-dash purple).
Probabilities are small given the index of stellar luminosity function measured for nearby galaxies (\bresolinlf; vertical blue) \cite{bresolinkennicuttferrarese98}, but become significantly larger for shallower power-law indices, such as that for the 30\,Doradus star-forming region in the LMC (vertical orange; approximate).  Here we have assumed \numobsaprseventeen\ visits by {\it HST}, the number of separate observations of MACS\,J1149 taken through 13 April 2017 after binning data by 10\,days.  The lower stellar luminosity limit used for these simulations is 10\,L$_{\odot}$.
\label{fig:lumprob}}
\end{figure}

\pagebreak

\begin{figure}[!htbp]
\centering
\includegraphics[angle=0,width=6.5in]{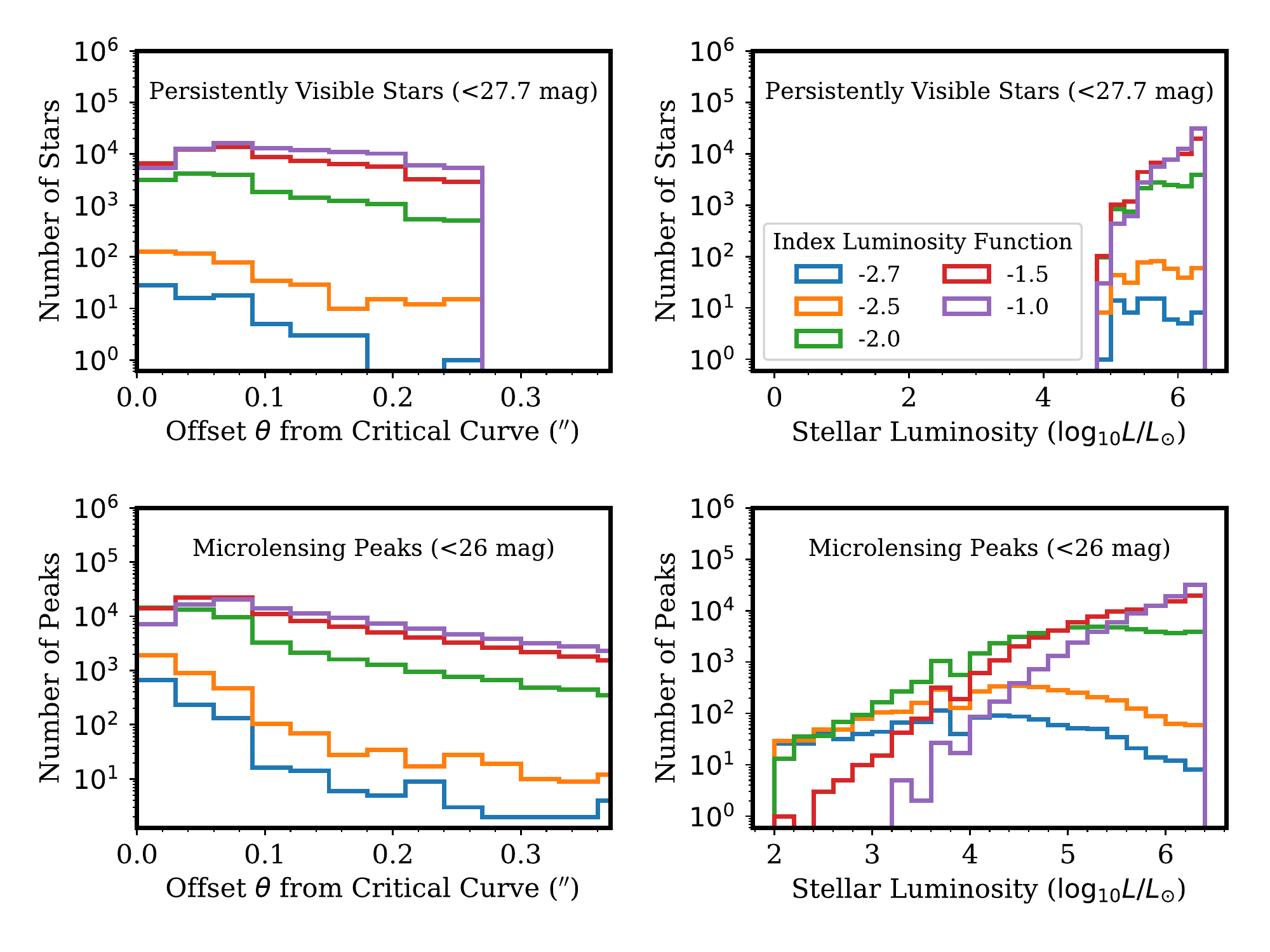}
\caption{Offsets from critical curve and luminosities of lensed stars for different luminosity functions from 10$^6$ Monte Carlo simulations.
We use the surface brightness (\sbalongarc) measured along the $0.2''$-wide arc to constrain the normalization of the stellar luminosity function, and then Poisson statistics to populate the source plane. The lower luminosity limit used for these simulations is 10\,L$_{\odot}$. Upper panels show that stars with $F125W\leq$\persistentmag over a period lasting many years should only appear within $\sim0.15''$ of the critical curve, and have luminosities of $\gtrsim10^{5.4}$\,L$_{\odot}$.
Lower panels show expected offset distribution of bright microlensing peaks ($F125W\leq26$\,mag) to $0.4''$, and of the luminosities of lensed stars. A stellar luminosity function similar to that measured in nearby galaxies (\bresolinlf) yields fewer events with less-luminous stars \cite{bresolinkennicuttferrarese98}. Left panels indicate the offset $\theta$ in arcseconds of stars from the critical curve, while right panels show the intrinsic luminosity of the stars in units of the solar luminosity (L$_{\odot}$).  
\label{fig:montecarlo}}
\end{figure}

\pagebreak

\begin{figure}[!htbp]
\centering
\includegraphics[angle=0,width=6.5in]{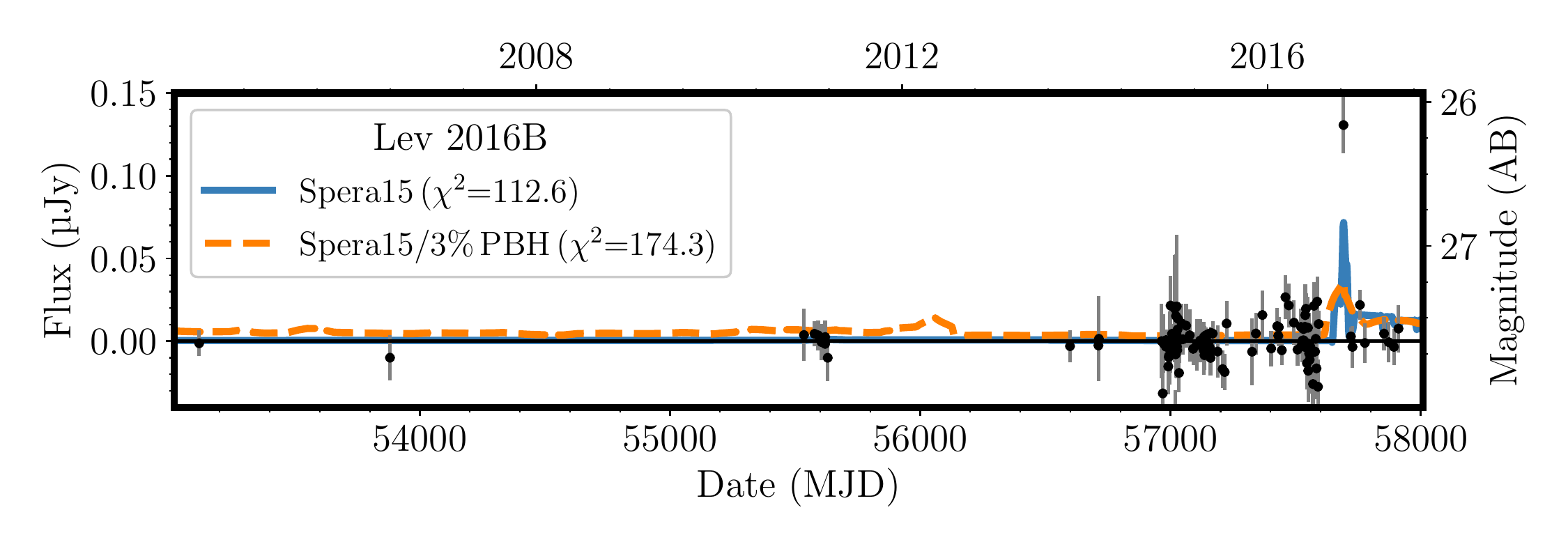}
\includegraphics[angle=0,width=6.5in]{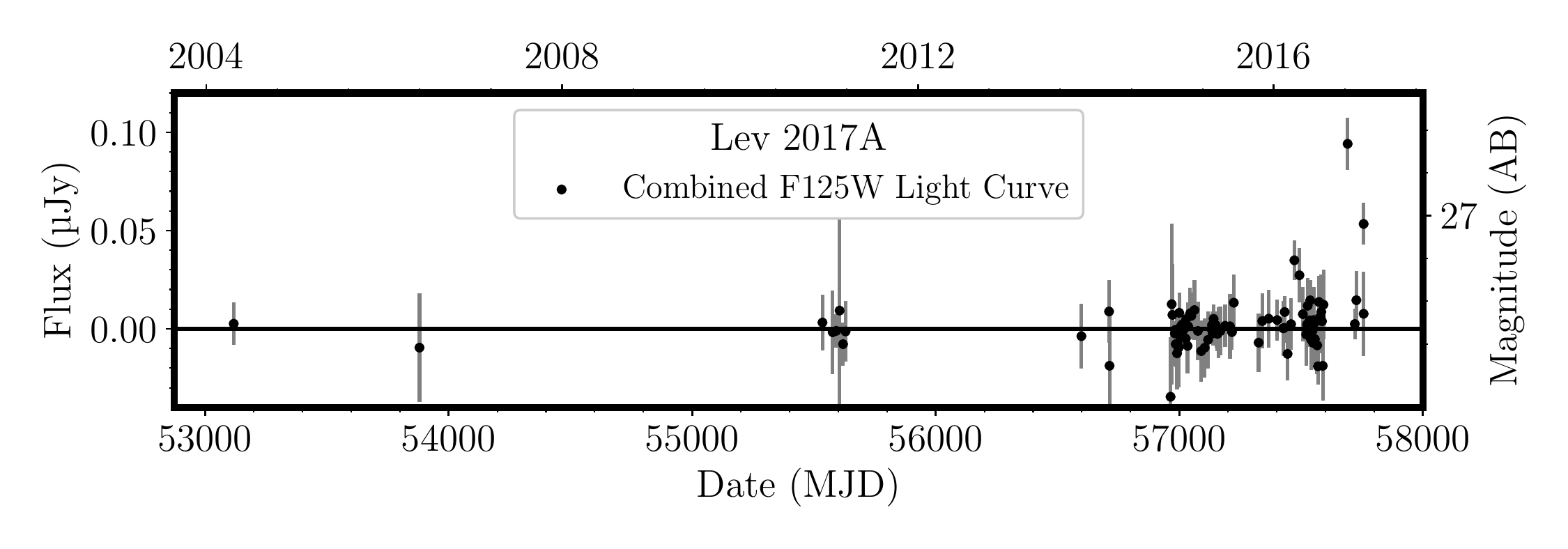}
\caption{Light curve at the positions of Lev16B (upper panel) detected on \iapyxdate\ and potential event Lev 2017A (lower panel) detected on \perdixdate. Fluxes measured through all wide-band {\it HST} filters are converted to $F125W$ using LS1's SED. Lev 2017A is only offset from Lev16B by \perdixiapyxoffset, so flux measurements at their positions are correlated. The first (higher) peak in Lev 2017A's light curve plotted here corresponds to flux from Lev16B. 
\label{fig:lightcurves_Lev_2016B_Lev_2017A}}
\end{figure}

\pagebreak

\begin{figure}[!htbp]
\centering
\includegraphics[angle=0,width=6.5in]{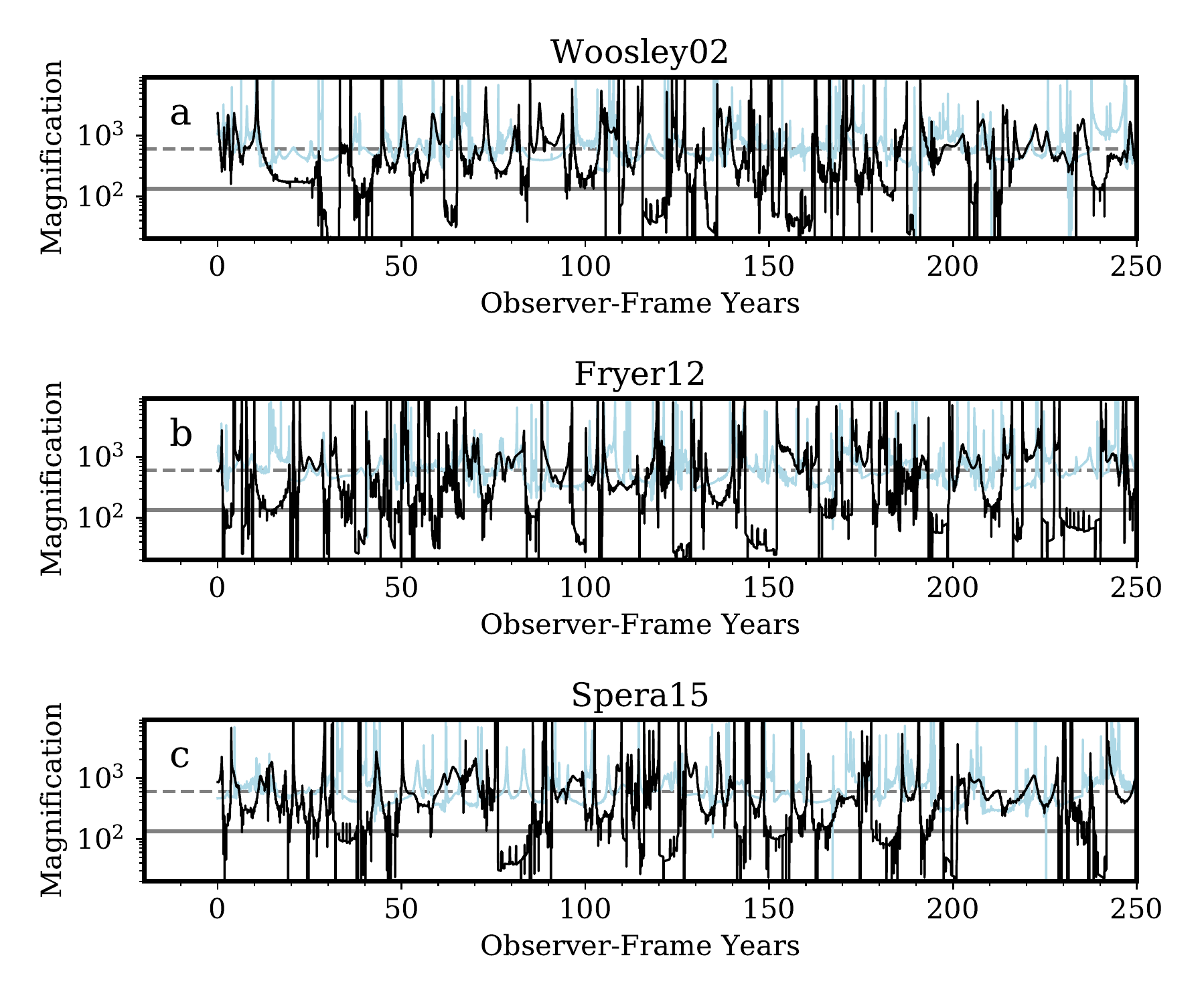}
\caption{Simulated light curves of a star at $\theta=0.13''$ from cluster critical curve for three stellar evolution and core-collapse models.  The mass functions are constructed using a Chabrier IMF and a prescription for binary fractions and mass ratios at low redshift \cite{duchenekraus13}.
The stellar evolution models used to determine the initial--final mass function for each star include the solar-metallicity, single stellar evolution models (Woosley02) \cite{woosleyhegerweaver02}, as well as single stars at subsolar metallicity ($Z=0.3\,{\rm Z}_{\odot}$; $Z=0.006$) where BHs with masses up to 30\,M$_{\odot}$ form from the collapse of massive stars (Fryer12) \cite{fryerbelczynskiwiktorowicz12}, and stars having initial masses greater than $\sim33$\,M$_{\odot}$ become BHs with masses within the range 20--50\,M$_{\odot}$ (Spera15) \cite{speramapellibressan15}. 
The Fryer12 and Spera15 models contain greater numbers of BH remnants, which may yield a higher frequency of decade-long intervals with low magnification.
\label{fig:binarysims}}
\end{figure}

\pagebreak

\begin{figure}[!htbp]
\centering
\includegraphics[angle=0,width=6.4in]{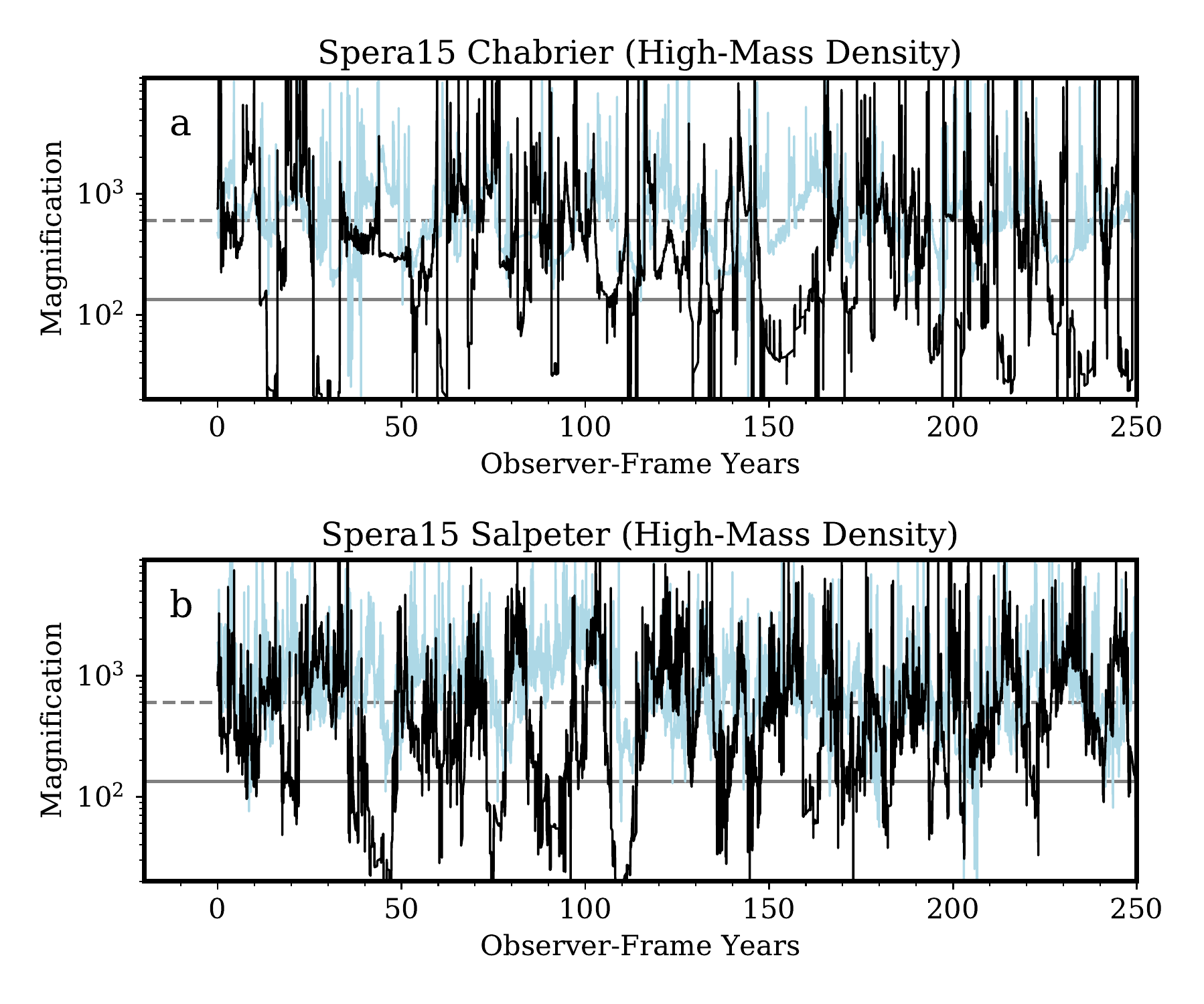}
\caption{Simulated light curves of a star at $\theta=0.13''$ from cluster critical curve for Chabrier (upper panel) and 
Salpeter (lower panel) IMFs.
The simulated light curve constructed from the model with a Salpeter IMF yields a greater number of peaks.
Plotted models are constructed using a prescription for binary fractions and mass ratios at low redshift \cite{duchenekraus13}.
The stellar-mass densities are \chabrierarcsecdensity\ and \salpeterarcsecdensity\ for the upper and lower plots, respectively,
and are the best-fitting values to the SED of the ICL for stellar-population synthesis models constructed using Chabrier and Salpeter IMFs.
\label{fig:imfs}}
\end{figure}

\pagebreak

\begin{figure}[!htbp]
\centering
\includegraphics[angle=0,width=6.4in]{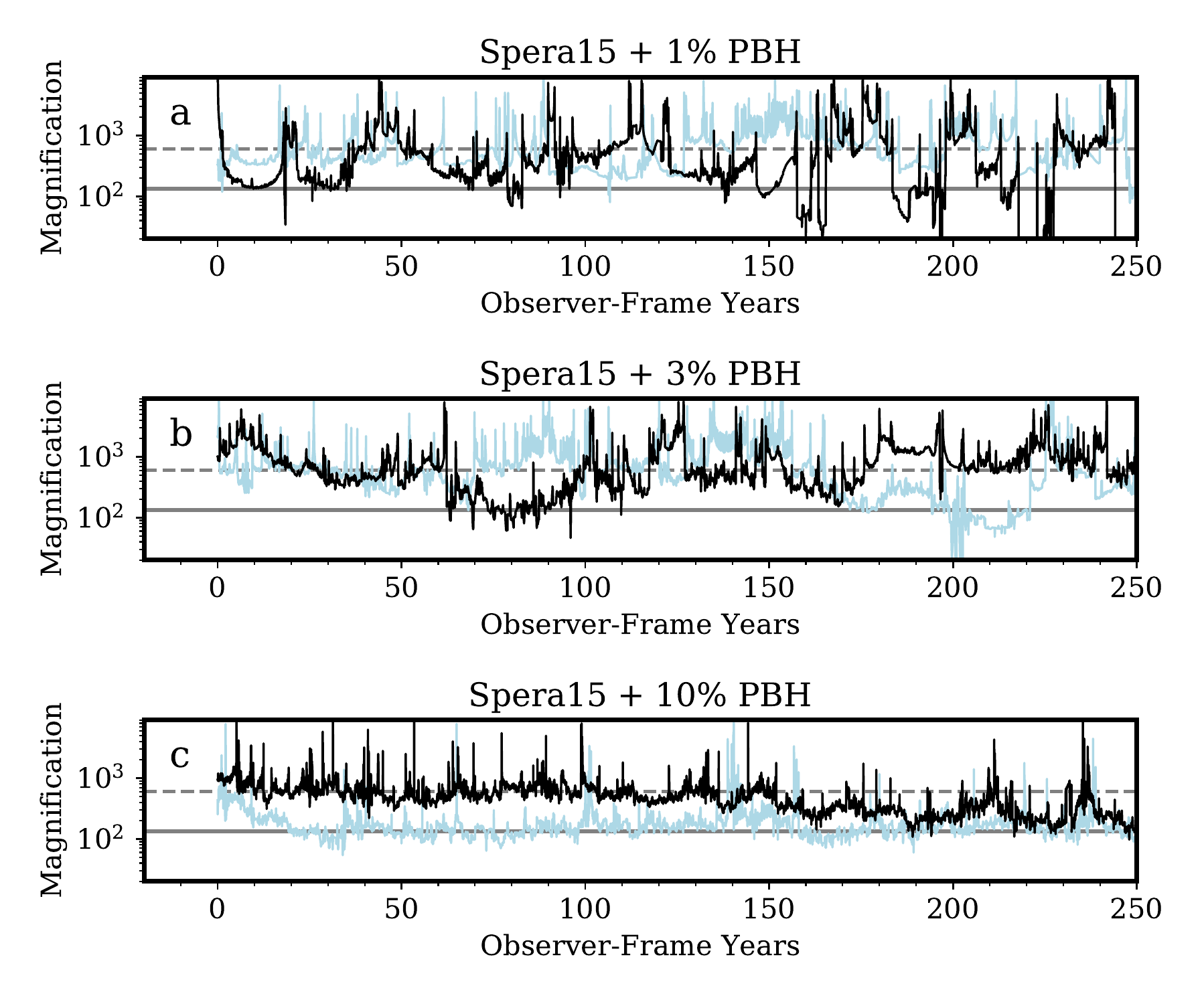}
\caption{Effect of increasing abundance of 30\,M$_{\odot}$ BHs on the simulated light curves of star at $\theta=0.13''$.  
Replacing 1\% (upper panel), 3\% (middle), and 10\% (bottom) of smooth DM with 30\,M$_{\odot}$ BHs yields light curves where the average magnification varies on an increasingly long timescales. 
An extended period of low magnification for one of the pair of images could help to explain why only a single image, LS1 / Lev16A, is persistently visible in {\it HST} imaging. 
\label{fig:addingpbhs}}
\end{figure}

\pagebreak

\begin{figure}[!htbp]
\centering
\includegraphics[angle=0,width=3.1in]{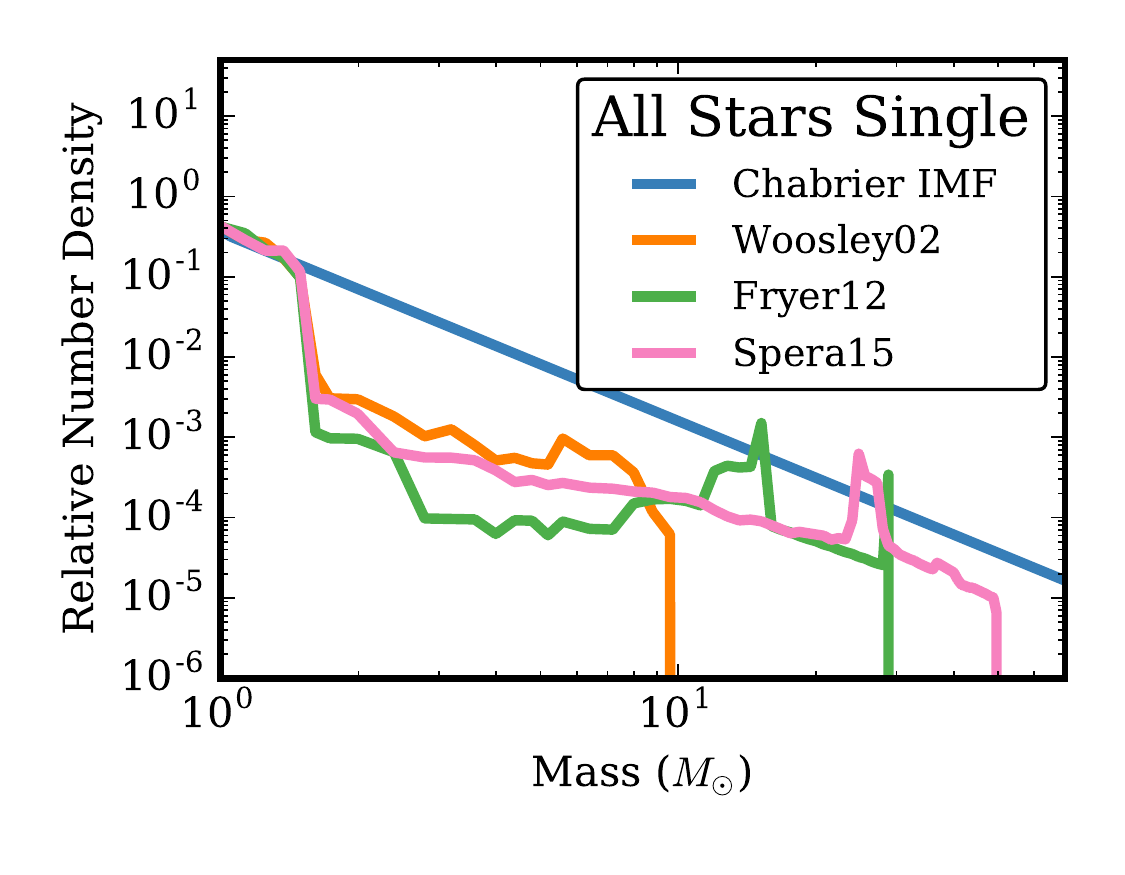}
\includegraphics[angle=0,width=3.1in]{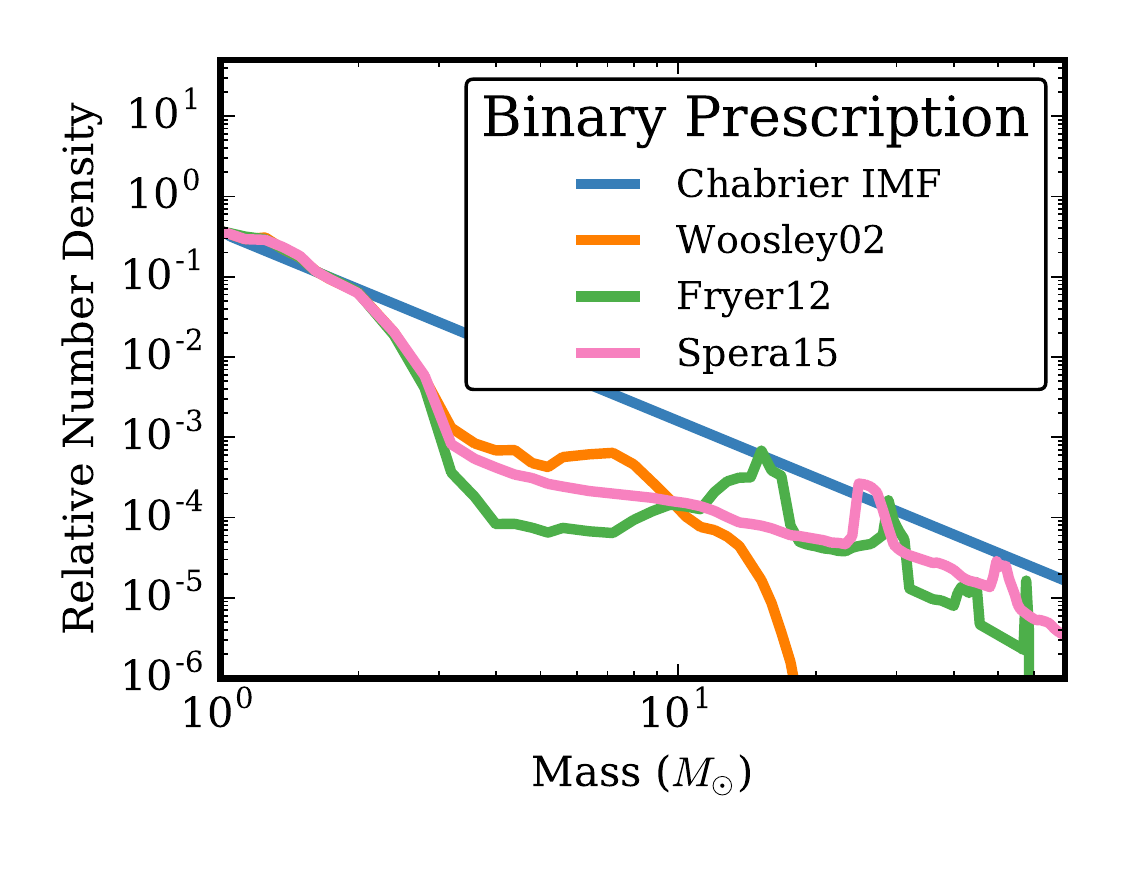}
\caption{Differences among the mass distributions of surviving stars and stellar remnants (i.e., white dwarf stars, neutron stars, and BHs) for the Woosley02, Fryer12, and Spera15 stellar evolution models.  Left panel plots the mass distributions assuming no stars have companions, and right panel shows mass functions assuming the mass-dependent binary fractions and mass ratios measured in the nearby universe \cite{duchenekraus13}.
\label{fig:singlebinary}}
\end{figure}

\pagebreak

\begin{figure}[!htbp]
\centering
\includegraphics[draft=\draft,angle=0,width=4.5in]{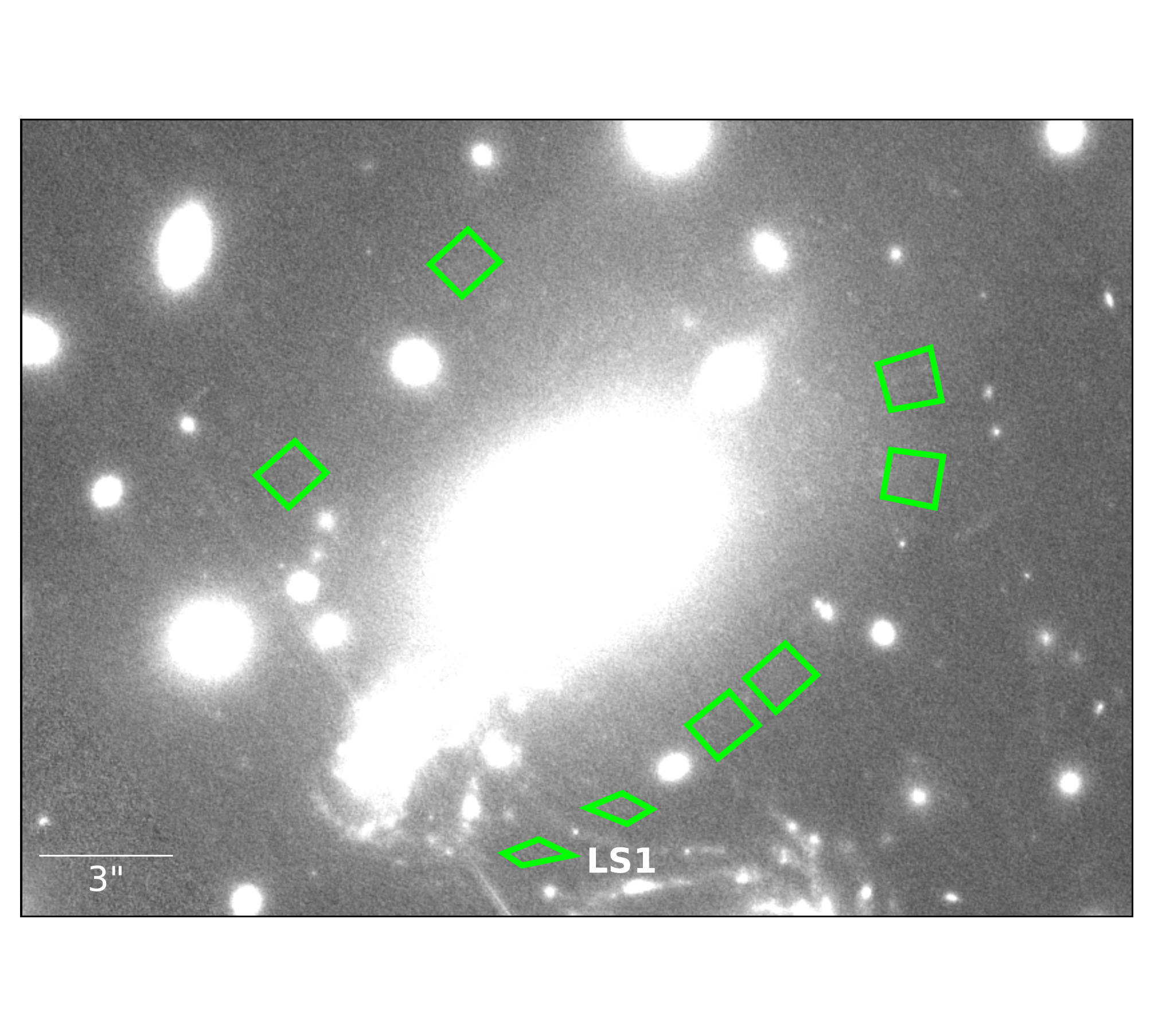} 
\caption{Apertures used for revised estimate of galaxy-cluster stellar-mass density along line-of-sight to LS1.
We first estimate the mean $M_{\ast}/L$ across all eight apertures using stellar-population synthesis models.
We next multiply the average $F140W$ surface brightness in the two apertures adjacent to LS1 by
the mean value of $M_{\ast}/L$ to estimate the stellar mass density along the line of sight to LS1.
\label{fig:icl_apertures}}
\end{figure}

\pagebreak

\begin{figure}[!htbp]
\centering
\includegraphics[angle=0,width=4.0in]{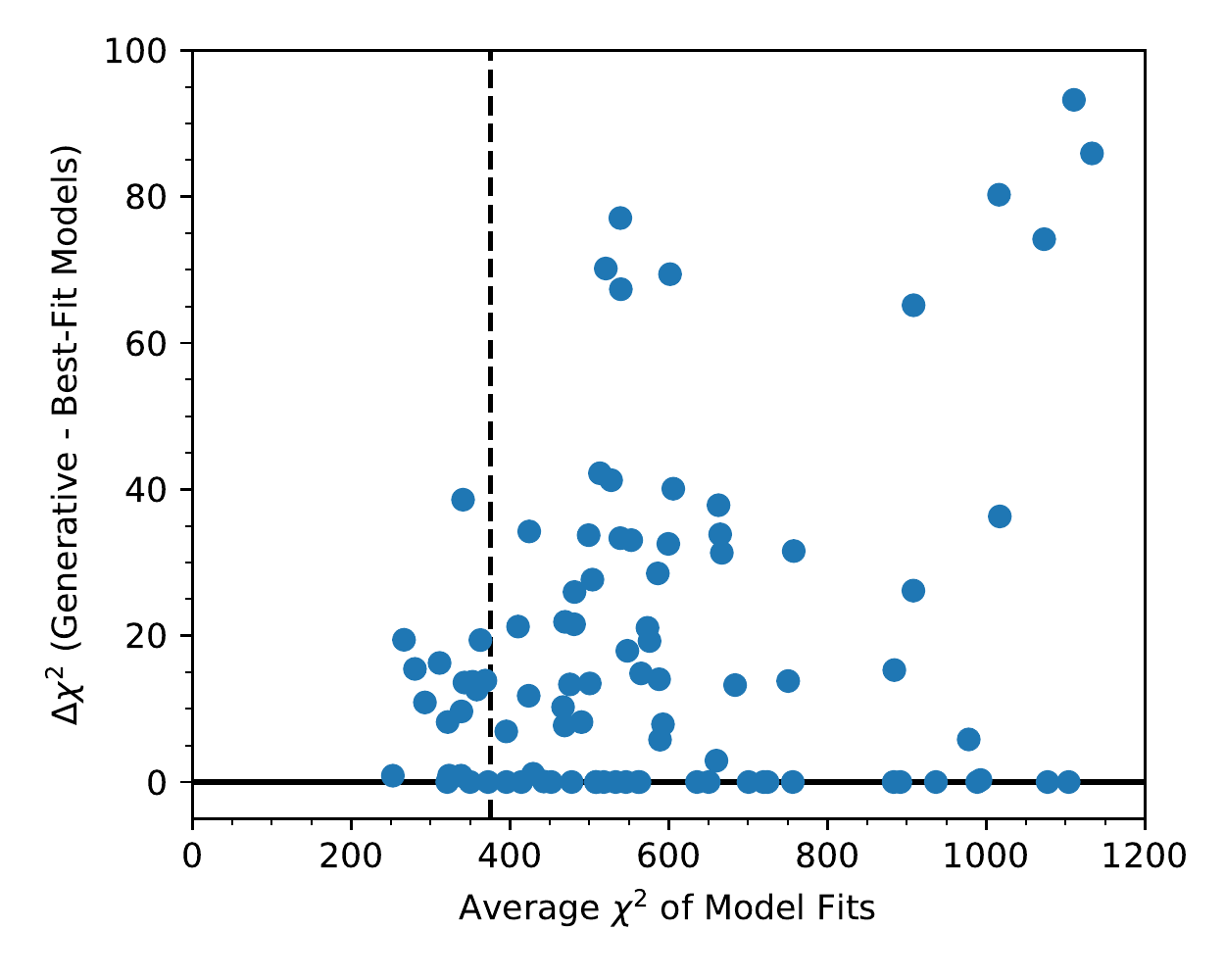}
\caption{Confidence intervals for $\langle \chi^2 \rangle$ statistics determined using simulated light curves.
For the models listed in Table~\ref{tab:chisq}, we generate simulated light curves for stars with $M_V = \{-8, -9, -10\}$, and fit them, allowing the lensed star to have an absolute magnitude within the range $-7.5 < M_V < -9.5$. 
The dashed black vertical shows the average of the $\langle \chi^2 \rangle$ statistics for the Table~\ref{tab:chisq} models for LS 1/ Lev16A and Lev16B.  
For all simulated light curves where the average $\langle \chi^2 \rangle$ value is within 100 of the vertical dashed line, we calculate the difference $\Delta$$\langle\chi^2\rangle$ values between the $\langle\chi^2\rangle$ values of the generative (``true'') model and of the best-fitting model.
For 68\% of simulated light curves, $\Delta$$\langle\chi^2\rangle \lesssim 13$, and, for 95\% of simulated light curves, $\Delta$$\langle\chi^2\rangle \lesssim 25$.
\label{fig:simstats}}
\end{figure}

\pagebreak

\begin{figure}[!htbp]
\centering
\includegraphics[angle=0,width=6.5in]{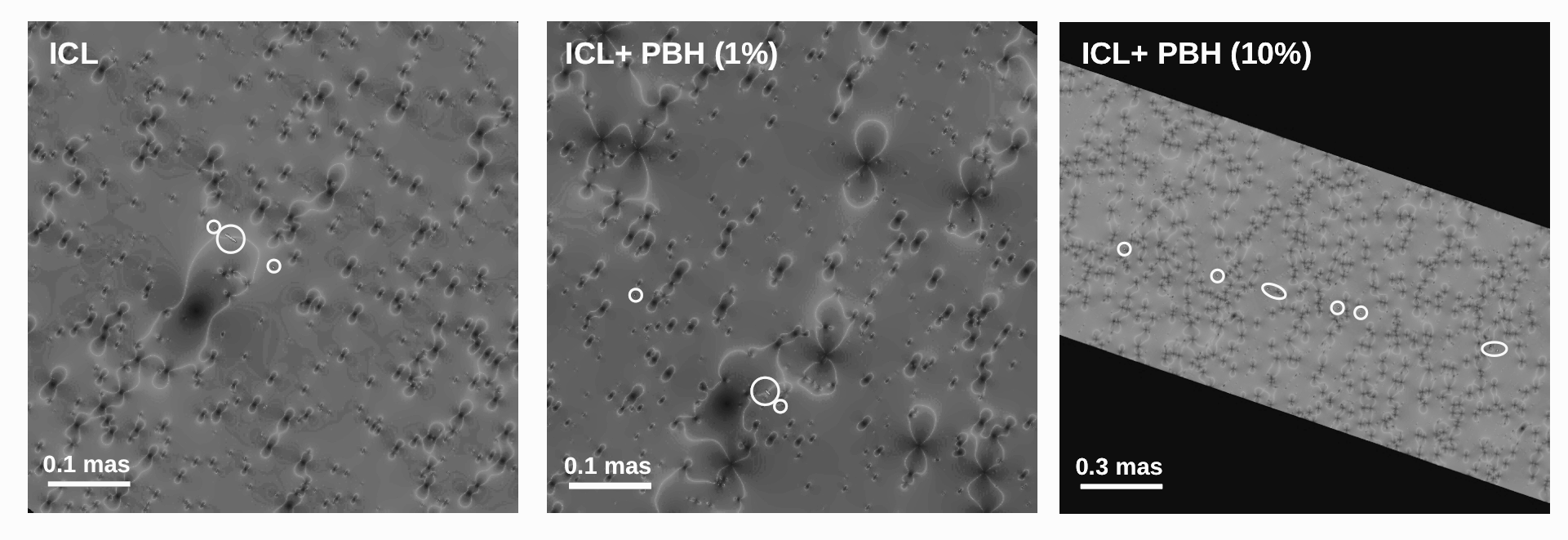}
\caption{Example of trains of multiple counterimages of a single background star as it traverses the region near the galaxy-cluster caustic. 
In general, there is a single counterimage per microlens \cite{paczynskilargeoptdepth86}, but most images have magnifications of order unity or even less and are not detectable. We show example of trains without PBHs (left panel), where 30\,M$_{\odot}$ PBHs account for 1\% of DM (middle panel), and where 30\,M$_{\odot}$ PBHs account for 10\% of DM (right panel). Increasing the PBH abundance yields more extended trains, although their extent ($\sim3$\,milliarcsec) when PBHs account for 10\% would be too small to detect in {\it HST} imaging. The simulation shown is of a 1000\,R$_{\odot}$ star whose image appears at an offset of $0.13''$ from the cluster critical curve. Near peak magnification, the star appears as a ``train'' of counterimages. Each cirlce in right panel encloses one image, and each ellipse encloses a set of two or three closely spaced images, in the train. The sizes of the circles and ellipses indicate the magnification of each image or set of images, respectively. In addition to replacing fractions of cluster DM with PBHs, we have populated the lens plane with stars and compact object remnants to match the mass density in surviving stars and remnants we infer for the ICM (\chabrierarcsecdensity).
\label{fig:train}}
\end{figure}

\pagebreak

\begin{longtable}[!htbp]{c c r r}
\multicolumn{1}{c|}{Date (MJD)} & \multicolumn{1}{c|}{Bandpass} & \multicolumn{1}{c}{Flux} & \multicolumn{1}{c}{$\sigma$} \\
\hline
\endhead
57538.38 & ACS F275W & -0.160 & 0.192 \\ 
57531.29 & ACS F336W & 0.348 & 0.242 \\ 
57538.33 & ACS F336W & 0.394 & 0.226 \\ 
57531.37 & ACS F475W & -0.005 & 0.098 \\ 
57531.46 & ACS F435W & -0.303 & 0.162 \\ 
57531.46 & ACS F435W & -0.033 & 0.184 \\ 
57524.28 & ACS F606W & 0.295 & 0.055 \\ 
57524.28 & ACS F606W & 0.329 & 0.044 \\ 
57531.71 & ACS F606W & 0.332 & 0.076 \\ 
57531.71 & ACS F606W & 0.438 & 0.054 \\ 
57534.29 & ACS F606W & 0.446 & 0.095 \\ 
57534.29 & ACS F606W & 0.522 & 0.063 \\ 
57536.10 & ACS F606W & 0.422 & 0.098 \\ 
57536.10 & ACS F606W & 0.470 & 0.068 \\ 
57537.09 & ACS F606W & 0.555 & 0.086 \\ 
57537.09 & ACS F606W & 0.588 & 0.070 \\ 
57524.38 & ACS F814W & 0.231 & 0.058 \\ 
57524.38 & ACS F814W & 0.304 & 0.038 \\ 
57531.49 & ACS F814W & 0.164 & 0.094 \\ 
57531.49 & ACS F814W & 0.189 & 0.062 \\ 
57524.47 & WFC3 F105W & 0.528 & 0.113 \\ 
57524.47 & WFC3 F105W & 0.681 & 0.095 \\ 
57524.47 & WFC3 F125W & 0.510 & 0.125 \\ 
57524.47 & WFC3 F125W & 0.624 & 0.097 \\ 
57524.61 & WFC3 F125W & 0.578 & 0.092 \\ 
57524.61 & WFC3 F125W & 0.669 & 0.085 \\ 
57527.18 & WFC3 F125W & 0.608 & 0.142 \\ 
57527.18 & WFC3 F125W & 0.748 & 0.122 \\ 
57532.02 & WFC3 F125W & 0.556 & 0.145 \\ 
57532.02 & WFC3 F125W & 0.696 & 0.138 \\ 
57534.27 & WFC3 F125W & 0.967 & 0.141 \\ 
57534.27 & WFC3 F125W & 1.174 & 0.129 \\ 
57536.06 & WFC3 F125W & 1.015 & 0.140 \\ 
57536.06 & WFC3 F125W & 1.211 & 0.124 \\ 
57537.05 & WFC3 F125W & 1.022 & 0.134 \\ 
57537.05 & WFC3 F125W & 1.216 & 0.128 \\ 
57538.31 & WFC3 F125W & 0.930 & 0.148 \\ 
57538.31 & WFC3 F125W & 1.154 & 0.121 \\ 
57524.54 & WFC3 F160W & 0.381 & 0.110 \\ 
57524.54 & WFC3 F160W & 0.489 & 0.093 \\ 
57527.20 & WFC3 F160W & 0.320 & 0.153 \\ 
57527.20 & WFC3 F160W & 0.427 & 0.155 \\
\hline
\caption{\captionmaysixteenpeak}
\label{tab:maysixteenpeak}
\end{longtable}

\pagebreak

\begin{longtable}[!htbp]{c r r}
\multicolumn{1}{c|}{Bandpass} & \multicolumn{1}{c}{Flux} & \multicolumn{1}{c}{$\sigma$} \\
\hline
\endhead
ACS $F225W$ & -0.007 & 0.026 \\ 
ACS $F275W$ & -0.005 & 0.019 \\ 
ACS $F336W$ & 0.019 & 0.010 \\ 
ACS $F435W$ & 0.024 & 0.005 \\ 
ACS $F606W$ & 0.050 & 0.006 \\ 
ACS $F814W$ & 0.072 & 0.003 \\ 
WFC3 $F105W$ & 0.143 & 0.006 \\ 
WFC3 $F125W$ & 0.141 & 0.008 \\ 
WFC3 $F140W$ & 0.113 & 0.004 \\ 
WFC3 $F160W$ & 0.127 & 0.011 \\ 
\hline
\caption{\captionunderlyingknot}
\label{tab:underlyingknot}
\end{longtable}

\pagebreak

\begin{longtable}[!htbp]{c r r}
\multicolumn{1}{c|}{Bandpass} & \multicolumn{1}{c}{Flux} & \multicolumn{1}{c}{$\sigma$} \\
\hline
\endhead
ACS $F225W$ & $-0.329$ & 0.011 \\ 
ACS $F275W$ & 0.110 & 0.054 \\ 
ACS $F336W$ & 0.100 & 0.027 \\ 
ACS $F435W$ & 0.089 & 0.016 \\ 
ACS $F606W$ & 0.098 & 0.014 \\ 
ACS $F814W$ & 0.091 & 0.008 \\ 
WFC3 $F105W$ & 0.095 & 0.008 \\ 
WFC3 $F125W$ & 0.097 & 0.009 \\ 
WFC3 $F140W$ & 0.097 & 0.011 \\ 
WFC3 $F160W$ & 0.107 & 0.010 \\
\hline
\caption{\captionadjacentarc}
\label{tab:adjacentarc}
\end{longtable}

\pagebreak

\begin{longtable}{c c r r r r r r}
\multicolumn{1}{c|}{Date} & \multicolumn{1}{c|}{Bandpass} & \multicolumn{2}{c|}{LS1/2016A} & \multicolumn{2}{c|}{2016B} & \multicolumn{2}{c}{2017A} \\
\multicolumn{1}{c|}{(MJD)} & \multicolumn{1}{c|}{} & \multicolumn{1}{c}{Flux} & \multicolumn{1}{c|}{$\sigma$} & \multicolumn{1}{c}{Flux} & \multicolumn{1}{c|}{$\sigma$} & \multicolumn{1}{c}{Flux} & \multicolumn{1}{c}{$\sigma$} \\
\hline
\endhead
55534.98 & WFC3 F125W & 0.079 [0.079] & 0.040 [0.040] & 0.019 & 0.055 & 0.021 & 0.044 \\
55629.90 & WFC3 F125W & 0.106 [0.106] & 0.045 [0.045] & -0.028 & 0.057 & 0.008 & 0.050 \\
56981.06 & WFC3 F125W & 0.148 [0.148] & 0.019 [0.019] & -0.009 & 0.014 & -0.014 & 0.017 \\
56982.12 & WFC3 F125W & 0.125 [0.125] & 0.020 [0.020] & -0.016 & 0.018 & -0.020 & 0.017 \\
56983.04 & WFC3 F125W & 0.131 [0.131] & 0.019 [0.019] & -0.017 & 0.018 & 0.006 & 0.014 \\
56983.31 & WFC3 F125W & 0.112 [0.112] & 0.014 [0.014] & -0.013 & 0.019 & -0.006 & 0.016 \\
56990.90 & WFC3 F125W & 0.198 [0.198] & 0.046 [0.046] & -0.038 & 0.055 & -0.017 & 0.056 \\
56992.95 & WFC3 F125W & 0.130 [0.130] & 0.059 [0.059] & -0.044 & 0.038 & -0.024 & 0.042 \\
56994.01 & WFC3 F125W & 0.146 [0.146] & 0.033 [0.033] & -0.029 & 0.047 & -0.045 & 0.051 \\
56996.73 & WFC3 F125W & 0.149 [0.149] & 0.045 [0.045] & 0.012 & 0.072 & -0.016 & 0.061 \\
56999.52 & WFC3 F125W & 0.150 [0.150] & 0.044 [0.044] & -0.028 & 0.049 & 0.015 & 0.044 \\
57000.11 & WFC3 F125W & 0.096 [0.096] & 0.073 [0.073] & 0.068 & 0.063 & 0.038 & 0.042 \\
57005.86 & WFC3 F125W & 0.148 [0.148] & 0.021 [0.021] & 0.012 & 0.016 & 0.005 & 0.012 \\
57019.94 & WFC3 F125W & 0.061 [0.061] & 0.076 [0.076] & -0.174 & 0.104 & -0.075 & 0.112 \\
57020.80 & WFC3 F125W & 0.006 [0.006] & 0.099 [0.099] & 0.005 & 0.088 & 0.080 & 0.079 \\
57021.80 & WFC3 F125W & 0.126 [0.126] & 0.093 [0.093] & 0.042 & 0.103 & -0.026 & 0.102 \\
57024.72 & WFC3 F125W & 0.100 [0.100] & 0.102 [0.102] & 0.058 & 0.119 & 0.010 & 0.044 \\
57025.78 & WFC3 F125W & 0.094 [0.094] & 0.082 [0.082] & -0.069 & 0.115 & -0.012 & 0.072 \\
57025.91 & WFC3 F125W & 0.055 [0.055] & 0.089 [0.089] & 0.034 & 0.104 & -0.007 & 0.098 \\
57026.90 & WFC3 F125W & 0.018 [0.018] & 0.080 [0.080] & -0.087 & 0.133 & -0.109 & 0.082 \\
57029.54 & WFC3 F125W & 0.151 [0.151] & 0.016 [0.016] & 0.037 & 0.015 & 0.029 & 0.014 \\
57033.94 & WFC3 F125W & 0.144 [0.144] & 0.044 [0.044] & -0.043 & 0.043 & -0.002 & 0.046 \\
57036.60 & WFC3 F125W & 0.109 [0.109] & 0.046 [0.046] & 0.010 & 0.050 & 0.045 & 0.041 \\
57044.69 & WFC3 F125W & 0.120 [0.120] & 0.036 [0.036] & 0.029 & 0.040 & 0.011 & 0.040 \\
57049.20 & WFC3 F125W & 0.099 [0.099] & 0.033 [0.033] & -0.006 & 0.030 & 0.023 & 0.039 \\
57062.36 & WFC3 F125W & 0.120 [0.120] & 0.040 [0.040] & 0.041 & 0.046 & 0.041 & 0.047 \\
57076.39 & WFC3 F125W & 0.111 [0.111] & 0.039 [0.039] & 0.001 & 0.055 & -0.011 & 0.047 \\
57090.39 & WFC3 F125W & 0.103 [0.103] & 0.037 [0.037] & -0.033 & 0.031 & -0.032 & 0.052 \\
57104.27 & WFC3 F125W & 0.132 [0.132] & 0.057 [0.057] & -0.017 & 0.050 & -0.028 & 0.045 \\
57118.14 & WFC3 F125W & 0.110 [0.110] & 0.040 [0.040] & -0.002 & 0.042 & -0.016 & 0.044 \\
57132.09 & WFC3 F125W & 0.144 [0.144] & 0.068 [0.068] & -0.027 & 0.071 & 0.001 & 0.079 \\
57149.06 & WFC3 F125W & 0.079 [0.079] & 0.055 [0.055] & 0.058 & 0.045 & 0.047 & 0.076 \\
57188.17 & WFC3 F125W & 0.148 [0.148] & 0.040 [0.040] & 0.004 & 0.049 & 0.009 & 0.031 \\
57216.20 & WFC3 F125W & 0.153 [0.153] & 0.049 [0.049] & -0.062 & 0.050 & 0.010 & 0.039 \\
57223.96 & WFC3 F125W & 0.095 [0.095] & 0.054 [0.054] & 0.030 & 0.044 & 0.033 & 0.044 \\
57325.82 & WFC3 F125W & 0.044 [0.044] & 0.060 [0.060] & -0.023 & 0.077 & 0.005 & 0.049 \\
57340.94 & WFC3 F125W & 0.109 [0.109] & 0.035 [0.035] & 0.039 & 0.043 & 0.026 & 0.050 \\
57367.04 & WFC3 F125W & 0.131 [0.131] & 0.051 [0.051] & 0.071 & 0.053 & 0.053 & 0.051 \\
57402.11 & WFC3 F125W & 0.212 [0.212] & 0.045 [0.045] & -0.027 & 0.034 & 0.011 & 0.039 \\
57426.21 & WFC3 F125W & 0.155 [0.155] & 0.037 [0.037] & 0.007 & 0.037 & -0.012 & 0.044 \\
57430.59 & WFC3 F125W & 0.195 [0.195] & 0.046 [0.046] & 0.024 & 0.038 & 0.049 & 0.037 \\
57444.57 & WFC3 F125W & 0.174 [0.174] & 0.033 [0.033] & 0.008 & 0.029 & -0.024 & 0.044 \\
57459.03 & WFC3 F125W & 0.243 [0.243] & 0.056 [0.056] & 0.107 & 0.046 & 0.063 & 0.057 \\
57472.53 & WFC3 F125W & 0.251 [0.251] & 0.049 [0.049] & 0.051 & 0.044 & 0.106 & 0.041 \\
57493.26 & WFC3 F125W & 0.323 [0.323] & 0.048 [0.048] & 0.018 & 0.044 & 0.072 & 0.041 \\
57507.53 & WFC3 F125W & 0.462 [0.462] & 0.033 [0.033] & -0.031 & 0.030 & 0.005 & 0.042 \\
57521.27 & WFC3 F125W & 0.328 [0.328] & 0.055 [0.055] & 0.010 & 0.036 & -0.047 & 0.052 \\
57524.47 & WFC3 F125W & 0.313 [0.313] & 0.056 [0.056] & -0.027 & 0.041 & 0.016 & 0.044 \\
57524.60 & WFC3 F125W & 0.399 [0.399] & 0.045 [0.045] & -0.020 & 0.044 & -0.013 & 0.049 \\
57524.61 & WFC3 F125W & 0.280 [0.280] & 0.041 [0.041] & -0.021 & 0.039 & 0.000 & 0.038 \\
57527.18 & WFC3 F125W & 0.345 [0.345] & 0.038 [0.038] & 0.010 & 0.046 & 0.031 & 0.044 \\
57532.02 & WFC3 F125W & 0.326 [0.326] & 0.048 [0.048] & 0.033 & 0.063 & 0.072 & 0.060 \\
57534.27 & WFC3 F125W & 0.563 [0.563] & 0.068 [0.068] & 0.001 & 0.071 & 0.027 & 0.056 \\
57536.06 & WFC3 F125W & 0.517 [0.517] & 0.064 [0.064] & -0.020 & 0.054 & 0.038 & 0.060 \\
57537.05 & WFC3 F125W & 0.516 [0.516] & 0.051 [0.051] & 0.051 & 0.054 & 0.039 & 0.048 \\
57538.31 & WFC3 F125W & 0.489 [0.489] & 0.066 [0.066] & 0.043 & 0.052 & 0.060 & 0.044 \\
57541.09 & WFC3 F125W & 0.233 [0.233] & 0.042 [0.042] & 0.068 & 0.028 & -0.011 & 0.049 \\
57545.07 & WFC3 F125W & 0.286 [0.286] & 0.071 [0.071] & -0.005 & 0.071 & 0.019 & 0.055 \\
57547.05 & WFC3 F125W & 0.214 [0.214] & 0.057 [0.057] & -0.002 & 0.049 & -0.004 & 0.048 \\
57549.00 & WFC3 F125W & 0.161 [0.161] & 0.062 [0.062] & 0.036 & 0.070 & -0.013 & 0.047 \\
57550.04 & WFC3 F125W & 0.141 [0.141] & 0.057 [0.057] & -0.034 & 0.063 & -0.019 & 0.035 \\
57551.67 & WFC3 F125W & 0.130 [0.130] & 0.071 [0.071] & 0.022 & 0.042 & 0.048 & 0.054 \\
57553.73 & WFC3 F125W & 0.204 [0.204] & 0.071 [0.071] & 0.018 & 0.087 & 0.023 & 0.053 \\
57555.91 & WFC3 F125W & 0.182 [0.182] & 0.044 [0.044] & -0.040 & 0.057 & 0.017 & 0.057 \\
57557.50 & WFC3 F125W & 0.100 [0.100] & 0.069 [0.069] & -0.007 & 0.056 & -0.042 & 0.055 \\
57566.20 & WFC3 F125W & 0.159 [0.159] & 0.055 [0.055] & 0.006 & 0.053 & 0.006 & 0.051 \\
57569.25 & WFC3 F125W & 0.181 [0.181] & 0.055 [0.055] & -0.070 & 0.047 & -0.051 & 0.026 \\
57573.22 & WFC3 F125W & 0.146 [0.146] & 0.041 [0.041] & 0.090 & 0.046 & 0.072 & 0.046 \\
57577.71 & WFC3 F125W & 0.170 [0.170] & 0.043 [0.043] & -0.011 & 0.053 & 0.017 & 0.034 \\
57580.18 & WFC3 F125W & 0.177 [0.177] & 0.053 [0.053] & 0.067 & 0.059 & 0.053 & 0.066 \\
57583.09 & WFC3 F125W & 0.168 [0.168] & 0.068 [0.068] & -0.039 & 0.072 & 0.017 & 0.072 \\
57586.01 & WFC3 F125W & 0.252 [0.252] & 0.060 [0.060] & 0.069 & 0.045 & 0.031 & 0.054 \\
57589.19 & WFC3 F125W & 0.232 [0.232] & 0.058 [0.058] & -0.053 & 0.059 & -0.047 & 0.060 \\
57592.04 & WFC3 F125W & 0.111 [0.111] & 0.071 [0.071] & 0.055 & 0.056 & 0.034 & 0.059 \\
57691.20 & WFC3 F125W & 0.257 [0.257] & 0.055 [0.055] & 0.391 & 0.055 & 0.260 & 0.040 \\
57720.81 & WFC3 F125W & 0.103 [0.103] & 0.038 [0.038] & 0.008 & 0.041 & 0.006 & 0.023 \\
57727.04 & WFC3 F125W & 0.155 [0.155] & 0.038 [0.038] & -0.001 & 0.038 & 0.046 & 0.046 \\
57756.90 & WFC3 F125W & 0.159 [0.159] & 0.037 [0.037] & 0.083 & 0.033 & 0.147 & 0.029 \\
55591.70 & WFC3 F105W & 0.031 [0.032] & 0.035 [0.037] & 0.009 & 0.032 & 0.011 & 0.028 \\
55619.67 & WFC3 F105W & 0.139 [0.145] & 0.045 [0.048] & 0.014 & 0.036 & -0.021 & 0.040 \\
56711.48 & WFC3 F105W & 0.297 [0.311] & 0.063 [0.067] & 0.011 & 0.068 & 0.018 & 0.065 \\
56711.94 & WFC3 F105W & 0.279 [0.291] & 0.094 [0.100] & -0.044 & 0.059 & -0.007 & 0.088 \\
56713.63 & WFC3 F105W & 0.260 [0.272] & 0.088 [0.094] & 0.004 & 0.071 & -0.052 & 0.077 \\
56964.16 & WFC3 F105W & 0.180 [0.188] & 0.078 [0.083] & 0.000 & 0.062 & -0.095 & 0.078 \\
56968.88 & WFC3 F105W & 0.138 [0.145] & 0.134 [0.143] & -0.091 & 0.105 & 0.034 & 0.106 \\
56972.06 & WFC3 F105W & 0.207 [0.217] & 0.107 [0.114] & 0.077 & 0.137 & 0.172 & 0.138 \\
56982.31 & WFC3 F105W & 0.109 [0.114] & 0.028 [0.029] & -0.014 & 0.032 & -0.004 & 0.029 \\
57002.87 & WFC3 F105W & 0.137 [0.143] & 0.014 [0.015] & 0.000 & 0.018 & -0.014 & 0.017 \\
57006.98 & WFC3 F105W & 0.134 [0.140] & 0.016 [0.017] & 0.003 & 0.015 & 0.017 & 0.019 \\
57011.89 & WFC3 F105W & 0.123 [0.129] & 0.015 [0.016] & -0.000 & 0.013 & 0.005 & 0.019 \\
57014.88 & WFC3 F105W & 0.157 [0.164] & 0.015 [0.016] & 0.014 & 0.019 & 0.009 & 0.018 \\
57015.81 & WFC3 F105W & 0.090 [0.094] & 0.016 [0.017] & 0.006 & 0.015 & -0.008 & 0.009 \\
57017.80 & WFC3 F105W & 0.126 [0.132] & 0.014 [0.015] & 0.000 & 0.017 & 0.002 & 0.014 \\
57020.58 & WFC3 F105W & 0.125 [0.130] & 0.013 [0.014] & -0.027 & 0.012 & -0.012 & 0.013 \\
57023.77 & WFC3 F105W & 0.135 [0.141] & 0.016 [0.017] & 0.015 & 0.016 & 0.014 & 0.014 \\
57025.56 & WFC3 F105W & 0.119 [0.125] & 0.012 [0.013] & -0.013 & 0.013 & -0.001 & 0.015 \\
57026.49 & WFC3 F105W & 0.139 [0.146] & 0.013 [0.014] & -0.015 & 0.012 & -0.016 & 0.016 \\
57027.81 & WFC3 F105W & 0.123 [0.128] & 0.014 [0.015] & 0.023 & 0.010 & 0.008 & 0.009 \\
57132.10 & WFC3 F105W & 0.118 [0.124] & 0.064 [0.068] & -0.014 & 0.069 & 0.008 & 0.061 \\
57149.07 & WFC3 F105W & 0.138 [0.145] & 0.048 [0.051] & 0.046 & 0.067 & 0.022 & 0.057 \\
57168.28 & WFC3 F105W & 0.070 [0.073] & 0.019 [0.021] & 0.012 & 0.022 & -0.006 & 0.034 \\
57208.06 & WFC3 F105W & 0.068 [0.071] & 0.054 [0.057] & -0.047 & 0.033 & 0.004 & 0.047 \\
57216.28 & WFC3 F105W & 0.129 [0.135] & 0.043 [0.046] & -0.029 & 0.042 & -0.000 & 0.034 \\
57430.59 & WFC3 F105W & 0.183 [0.192] & 0.030 [0.032] & 0.005 & 0.019 & -0.023 & 0.024 \\
57432.75 & WFC3 F105W & 0.157 [0.164] & 0.023 [0.025] & 0.042 & 0.019 & 0.028 & 0.022 \\
57524.47 & WFC3 F105W & 0.367 [0.384] & 0.031 [0.033] & 0.005 & 0.035 & -0.002 & 0.025 \\
55535.00 & WFC3 F160W & 0.000 [0.000] & 0.078 [0.100] & -0.005 & 0.072 & -0.042 & 0.068 \\
55577.06 & WFC3 F160W & 0.062 [0.078] & 0.063 [0.082] & 0.004 & 0.059 & -0.003 & 0.058 \\
55619.12 & WFC3 F160W & 0.099 [0.125] & 0.059 [0.076] & -0.052 & 0.062 & -0.057 & 0.061 \\
55629.85 & WFC3 F160W & 0.116 [0.146] & 0.058 [0.075] & -0.035 & 0.052 & -0.036 & 0.063 \\
56598.14 & WFC3 F160W & 0.061 [0.077] & 0.026 [0.033] & -0.010 & 0.027 & -0.010 & 0.035 \\
56990.91 & WFC3 F160W & 0.126 [0.159] & 0.086 [0.111] & -0.067 & 0.087 & -0.104 & 0.088 \\
56992.97 & WFC3 F160W & 0.038 [0.048] & 0.096 [0.124] & 0.002 & 0.048 & -0.004 & 0.108 \\
56994.03 & WFC3 F160W & 0.076 [0.095] & 0.033 [0.042] & -0.016 & 0.066 & 0.041 & 0.061 \\
56996.75 & WFC3 F160W & 0.075 [0.094] & 0.058 [0.075] & -0.086 & 0.113 & -0.080 & 0.106 \\
56999.56 & WFC3 F160W & 0.048 [0.060] & 0.067 [0.087] & 0.052 & 0.085 & 0.036 & 0.083 \\
57000.13 & WFC3 F160W & 0.146 [0.185] & 0.061 [0.079] & 0.057 & 0.081 & -0.048 & 0.081 \\
57002.89 & WFC3 F160W & 0.118 [0.149] & 0.022 [0.028] & 0.002 & 0.023 & -0.003 & 0.019 \\
57007.00 & WFC3 F160W & 0.111 [0.140] & 0.023 [0.030] & -0.021 & 0.019 & -0.011 & 0.014 \\
57011.91 & WFC3 F160W & 0.124 [0.157] & 0.023 [0.030] & 0.032 & 0.023 & 0.008 & 0.027 \\
57014.87 & WFC3 F160W & 0.020 [0.025] & 0.026 [0.034] & 0.005 & 0.031 & 0.025 & 0.027 \\
57015.82 & WFC3 F160W & 0.124 [0.157] & 0.025 [0.032] & 0.014 & 0.023 & 0.015 & 0.021 \\
57016.79 & WFC3 F160W & 0.072 [0.091] & 0.057 [0.074] & 0.069 & 0.089 & -0.034 & 0.108 \\
57017.82 & WFC3 F160W & 0.097 [0.122] & 0.017 [0.022] & 0.006 & 0.017 & -0.008 & 0.019 \\
57018.78 & WFC3 F160W & 0.076 [0.096] & 0.099 [0.128] & 0.073 & 0.084 & 0.041 & 0.105 \\
57019.64 & WFC3 F160W & -0.082 [-0.103] & 0.094 [0.121] & -0.196 & 0.164 & -0.048 & 0.128 \\
57020.60 & WFC3 F160W & 0.187 [0.236] & 0.028 [0.037] & -0.014 & 0.028 & -0.028 & 0.028 \\
57020.94 & WFC3 F160W & -0.004 [-0.006] & 0.143 [0.185] & 0.230 & 0.134 & 0.160 & 0.166 \\
57023.78 & WFC3 F160W & 0.085 [0.107] & 0.020 [0.026] & 0.008 & 0.020 & 0.012 & 0.025 \\
57025.57 & WFC3 F160W & 0.108 [0.136] & 0.023 [0.029] & -0.007 & 0.013 & -0.013 & 0.023 \\
57026.50 & WFC3 F160W & 0.116 [0.147] & 0.020 [0.026] & -0.019 & 0.027 & 0.002 & 0.021 \\
57027.83 & WFC3 F160W & 0.114 [0.143] & 0.025 [0.032] & 0.004 & 0.026 & 0.008 & 0.019 \\
57033.96 & WFC3 F160W & 0.116 [0.147] & 0.067 [0.086] & -0.083 & 0.050 & -0.072 & 0.053 \\
57036.61 & WFC3 F160W & 0.082 [0.103] & 0.039 [0.051] & -0.012 & 0.061 & -0.092 & 0.048 \\
57044.71 & WFC3 F160W & 0.102 [0.129] & 0.057 [0.074] & 0.038 & 0.058 & 0.067 & 0.063 \\
57049.21 & WFC3 F160W & 0.108 [0.136] & 0.034 [0.044] & 0.034 & 0.048 & 0.007 & 0.047 \\
57062.40 & WFC3 F160W & 0.077 [0.097] & 0.074 [0.096] & -0.002 & 0.061 & -0.036 & 0.073 \\
57076.41 & WFC3 F160W & 0.117 [0.147] & 0.065 [0.083] & 0.030 & 0.071 & 0.025 & 0.067 \\
57090.42 & WFC3 F160W & 0.149 [0.188] & 0.074 [0.096] & 0.056 & 0.053 & -0.028 & 0.060 \\
57104.31 & WFC3 F160W & 0.072 [0.091] & 0.070 [0.091] & 0.027 & 0.080 & -0.017 & 0.079 \\
57118.22 & WFC3 F160W & 0.091 [0.115] & 0.074 [0.096] & 0.014 & 0.072 & -0.015 & 0.074 \\
57132.11 & WFC3 F160W & 0.029 [0.036] & 0.067 [0.086] & 0.001 & 0.081 & 0.065 & 0.091 \\
57149.08 & WFC3 F160W & 0.127 [0.161] & 0.086 [0.112] & 0.051 & 0.077 & -0.027 & 0.086 \\
57168.29 & WFC3 F160W & 0.059 [0.074] & 0.066 [0.085] & 0.026 & 0.057 & 0.021 & 0.089 \\
57188.19 & WFC3 F160W & 0.035 [0.045] & 0.096 [0.123] & -0.126 & 0.095 & -0.057 & 0.087 \\
57208.09 & WFC3 F160W & 0.094 [0.119] & 0.088 [0.113] & -0.069 & 0.094 & 0.002 & 0.079 \\
57216.22 & WFC3 F160W & 0.109 [0.138] & 0.052 [0.067] & -0.137 & 0.079 & -0.072 & 0.055 \\
57224.00 & WFC3 F160W & 0.019 [0.024] & 0.062 [0.080] & 0.036 & 0.070 & 0.053 & 0.072 \\
57325.84 & WFC3 F160W & 0.085 [0.108] & 0.064 [0.083] & -0.014 & 0.081 & -0.078 & 0.059 \\
57340.95 & WFC3 F160W & 0.125 [0.158] & 0.081 [0.105] & -0.044 & 0.058 & -0.011 & 0.047 \\
57367.06 & WFC3 F160W & 0.082 [0.104] & 0.075 [0.097] & 0.009 & 0.062 & -0.055 & 0.052 \\
57402.15 & WFC3 F160W & 0.155 [0.196] & 0.061 [0.078] & 0.044 & 0.061 & 0.013 & 0.033 \\
57426.23 & WFC3 F160W & 0.157 [0.198] & 0.071 [0.092] & 0.083 & 0.055 & 0.043 & 0.062 \\
57432.75 & WFC3 F160W & 0.147 [0.185] & 0.046 [0.059] & -0.074 & 0.042 & -0.015 & 0.052 \\
57444.61 & WFC3 F160W & 0.084 [0.106] & 0.059 [0.076] & -0.090 & 0.045 & -0.064 & 0.054 \\
57459.09 & WFC3 F160W & 0.237 [0.299] & 0.056 [0.072] & 0.023 & 0.059 & -0.032 & 0.036 \\
57472.54 & WFC3 F160W & 0.166 [0.210] & 0.064 [0.083] & 0.102 & 0.068 & 0.088 & 0.030 \\
57493.30 & WFC3 F160W & 0.229 [0.289] & 0.070 [0.090] & 0.079 & 0.069 & 0.094 & 0.074 \\
57507.57 & WFC3 F160W & 0.269 [0.339] & 0.062 [0.080] & 0.094 & 0.067 & 0.088 & 0.069 \\
57521.30 & WFC3 F160W & 0.253 [0.319] & 0.060 [0.077] & 0.071 & 0.055 & 0.089 & 0.063 \\
57524.54 & WFC3 F160W & 0.259 [0.327] & 0.047 [0.061] & -0.020 & 0.043 & 0.038 & 0.035 \\
57527.20 & WFC3 F160W & 0.244 [0.309] & 0.053 [0.069] & -0.018 & 0.058 & 0.037 & 0.064 \\
57541.13 & WFC3 F160W & 0.168 [0.212] & 0.054 [0.070] & -0.021 & 0.062 & -0.030 & 0.070 \\
57545.10 & WFC3 F160W & 0.301 [0.380] & 0.081 [0.105] & -0.070 & 0.055 & -0.022 & 0.075 \\
57547.09 & WFC3 F160W & 0.212 [0.268] & 0.075 [0.097] & -0.006 & 0.061 & 0.003 & 0.078 \\
57549.04 & WFC3 F160W & 0.139 [0.176] & 0.060 [0.077] & 0.009 & 0.075 & -0.029 & 0.071 \\
57550.10 & WFC3 F160W & 0.104 [0.131] & 0.077 [0.099] & -0.102 & 0.089 & -0.062 & 0.092 \\
57551.71 & WFC3 F160W & 0.176 [0.222] & 0.076 [0.098] & -0.103 & 0.071 & -0.084 & 0.068 \\
57553.79 & WFC3 F160W & 0.121 [0.153] & 0.107 [0.138] & -0.096 & 0.106 & -0.053 & 0.083 \\
57555.95 & WFC3 F160W & 0.188 [0.237] & 0.087 [0.112] & -0.010 & 0.091 & 0.011 & 0.078 \\
57557.53 & WFC3 F160W & 0.139 [0.175] & 0.043 [0.056] & -0.011 & 0.055 & -0.069 & 0.098 \\
57566.22 & WFC3 F160W & 0.169 [0.214] & 0.079 [0.103] & -0.074 & 0.066 & -0.072 & 0.052 \\
57569.26 & WFC3 F160W & 0.133 [0.168] & 0.067 [0.086] & -0.093 & 0.081 & -0.085 & 0.082 \\
57573.24 & WFC3 F160W & 0.155 [0.196] & 0.051 [0.066] & -0.036 & 0.077 & -0.021 & 0.047 \\
57577.74 & WFC3 F160W & 0.226 [0.286] & 0.077 [0.100] & -0.038 & 0.073 & 0.010 & 0.069 \\
57583.13 & WFC3 F160W & 0.163 [0.206] & 0.079 [0.102] & -0.066 & 0.073 & 0.032 & 0.060 \\
57586.05 & WFC3 F160W & -0.005 [-0.006] & 0.068 [0.088] & 0.059 & 0.113 & -0.038 & 0.064 \\
57589.20 & WFC3 F160W & 0.018 [0.023] & 0.073 [0.094] & -0.135 & 0.069 & -0.061 & 0.063 \\
57691.21 & WFC3 F160W & 0.207 [0.261] & 0.088 [0.113] & 0.352 & 0.088 & 0.258 & 0.065 \\
57727.07 & WFC3 F160W & 0.112 [0.142] & 0.056 [0.072] & -0.053 & 0.077 & 0.016 & 0.070 \\
55591.71 & WFC3 F140W & 0.072 [0.092] & 0.043 [0.054] & 0.016 & 0.040 & -0.026 & 0.030 \\
55619.65 & WFC3 F140W & 0.009 [0.011] & 0.038 [0.047] & -0.002 & 0.043 & -0.013 & 0.045 \\
56711.77 & WFC3 F140W & 0.208 [0.265] & 0.071 [0.090] & 0.022 & 0.063 & 0.048 & 0.056 \\
56971.93 & WFC3 F140W & 0.160 [0.204] & 0.059 [0.074] & -0.111 & 0.069 & -0.101 & 0.081 \\
56972.13 & WFC3 F140W & 0.261 [0.332] & 0.080 [0.100] & 0.109 & 0.081 & 0.107 & 0.109 \\
56981.85 & WFC3 F140W & 0.084 [0.107] & 0.016 [0.020] & 0.001 & 0.019 & -0.004 & 0.016 \\
56981.98 & WFC3 F140W & 0.110 [0.139] & 0.014 [0.018] & 0.019 & 0.011 & 0.014 & 0.012 \\
56982.32 & WFC3 F140W & 0.115 [0.147] & 0.019 [0.024] & -0.007 & 0.015 & -0.013 & 0.015 \\
56983.18 & WFC3 F140W & 0.127 [0.161] & 0.009 [0.011] & -0.004 & 0.014 & -0.012 & 0.011 \\
56984.84 & WFC3 F140W & 0.101 [0.128] & 0.013 [0.017] & -0.009 & 0.015 & 0.016 & 0.019 \\
55576.98 & ACS F606W & 0.000 [0.002] & 0.026 [0.109] & 0.047 & 0.023 & -0.008 & 0.023 \\
55619.52 & ACS F606W & -0.022 [-0.074] & 0.027 [0.113] & 0.002 & 0.020 & 0.016 & 0.028 \\
57149.51 & ACS F606W & 0.028 [0.097] & 0.006 [0.026] & -0.013 & 0.009 & -0.001 & 0.011 \\
57150.57 & ACS F606W & 0.042 [0.142] & 0.011 [0.046] & 0.046 & 0.008 & 0.008 & 0.007 \\
57151.49 & ACS F606W & 0.025 [0.086] & 0.008 [0.034] & -0.019 & 0.008 & -0.002 & 0.010 \\
57155.60 & ACS F606W & 0.035 [0.119] & 0.012 [0.048] & -0.063 & 0.009 & -0.012 & 0.010 \\
57161.36 & ACS F606W & 0.035 [0.118] & 0.009 [0.038] & 0.047 & 0.010 & 0.004 & 0.011 \\
57524.28 & ACS F606W & 0.108 [0.369] & 0.013 [0.053] & 0.027 & 0.016 & 0.007 & 0.015 \\
57531.40 & ACS F606W & 0.113 [0.387] & 0.028 [0.114] & 0.062 & 0.022 & -0.001 & 0.021 \\
57720.74 & ACS F606W & 0.014 [0.046] & 0.025 [0.103] & 0.025 & 0.021 & -0.014 & 0.031 \\
57720.89 & ACS F606W & 0.023 [0.078] & 0.025 [0.104] & -0.035 & 0.031 & -0.008 & 0.024 \\
53117.80 & ACS F814W & 0.038 [0.051] & 0.020 [0.031] & -0.005 & 0.021 & 0.007 & 0.019 \\
53880.47 & ACS F814W & 0.153 [0.203] & 0.035 [0.056] & -0.036 & 0.038 & -0.026 & 0.048 \\
57131.53 & ACS F814W & 0.074 [0.098] & 0.023 [0.036] & -0.031 & 0.020 & -0.007 & 0.021 \\
57132.59 & ACS F814W & 0.073 [0.096] & 0.018 [0.028] & 0.004 & 0.021 & -0.001 & 0.021 \\
57133.79 & ACS F814W & 0.057 [0.076] & 0.024 [0.038] & 0.010 & 0.025 & 0.014 & 0.020 \\
57134.65 & ACS F814W & 0.084 [0.112] & 0.020 [0.031] & 0.006 & 0.021 & 0.010 & 0.020 \\
57135.58 & ACS F814W & 0.067 [0.089] & 0.026 [0.041] & -0.031 & 0.025 & -0.010 & 0.022 \\
57137.09 & ACS F814W & 0.090 [0.120] & 0.017 [0.026] & 0.008 & 0.024 & -0.024 & 0.025 \\
57137.63 & ACS F814W & 0.081 [0.107] & 0.025 [0.040] & -0.010 & 0.017 & 0.004 & 0.017 \\
57137.83 & ACS F814W & 0.089 [0.118] & 0.021 [0.033] & 0.006 & 0.023 & -0.005 & 0.027 \\
57138.09 & ACS F814W & 0.046 [0.061] & 0.020 [0.031] & 0.012 & 0.025 & -0.011 & 0.021 \\
57140.34 & ACS F814W & 0.099 [0.132] & 0.020 [0.032] & 0.036 & 0.027 & 0.015 & 0.027 \\
57140.61 & ACS F814W & 0.076 [0.101] & 0.021 [0.033] & 0.009 & 0.021 & 0.011 & 0.024 \\
57141.60 & ACS F814W & 0.114 [0.152] & 0.016 [0.025] & 0.047 & 0.018 & 0.019 & 0.021 \\
57142.46 & ACS F814W & 0.073 [0.097] & 0.020 [0.031] & -0.010 & 0.024 & -0.009 & 0.012 \\
57143.39 & ACS F814W & 0.076 [0.101] & 0.021 [0.033] & 0.002 & 0.018 & 0.001 & 0.025 \\
57143.66 & ACS F814W & 0.067 [0.089] & 0.012 [0.019] & -0.001 & 0.020 & 0.003 & 0.022 \\
57149.49 & ACS F814W & 0.094 [0.124] & 0.022 [0.035] & -0.023 & 0.021 & 0.000 & 0.020 \\
57150.55 & ACS F814W & 0.096 [0.127] & 0.026 [0.041] & -0.000 & 0.022 & 0.002 & 0.015 \\
57151.48 & ACS F814W & 0.062 [0.082] & 0.028 [0.045] & 0.021 & 0.027 & -0.000 & 0.020 \\
57157.39 & ACS F814W & 0.118 [0.157] & 0.022 [0.035] & -0.014 & 0.027 & -0.005 & 0.018 \\
57159.84 & ACS F814W & 0.125 [0.166] & 0.026 [0.042] & -0.037 & 0.029 & -0.017 & 0.034 \\
57524.34 & ACS F814W & 0.271 [0.360] & 0.037 [0.059] & -0.017 & 0.037 & 0.015 & 0.043 \\
57524.41 & ACS F814W & 0.284 [0.377] & 0.049 [0.077] & -0.036 & 0.029 & 0.006 & 0.032 \\
57531.49 & ACS F814W & 0.246 [0.327] & 0.043 [0.068] & 0.029 & 0.053 & -0.033 & 0.057 \\
57720.76 & ACS F814W & 0.104 [0.139] & 0.079 [0.126] & 0.084 & 0.072 & 0.031 & 0.076 \\
57720.87 & ACS F814W & 0.062 [0.082] & 0.071 [0.113] & 0.056 & 0.057 & 0.028 & 0.067 \\
55605.27 & ACS F435W & 0.038 [0.267] & 0.028 [0.230] & -0.011 & 0.030 & 0.026 & 0.033 \\
55619.53 & ACS F435W & 0.011 [0.079] & 0.034 [0.282] & -0.096 & 0.051 & -0.009 & 0.040 \\
57131.54 & ACS F435W & 0.009 [0.061] & 0.013 [0.112] & -0.061 & 0.016 & -0.016 & 0.017 \\
57137.11 & ACS F435W & 0.013 [0.090] & 0.015 [0.124] & -0.087 & 0.017 & -0.010 & 0.011 \\
57138.10 & ACS F435W & 0.039 [0.271] & 0.020 [0.163] & -0.030 & 0.013 & 0.001 & 0.016 \\
57140.36 & ACS F435W & -0.015 [-0.102] & 0.018 [0.147] & -0.014 & 0.019 & 0.022 & 0.015 \\
57140.62 & ACS F435W & 0.011 [0.077] & 0.019 [0.156] & 0.066 & 0.016 & -0.011 & 0.014 \\
57141.62 & ACS F435W & 0.039 [0.271] & 0.020 [0.170] & -0.109 & 0.018 & -0.007 & 0.016 \\
57142.48 & ACS F435W & 0.011 [0.080] & 0.017 [0.142] & -0.026 & 0.018 & -0.024 & 0.018 \\
57143.41 & ACS F435W & 0.022 [0.154] & 0.016 [0.135] & 0.196 & 0.019 & 0.033 & 0.019 \\
57143.67 & ACS F435W & 0.029 [0.203] & 0.009 [0.073] & -0.114 & 0.016 & -0.005 & 0.015 \\
57531.46 & ACS F435W & 0.056 [0.391] & 0.059 [0.494] & 0.215 & 0.053 & 0.010 & 0.047 \\
56985.07 & WFC3 F606W & 0.064 [0.140] & 0.017 [0.036] & 0.005 & 0.024 & -0.005 & 0.014 \\
56985.80 & WFC3 F606W & 0.059 [0.129] & 0.017 [0.038] & -0.001 & 0.020 & -0.019 & 0.016 \\
57532.03 & WFC3 F606W & 0.104 [0.228] & 0.039 [0.085] & -0.009 & 0.026 & -0.015 & 0.029 \\
57534.29 & WFC3 F606W & 0.144 [0.314] & 0.026 [0.056] & -0.001 & 0.017 & 0.014 & 0.031 \\
57536.10 & WFC3 F606W & 0.170 [0.371] & 0.029 [0.062] & -0.013 & 0.030 & -0.007 & 0.034 \\
57537.09 & WFC3 F606W & 0.208 [0.453] & 0.032 [0.069] & 0.038 & 0.027 & 0.011 & 0.022 \\
57580.19 & WFC3 F606W & 0.099 [0.217] & 0.023 [0.051] & -0.059 & 0.041 & 0.009 & 0.025 \\
57592.07 & WFC3 F606W & 0.033 [0.072] & 0.034 [0.074] & 0.053 & 0.023 & 0.051 & 0.035 \\
57756.92 & WFC3 F606W & 0.042 [0.092] & 0.035 [0.075] & 0.060 & 0.039 & 0.031 & 0.031 \\
57776.68 & WFC3 F606W & 0.036 [0.079] & 0.030 [0.065] & -0.007 & 0.033 & -0.025 & 0.025 \\
57853.22 & WFC3 F606W & 0.047 [0.102] & 0.035 [0.075] & 0.028 & 0.028 & 0.008 & 0.024 \\
57872.02 & WFC3 F606W & 0.089 [0.193] & 0.035 [0.075] & -0.004 & 0.034 & -0.023 & 0.040 \\
57892.57 & WFC3 F606W & 0.083 [0.180] & 0.033 [0.072] & -0.021 & 0.030 & 0.014 & 0.034 \\
\hline
\caption{\captionphottab}
\label{tab:photometry}
\end{longtable}

\end{document}